\documentclass[useAMS,usenatbib]{mn2e}
\usepackage{natbib}
\usepackage{amsmath,amsfonts}
\usepackage{epsfig}
\usepackage{multirow}
\usepackage{mathptmx}
\usepackage{longtable}
\usepackage{graphicx}
\usepackage{subfig}
\usepackage{float}
\usepackage{lscape}
\usepackage{rotating}
\usepackage{pdflscape}
\usepackage[T1]{fontenc}
\usepackage{aecompl}

\def\reff@jnl#1{{\rm#1\/}}

\def\aj{\reff@jnl{AJ}}                  % Astronomical Journal
\def\araa{\reff@jnl{ARA\&A}}            % Annual Review of Astron and Astrophys
\def\apj{\reff@jnl{ApJ}}                        % Astrophysical Journal
\def\apjl{\reff@jnl{ApJ}}               % Astrophysical Journal, Letters
\def\apjs{\reff@jnl{ApJS}}              % Astrophysical Journal, Supplement
\def\ao{\reff@jnl{Appl.Optics}}         % Applied Optics
\def\apss{\reff@jnl{Ap\&SS}}            % Astrophysics and Space Science
\def\aap{\reff@jnl{A\&A}}                       % Astrophysical Journal
\def\apjl{\reff@jnl{ApJ}}               % Astronomy and Astrophysics
\def\aapr{\reff@jnl{A\&A~Rev.}}         % Astronomy and Astrophysics Reviews
\def\aaps{\reff@jnl{A\&AS}}             % Astronomy and Astrophysics, Supplement
\def\azh{\reff@jnl{AZh}}                        % Astronomicheskii Zhurnal
\def\baas{\reff@jnl{BAAS}}              % Bulletin of the AAS
\def\jrasc{\reff@jnl{JRASC}}            % Journal of the RAS of Canada
\def\memras{\reff@jnl{MmRAS}}           % Memoirs of the RAS
\def\mnras{\reff@jnl{MNRAS}}            % Monthly Notices of the RAS
\def\pra{\reff@jnl{Phys. Rev. A}}         % Physical Review A: General Physics
\def\prb{\reff@jnl{Phys. Rev. B}}         % Physical Review B: Solid State
\def\prc{\reff@jnl{Phys. Rev. C}}         % Physical Review C
\def\prd{\reff@jnl{Phys. Rev. D}}         % Physical Review D
\def\prl{\reff@jnl{Phys. Rev. Lett}}      % Physical Review Letter
\def\pasp{\reff@jnl{PASP}}              % Publications of the ASP
\def\pasj{\reff@jnl{PASJ}}              % Publications of the ASJ
\def\qjras{\reff@jnl{QJRAS}}            % Quarterly Journal of the RAS
\def\skytel{\reff@jnl{S\&T}}            % Sky and Telescope
\def\solphys{\reff@jnl{Solar~Phys.}}    % Solar Physics
\def\sovast{\reff@jnl{Soviet~Ast.}}     % Soviet Astronomy
\def\ssr{\reff@jnl{Space~Sci.Rev.}}     % Space Science Reviews
\def\zap{\reff@jnl{ZAp}}                        % Zeitschrift fuer Astrophysik
\def\nat{\reff@jnl{Nature}}             % Nature

\def\p#1by#2{{\partial{#1} \over \partial{#2}}}
\def\pp#1by#2#3{{\partial^2{#1} \over \partial{#2}\partial{#3}}}
\def\d#1by#2{{{\rm d}{#1} \over {\rm d}{#2}}}
\def\dd#1by#2#3{{{\rm d}^2{#1} \over {\rm d}{#2}{\rm d}{#3}}}

\def\HI{H\,{\sc i}}

\begin{document}
\title{The Arecibo Galaxy Environment Survey X: The Structure of Halo Gas Around M33}

\label{firstpage}

%\author[O.C.~Keenan, J.I.~Davies, R.~Taylor, R.F.~Minchin]{
 %O.C.~Keenan$^1$, J.I.~Davies$^1$, R.~Taylor$^2$, R.F.~Minchin$^3$}

%{\parbox{$^1$ School of Physics and Astronomy, Cardiff University, Queens Buildings, The Parade, %Cardiff, CF24 3AA, U.K. \\
%$^2$ Astronomical Institute of the ASCR, Bo\v{c}n\'{i} II 1401, 14100, Prague, Czech Republic \\
%$^3$ Arecibo Observatory, HC03 Box 53995, Arecibo, Puerto Rico 00612 }

\author[O.C. Keenan et al.] {O.C.~Keenan,$^1$ J.I.~Davies,$^1$ R.~Taylor,$^2$ R.F.~Minchin,$^3$\\
$^1$School of Physics and Astronomy, Cardiff University, Queens Buildings, The Parade, Cardiff, CF24 3AA, U.K.\\ 
$^2$Astronomical Institute of the ASCR, Bo\v{c}n\'{i} II 1401, 14100, Prague, Czech Republic\\ 
$^3$Arecibo Observatory, HC03 Box 53995, Arecibo, Puerto Rico 00612}

\date{Accepted ---; received ---; in original form \today}

\pagerange{\pageref{firstpage}--\pageref{lastpage}}

\pubyear{2014}

%------------------------------------------------------------------------------%
\maketitle

\begin{abstract}
As part of the \HI\ Arecibo Galaxy Environments Survey (AGES) we have observed 5$\times$4 degrees of sky centred on M33, reaching a limiting column density of $\sim 1.5 \times 10^{17}$ cm$^{-2}$ (line width of 10 km s$^{-1}$ and resolution 3.5\arcmin).  We particularly investigate the absence of optically detected dwarf galaxies around M33, something that is contrary to galaxy formation models. We identify 22 discrete \HI\ clouds, 11 of which are new detections. The number of objects detected and their internal velocity dispersion distribution is consistent with expectations from standard galaxy formation models. However, the issue remains open as to whether the observed velocity dispersions can be used as a measure of the \HI\ clouds total mass i.e. are the velocities indicative of virialised structures or have they been influenced by tidal interactions with other structures in the Local Group? We identify one particularly interesting \HI\ cloud, AGESM33-31, that has many of the characteristics of \HI\ distributed in the disc of a galaxy, yet there is no known optical counterpart associated with it. This object has a total \HI\ mass of $1.22 \times 10^{7}$ M$_{\odot}$ and a diameter of 18 kpc if at the distance of M33 ($D_{M33}=840$ kpc). However, we also find that there are numerous other \HI\ clouds in this region of sky that have very similar velocities and so it is plausible that all these clouds are actually associated with debris from the Magellanic stream.

\end{abstract}


\section{Introduction}
\label{sec:intro}
M33 is the third most massive galaxy in the Local Group (after Andromeda (M31) and the Milky Way). It is generally thought that M31 and M33 are closely associated with each other, for example they may have gravitationally interacted with each other in the relatively recent past (e.g. \citealt{2008MNRAS.390L..24B}). The M31-33 system collectively has around 30 known satellite (dwarf galaxy) companions (\citealt{2009ApJ...705..758M}), almost all of which are confirmed satellites of M31 with no obvious association with M33.  However, \cite{2009ApJ...705..758M} recently announced the discovery of two new dwarf galaxies in close proximity to both M31 and M33. One of these they classified as a definite M31 satellite, but the other, Andromeda XXII, is the first proposed satellite of M33 (e.g.\citealt{2013MNRAS.430...37C}). And XXII is a lot closer in projection to M33 ($\sim$40 kpc) than to M31 ($\sim$220 kpc) and lies in a region where M33's dark matter halo would appear to dominate gravitationally over that of M31's. The two newly discovered dwarf galaxies were found using the PAndAS CUBS optical survey by searching for spatial over densities of stars. In fact almost all previous surveys for faint Local Group dwarf galaxies have relied upon optical searches for these stellar over densities. Given the very low optical surface brightness of the objects detected there may come a point where observations at other wavelengths, like those described here, may become even more productive at finding dwarf galaxies.

With specific regard to M33, extended star clusters have also been found, which \cite{2013MNRAS.430...37C} propose may be tidally stripped dwarf galaxies because of their structure - they are extremely sparse star clusters with elongated distributions of stars. There are also other detections of objects in the halo and around M33. \cite{2009ApJ...698L..77H} show that the globular cluster population of M33 is weighted towards the far side of M33 with respect to M31, which they cite as further evidence of an interaction with M31 having affected M33's halo population. \cite{2006AJ....132.1361S} have identified RR Lyrae stars in the halo of M33. RR Lyrae stars are found in stellar populations that are older than 10 Gyrs. This helps to determine the age of the halo population of M33 and therefore place a limit on when an interaction with M31 may have occurred. As we will see below these stars are considerably older than derived ages for a M31/M33 interaction suggesting that M33's halo may well have been rather unperturbed over quite long timescales.

\begin{figure*}
\centering
\includegraphics[scale=0.5]{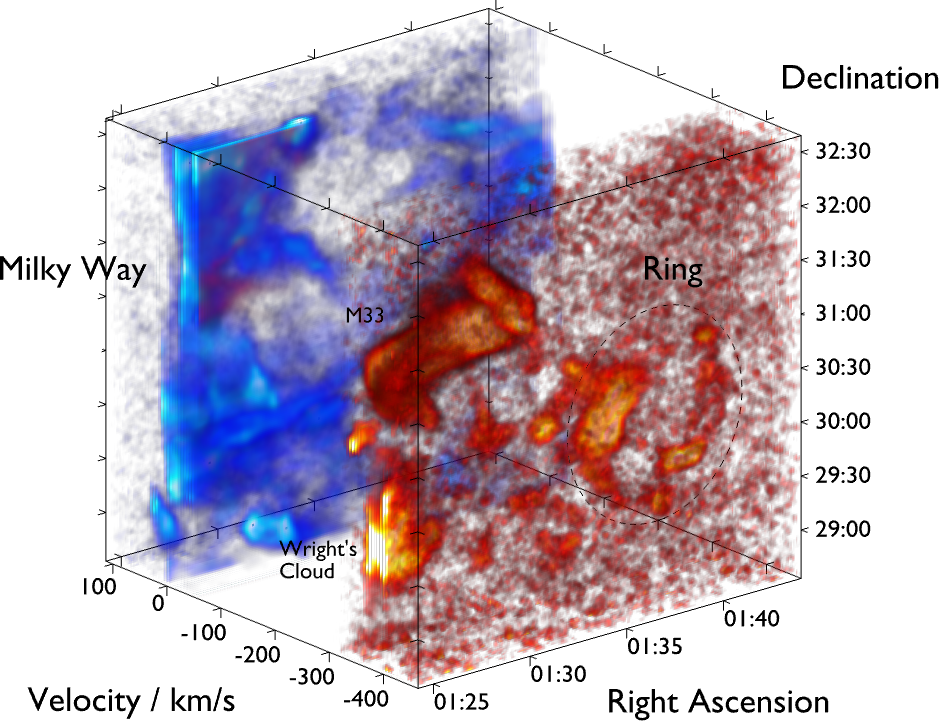}
\caption{Position-Velocity plot of the AGES HI data cube. The HI from M33 and the Milky Way are confused over the approximate  velocity range of $-75 < v < 0$ km s$^{-1}$ so we have used blue for velocities $v > -75$ km s$^{-1}$ (Milky Way) and red for everything else. Some of the distinct HI structures referenced in the text are labelled. The figure shows just a single, representative still image. A full movie of the rotating cube is available online.}
\label{fig:pv}
\end{figure*}

Given the proximity of M33 (we will use a distance of 840 kpc throughout this paper (\citealt{1991ApJ...372..455F})) and the past optical searches for stars, e.g. \cite{2009ApJ...705..758M}, it is surprising that currently there is only one known possible dwarf galaxy companion of M33. It is even more surprising given the standard galaxy formation models that predict many hundreds of dwarf satellite galaxies surrounding larger galaxies like M33 (\citealt{2009MNRAS.398L..21S} and references therein). So, do the dwarf companions exist, but are still not detected by optical surveys (maybe because of their extremely low surface brightness), or have the dwarfs been destroyed by a M33/M31 interaction? In this paper we will investigate the halo of M33 not by observing the emission from stars, but by using the 21cm emission from atomic hydrogen. Atomic hydrogen is the fuel for dwarf galaxy (star) formation, and is also easily drawn out of galaxies when they gravitationally interact, and as such should offer us some insight into the current nature of M33's halo.

There has been considerable previous work, utilising 21cm observations, that has been used to investigate the halo of M33 and the nature of its relationship with M31. \cite{2004A&A...417..421B} (hereafter BT04) used the Westerbork array to detect \HI\ over a 1800 deg$^2$ region of the sky that covered both M31 and M33. Their survey had a very low \HI\ column density sensitivity of $1.5 \times 10^{17}$ cm$^{-2}$ over a 30 km s$^{-1}$ velocity width, but a rather large spatial resolution of 49 arcmin (FWHM). They detected 95 \HI\ clouds almost all of which seem to be clearly associated with M31. They also identified a faint \HI\ stream seemingly joining M33 and M31 however, they were unable to trace this all the way to M33 due to confusion with \HI\ in the outskirts of the Milky Way. They detected one \HI\ cloud in the region of M33 with a \HI\ mass of a few times 10$^5$ M$_{\odot}$ - this cloud has properties similar to the High Velocity Clouds (HVCs) found around the Milky Way. HVCs are clouds of \HI\ with velocities than cannot be explained with a simple model of galactic rotation - Galactic HVCs typically have velocities which deviate from expectations of rotation by $>$ 50 kms$^{-1}$ (\citealt{1991A&A...250..499W}). \cite{2005A&A...436..101W} present high-resolution ($\sim$2 arcmin) follow up observations of the previously identified M31/M33  \HI\ clouds and again conclude that the single M33 HVC displays very similar characteristics to Galactic HVCs. A possible link between HVCs and the missing dwarf galaxy population as either progenitors of, or failed, dwarf galaxies has previously been proposed by  \cite{1999ApJ...514..818B} and \cite{2004ASSL..312..297B}.

BT04 propose two possible mechanisms for the origin of the \HI\ stream between M31 and M33: (1) that it is a pre-existing filament extending between the two galaxies, filaments like this are a prediction arising from high resolution numerical models of structure formation (\citealt{1999ApJ...511..521D}, \citealt{2001ApJ...547..574D}) or (2) the stream has arisen from a past tidal interaction between M31 and M33. Using the measured systemic velocities of M33 and M31 BT04 concluded that the two galaxies are currently moving towards each other, indicating that no recent ($<2$ Gyr) interaction has taken place.
\cite{2009ApJ...703.1486P} have used similar methods to BT04 (an assessment of possible orbits given the observed velocities) and conclude that the galaxies are likely to have interacted between 1 and 3 Gyr ago, which is roughly consistent with BT04. \cite{2009ApJ...703.1486P} list distinct features of the gaseous disc of M33 including: an extended warp in the disc, a filament from the northern part of the disc, diffuse gas surrounding the galaxy,  and a cloud with a filament extending from the galaxy. They say that all of these features strongly indicate a past interaction between the two galaxies and that this interaction is the probable origin of the M31-M33 stream. \cite{2008MNRAS.390L..24B} have carried out an ensemble of simulations of possible orbits for M33 and M31 and conclude that an interaction occurred even longer ago than proposed above - between 4 and 8 Gyr.

However, using observations of the stream down to an almost identical BT04 column density limit of $2.7 \times 10^{17}$ cm$^{-2}$ for a line width of 25 km s$^{-1}$  \cite{2013Natur.497..224W} report that about 50\% of the atomic hydrogen in the stream is composed of distinct clouds whilst what remains is made up of an extended, diffuse component. They are able to identify these distinct clouds because of their much improved spatial resolution of 9 arcmin compared to that of BT04 (49 arcmin).  \cite{2013Natur.497..224W} propose an intergalactic filamentary origin of the stream. Basically they argue that the collapse timescale for the individual clouds they identify is much shorter ($\sim$400 Myr) than the likely time since an interaction (a few Gyr).  \cite{2013Natur.497..224W} propose that the clouds may have several possible origins: (1) they may be primordial gas rich objects similar to dwarf spheroidal or dwarf irregular galaxies, (2) the clouds may be gas accreting into sub-halos, (3) the clouds may be tidal dwarf galaxies in formation and (4) the clouds may be transient objects condensing from a pre-existing intergalactic filament. 

Clearly, there is still some lack of agreement as to the origin of the \HI\ stream/filament. If it is tidal then at first sight it is difficult to see how distinct clouds along its length will not have already collapsed into much smaller denser structures. However, this is based on the assumption that the observed velocity widths measure the 'local' gravitational field and are not highly biased by streaming motions along the filament - something that cannot be ruled out. The question also remains open as to whether any of the \HI\ condensations could eventually become dwarf galaxies and so help explain the discrepancy between the numbers that theory predicts and the observations. 

In this paper we intend to explore these issues further, with a particular emphasis on whether any of the complex \HI\ structures seen around and in the halo of M33 could be part of a significant population of M33 dwarf galaxies, and so account for the lack of optically detected companions.

\begin{figure}
\centering
\includegraphics[scale=0.13]{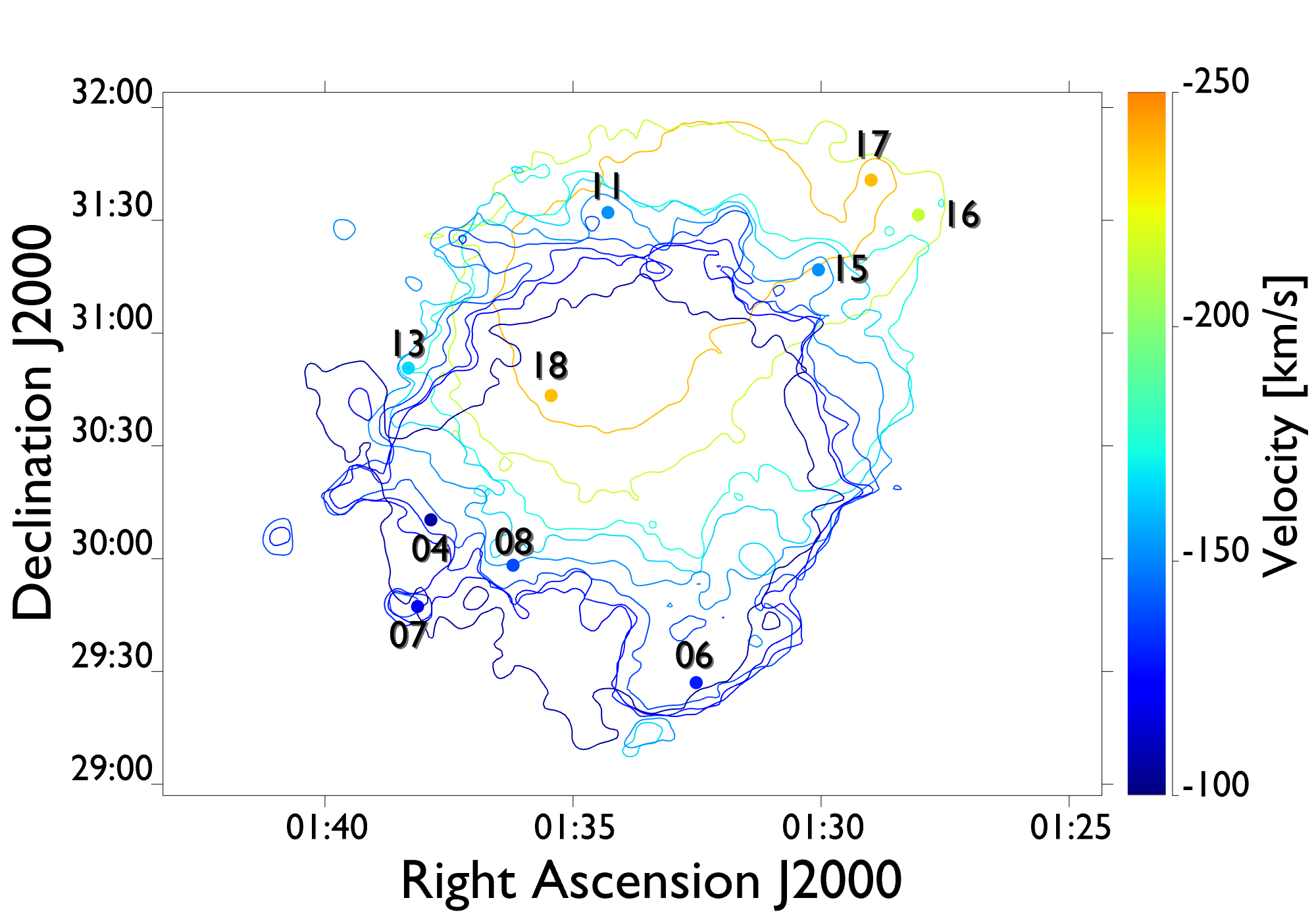}
\caption{A Renzogram of M33 across a velocity range of 150kms$^{-1}$. Labelled dots indicate the positions of clouds which appear to form part of the disk of M33 (table 2). Colours are indicative of velocity.}
\label{velM33cloudssep}
\end{figure}

The data described in this paper extends both the column density limit and spatial resolution, of the surveys described above - we reach $\sim 1.5 \times 10^{17}$ cm$^{-2}$ for a line width of 10 km s$^{-1}$ with a spatial resolution 3.5 arcmin. Our survey also extends that of  \cite{2008A&A...487..161G} (hereafter G08) who used the Arecibo Legacy Fast ALFA (ALFALFA) survey observations of a differently shaped area to us around M33 and with lower sensitivity (see below). 
Our deeper Arecibo observations form part of the Arecibo Galaxy Environment Survey (AGES). The survey is fully described in \cite{2006MNRAS.371.1617A}.

In the following sections we will describe the observational data and its reduction (section 2), our method of source identification and measurement (section 3), an analysis of our results (section 4), a discussion putting our results in context with regard to the astrophysical problems discussed above (section 5) and finally a summary of our work and conclusions (section 6).

\section{Observations and Data Reduction}
The observations and data reduction techniques are described in depth in \cite{2006MNRAS.371.1617A}, and also in \cite{2008MNRAS.383.1519C}, \cite{2010AJ....140.1093M} and \cite{2011MNRAS.415.1883D}. Therefore, they have only been summarised here. Observations of this field began in August 2008 and were completed in August 2013. The field observed to full depth spans approximately 5 degrees of R.A. by 4 degrees of declination centred on M33. The full spatial range of the data considered here is from 1:23:36 to 1:44:39 in R.A, and from +28:23:50 to 32:53:54 in declination (J2000), see Fig. 1. 

Observations were performed with the telescope in drift scan mode, which means that the array is kept at a fixed azimuth and elevation whilst the sky drifts overhead. Each drift lasts for 20 minutes and covers about 5 degrees of RA. The 3.5 arcmin resolution element takes about 13 seconds 
to cross the beam when observing at 1.4 GHz, so 25 scans are required for an integration time of $\sim$330 sec. Each beam records every second data in two polarisations with 4096 channels, spanning a velocity range of approximately -2,000 km/s to +20,000 km/s. The two polarisations are combined and then the data are gridded into 1' spatial pixels (the Arecibo beam has a FWHM of 3.5' at this wavelength) and into velocity channels of width 5.2 kms$^{-1}$.

The data was reduced using the {\sc aips++} packages {\sc livedata} and {\sc gridzilla}, which are extensively described in \cite{2001MNRAS.322..486B}. In brief, {\sc livedata} is used to perform bandpass estimation and removal, Doppler tracking and it calibrates the residual spectrum. We also fit and subtract a sigma-clipped second-order polynomial to each spectrum, which, as shown in \cite{2014MNRAS.443.2634T} (hereafter AGES VII), can significantly improve the baseline quality.  {\sc gridzilla} is a gridding package which co-adds all of the spectra to produce the data cube described above. It also performs side lobe correction so that in the final gridded data cube each beam contributes equally to each pixel, with the exception of the edges of the map where sky coverage is incomplete. 

We obtain an approximate 1$\sigma$ \HI\ column density sensitivity of 1.5$\times$10$^{17}$cm$^{-2}$ over a velocity of 10 kms$^{-1}$. Our column density sensitivity is around three times better than the ALFALFA data of G08. Sources of any given mass will be detected with lower S/N levels if their velocity width is large or equally, if they have the same intrinsic velocity widths but are discs with higher inclination angles, since their flux is spread out over more channels.

\begin{figure*}
\centering
\includegraphics[scale=0.23]{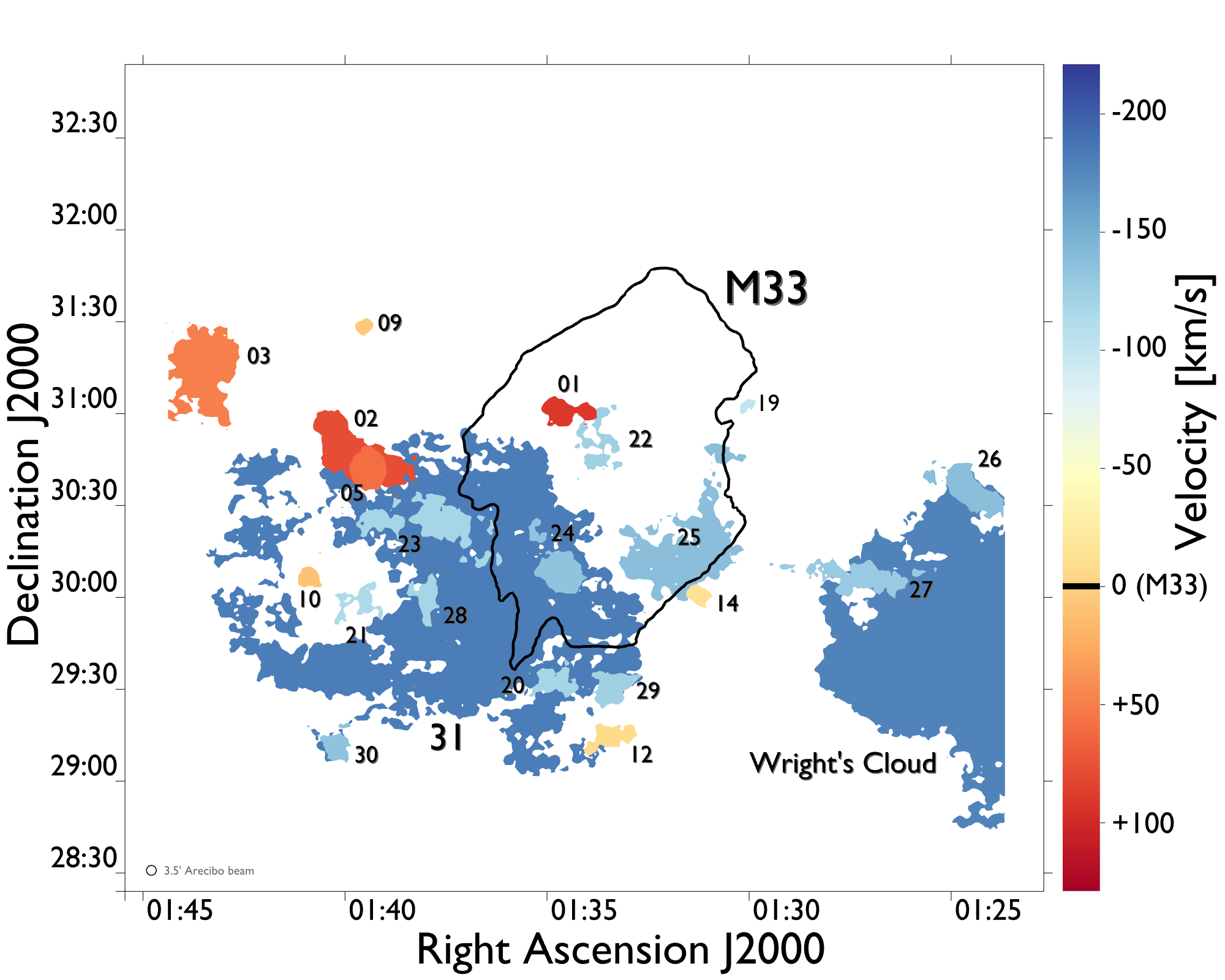}
\caption{The outline of M33's HI disk is shown, with all of the discrete detected clouds overlaid. The figure was made by integrating the flux over the velocity range of each cloud, then filling in the lowest contour. The central velocity of each cloud was used so the colours are indicative of this velocity. The black line shows the approximate extent of M33. Velocities shown are relative to M33.}
\label{fig:velM33clouds}
\end{figure*}

\section{Source Extraction}
To isolate the \HI\ associated with M33 we selected a region of the cube spanning a velocity range of $\sim$600 kms$^{-1}$. The systemic velocity of M33 is $\sim$-180 kms$^{-1}$ and as the Milky Way is centred at 0 kms$^{-1}$ distinguishing \HI\ in M33 from that in the Milky Way is not straightforward (all velocities given are barycentric except where stated otherwise). Each channel was inspected to determine the extent of the Milky Way contamination. \HI\ from the Milky Way appears as extended emission spread across the spatial extent of the cube and is continuously connected, it also appears outside of the region in which one would expect to find M33. The systemic velocities of the two galaxies, along this line of sight, differ by only about 200 kms$^{-1}$, which is of the order of their internal velocities, so they 'bleed together' in the cube (Fig. \ref{fig:pv} ). This means that any sources in this crossover area (from around 0 kms$^{-1}$ to 75 kms$^{-1}$) must be inspected by eye to determine whether they are likely to be a cloud associated with M33. If we find a compact source as opposed to extended emission across the field then we conclude that the source does not appear to be associated with the Milky Way.This process led to a final data cube analysed here that spanned a velocity range of -500 to 100 km s$^{-1}$, see Fig\ref{fig:pv}. This cube is available as an MP4 movie through the online materials.
Initially source identification was carried out by visual inspection of the data cube. The cube was split into moment 0 maps each of width $\sim$50 kms$^{-1}$ and these were inspected for individual sources.  Visual examination is also preferable to automated source extraction when attempting to distinguish between real sources and radio frequency interference (RFI) (see \citealt{2013MNRAS.428..459T} - AGES VI). However, visual inspection does suffer from our subjectivity and is generally slower than most algorithms. In AGES VII we describe a new data cube viewer, which has been specifically designed to assist visual extraction of sources from \HI\ data cubes\footnote{The data cube viewer is called {\sc frelled} ({\sc fits} Realtime Explorer of Low Latency in Every Dimension). Full details are given in AGES VII appendix A and the source code is available through our website at www.naic.edu/$\sim$ages/.}. {\sc frelled} was used to visualise and catalogue all of our detections. 

We then used the {\sc miriad}\footnote{{Miriad is a radio interferometry data reduction package which can be used for reduction of observations through to image synthesis, analysis and display  (\citealt{1995ASPC...77..433S}).}} task {\sc mbspect} to view the spectra of each identified \HI\ cloud and confirm our detections. {\sc mbspect} is particularly useful in this case as it can be used to determine whether a cloud is distinct from the Milky Way (and possibly M33) by confirming that they have separate spectral peaks i.e. whether our identified \HI\ clouds appear to be distinct both spatially and in velocity. We define a cloud as an object detected after a 3$\sigma$ clip has taken place on the data cube and channel smoothing has been applied (the channels are smoothed to 10 kms$^{-1}$). After this the detection is checked with both {\sc frelled} and {\sc mbspect}- for it to be included in our detections it must appear in {\sc mbspect} as a single peak at 3$\sigma$, but may separate into more peaks above this level. {\sc mbspect} was also used to measure the velocity widths (at 50 and 20\% of peak flux density) of the identified \HI\ clouds. We then used the image display and processing package {\sc ds9} to fit ellipses to each of the identified sources. From this we obtained the central coordinates of each cloud and the lengths of the semi-minor and semi-major axes of the fitted ellipse as measured at a column density of 3$\sigma$ - the noisiest pixel has a 3$\sigma$ noise value of $\sim$0.1Jy. The cloud sizes given were calculated using the measured values of the semi-major and semi-minor axes. We used these positions and sizes to measure the total flux density and \HI\ mass within the detection aperture using the standard formula: \\
\begin{center}$M = 2.36 \times 10^{5} d^{2}_{Mpc} \sum f_{v} \Delta V$ \end{center}
$d_{Mpc}$ is the distance to the source in Mpc, which we assume to be the same as M33 i.e 0.84 Mpc (\citealt{1991ApJ...372..455F}), $f_{v}$ is the flux density in Jy and $\Delta V$ is the velocity width in kms$^{-1}$. In addition we can create integrated flux maps by summing pixel values over velocity. Full details of the sources detected and their derived parameters are given in Tables 1 and 2.

\section{Results}
As a check of our calibration compared to other surveys we have measured the \HI\ mass of M33 in the same way as the other \HI\ clouds we have identified - we obtain a total \HI\ mass for M33 of 4.7 $\times$ 10$^{9}$ M$_{\odot}$. To compare our results with a previously derived value we reduce our aperture to approximately the same size as that given in \cite{2009ApJ...703.1486P}. When this is done we obtain a \HI\ mass of 2.0 $\times$ 10$^{9}$ M$_{\odot}$, this compares well with the value given in  \cite{2009ApJ...703.1486P}, which is 1.4 $\times$ 10$^{9}$ M$_{\odot}$ - we agree to within 30\%, though defining exactly the same apertures both spatially and in velocity is not possible. 

Using the source extraction methods described in section 3 we have identified 32 clouds that have a probable association with M33. In the appendix we show moment zero (integrated flux maps), moment 1 (velocity field) and spectra for each detected source. The sources are split into two groups: those that appear discrete and those that clearly show an extended connection to M33.
One clear initial result is that some of the features identified by G08 as discrete clouds are actually connected to the \HI\ disc of M33 by areas of low column density \HI\ in our data, which has a better column density sensitivity. This is demonstrated in Fig. \ref{velM33cloudssep} which shows the positions of clouds, which fall within the disk of M33.

We employ a similar distinguishing criterion as Braun and Thilker (2002) to separate what we describe as discrete objects from those that are clearly linked to M33 by low column density \HI\ i.e they describe them as discrete if there is no connection observed down to a column density of $1.5 \times 10^{17}$ cm$^{-2}$. Table 1 is a full list of the 22 \HI\ clouds that we identify as being discrete down to our column density sensitivity limit of $\sim$1.5$\times$10$^{17}$cm$^{-2}$. Table 2 lists 10 additional clouds from G08, which we detect but are either clearly part of the disc of M33 or joined to M33 by low column density \HI\ . These latter objects have been listed in a separate table as we argue that they cannot be classified as independent clouds due to their obvious connection to M33. These objects may be of an entirely different nature to those in Table 1, for example the result of tidal interactions rather than potential primordial objects. Of course, subsequent even better sensitivity observations may reveal even more connectivity between objects we consider to be discrete. 

We detect all clouds listed in G08, with the exception of their clouds AA3, AA22 and AA23, which do not appear in our data at the coordinates and velocities specified. G08 state that their clouds AA4, AA6, AA12, AA13 and AA14 are connected to M33, we extend this list to include AA7, AA8, AA11, AA15 and AA16.  In addition to this, the clouds G08 designate as AA21a and AA21b appear in our data to be a single ring like structure (AGESM33-31, see below). A moment zero map of our data is shown in Fig. \ref{fig:velM33clouds}. The outline of M33's \HI\ disc is shown, with all discrete clouds overlaid. The figure was made by integrating the flux over the velocity range of each cloud, then filling in the lowest contour. The colour indicates the central velocity of each cloud. The large structure to the bottom right of Fig. \ref{fig:velM33clouds} has previously been identified and is known as Wright's cloud (\citealt{1979ApJ...233...35W}).

\begin{figure}
\centering
\includegraphics[scale=0.45]{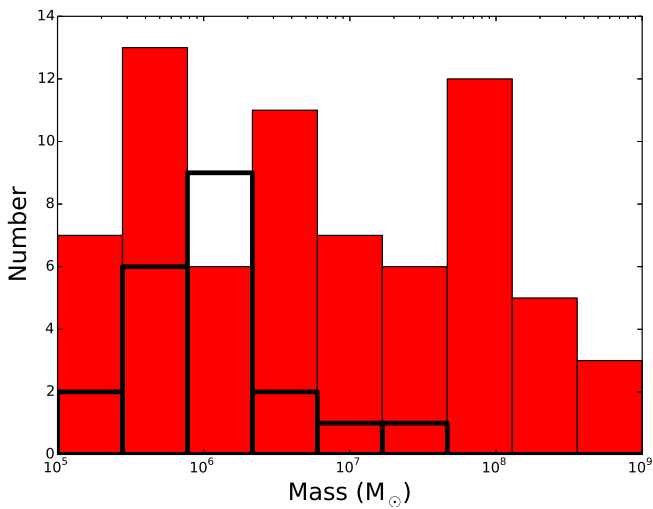}
\caption{A histogram showing the baryonic masses of our detected HI clouds (black outline) compared to Local Group dwarf galaxies (red). The baryonic masses of the Local Group dwarf galaxies were obtained from table 4 in \citealt{2012AJ....144....4M} - from this table we have multiplied the HI mass by 1.4 to account for the primordial fraction of helium and then added the stellar mass to get the total baryonic mass (this ignores molecular and ionised gas).}
\label{bar_mass}
\end{figure}

\begin{figure*}
\subfloat[]{\includegraphics[width = 3.2in]{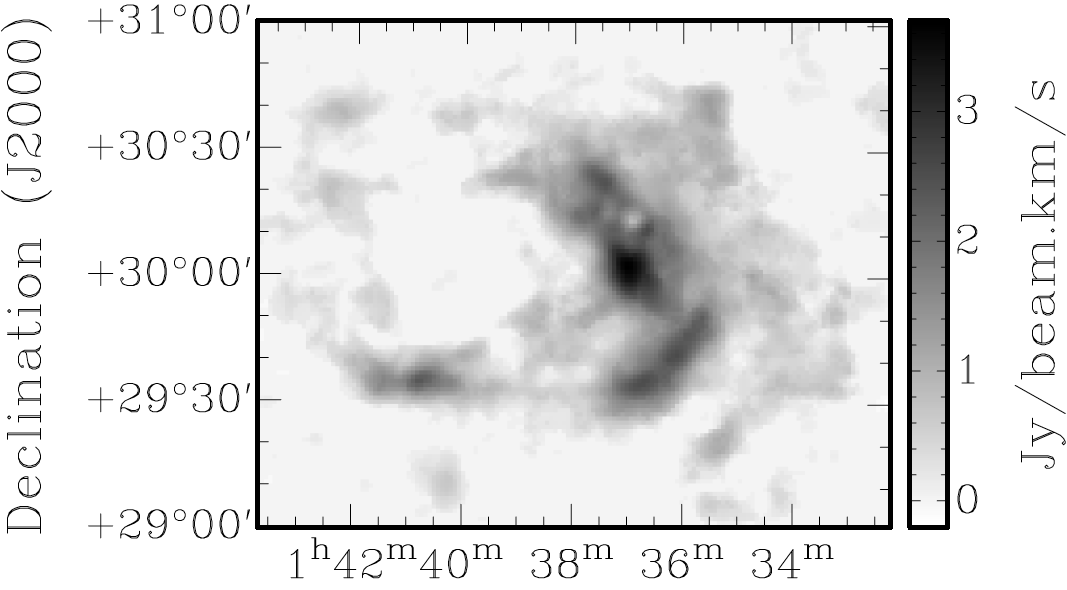}}\
\subfloat[]{\includegraphics[width = 3in]{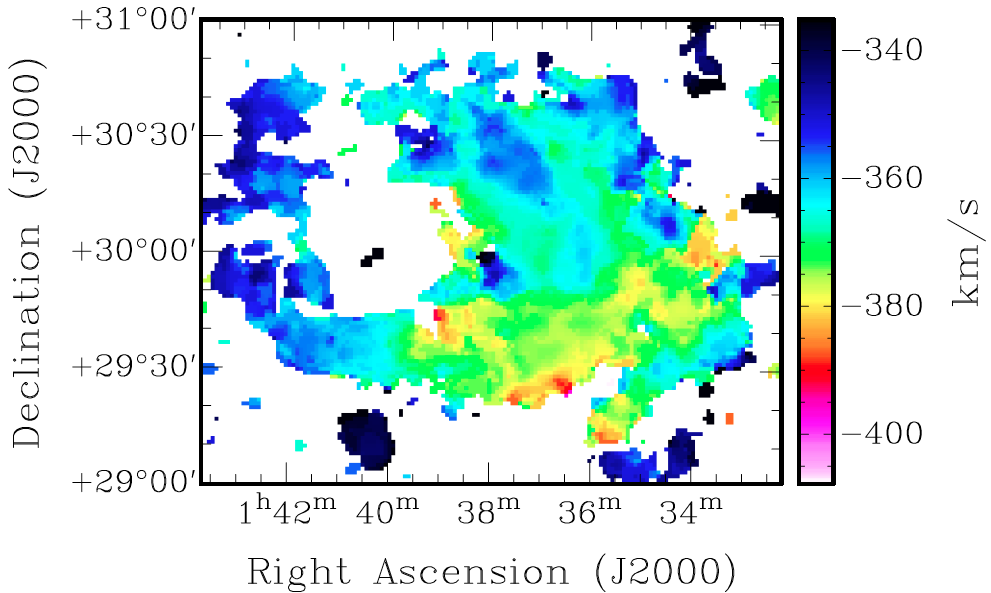}}
\caption{(a) A \HI\ integrated flux map and (b) a velocity map of AGESM33-31. The cloud is listed as two distinct objects by G08, but it is clearly a ring-like structure. The velocity map shows the cloud has a velocity gradient across it of about 30 km s$^{-1}$. To convert the integrated flux map to a column density map the conversion factor is 2.5$\times$10$^{19}$cm$^{-2}$(Jy/beam.km/s)$^{-1}$.}
\label{OK31}
\end{figure*}

We can also compare our results with those of BT04. As stated above their survey covers the much larger M33/M31 region, but is somewhat compromised compared to our survey by the large spatial resolution of 49 arcmin. Using the BT04 sensitivity limits we calculate that their minimum  detectable \HI\  mass at the distance of M33 is about 10$^5$M$_{\odot}$, which is smaller than the mass of most of the clouds we detect (see Table 1 and 2). BT04 only detect two discrete clouds over the region where our data overlaps (as listed in \citealt{2002ASPC..276..370T}), so the limitation of their data compared to ours must be their spatial resolution. We detect both of the clouds described in BT04 using the AGES data. Wright's cloud (AGESM33-32), which is a previously known \HI\ cloud, is either associated with M33 or possibly extended debris from the Magellanic stream. Wright's cloud has a total mass of $\sim$10$^{8}$M$_{\odot}$ (\citealt{1979ApJ...233...35W}), but our cube only contains part of it (Fig. \ref{fig:velM33clouds}), it is fully described in {\citealt{1979ApJ...233...35W}}. The second cloud, AGESM33-31, will be discussed in detail in section 5.

The clouds we detect range in \HI\ mass from 1.0$\times$10$^5$M$_{\odot}$ to 4.5$\times$10$^7$M$_{\odot}$ and have velocity widths of between $\sim$17 and $\sim$72 kms$^{-1}$ at 50$\%$ of their peak flux density. The smallest of the clouds is just over 1 kpc in size and, as the Arecibo beam size is about 1 kpc on the sky at the distance of M33 (840 kpc), we wouldn't expect to measure sizes any smaller than this. 

\section{Discussion}
We have identified 32 \HI\ structures around the nearby galaxy M33. 22 of these appear to be discrete clouds (Table 1) while a further 10 appear to have low column density structures that connect them to M33 (Fig. \ref{velM33cloudssep}). 11 of the clouds appear to be previously undetected which is due to our area coverage, sensitivity and resolution compared to other surveys. 

The  \HI\ masses of known dwarf galaxies in the Local Group range from $1.2\times$10$^{5}$M$_{\odot}$ to 1.8$\times$10$^{8}$M$_{\odot}$ (data from tables 1 and 4 in \citealt{2012AJ....144....4M}), which means that the majority of our detected \HI\ clouds have \HI\ masses that are typical of known dwarf galaxies. If our detected objects are progenitor dwarf galaxies that currently lack a stellar component then a better comparison is with the total baryonic mass of other Local Group dwarf galaxies. The baryonic masses of Local Group dwarfs range from 3.4$\times$10$^{2}$M$_{\odot}$  to 4.6$\times$10$^{9}$M$_{\odot}$ (also calculated using data in \citealt{2012AJ....144....4M} by multiplying the \HI\ mass by 1.4 to account for the primordial fraction of helium and then adding the stellar mass to get the total baryonic mass, where no \HI\ mass is given only stellar mass is used). In Fig. \ref{bar_mass} we show a comparison of these baryonic masses with the \HI\ masses of our detections - clearly our detections have masses consistent with those of dwarf galaxies, though at the lower end of the mass scale.

\begin{figure}
\centering
\includegraphics[scale=0.45]{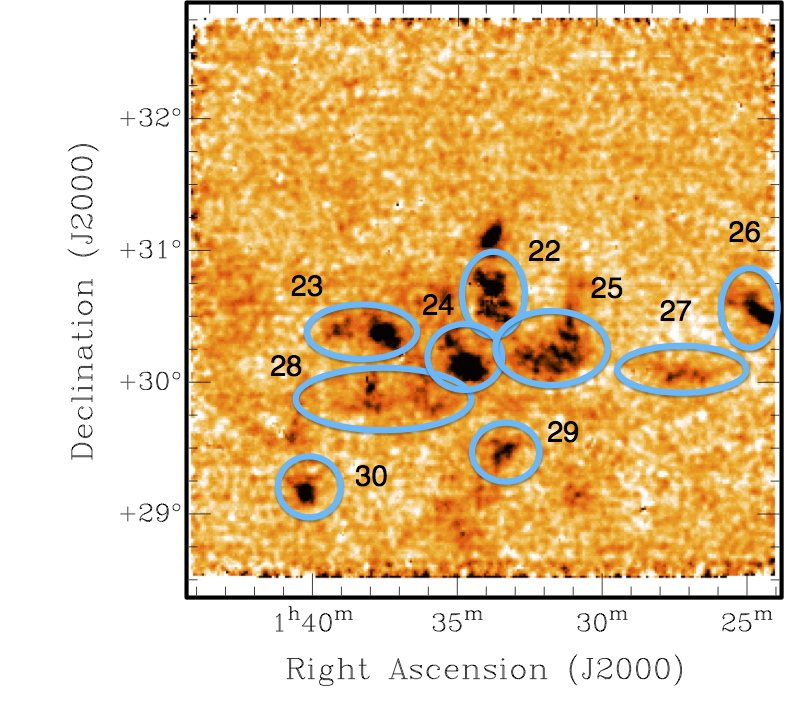}
\caption{An integrated flux map of a section of the cube over the full RA and Dec range, but from -320 to -346 km s$^{-1}$ in velocity, showing a group of clouds with very similar central velocities. AGESM33 cloud numbers are shown. The uncircled patch above cloud 22 is the northern arc of M33.}
\end{figure}

\begin{figure}
\centering
\includegraphics[scale=0.45]{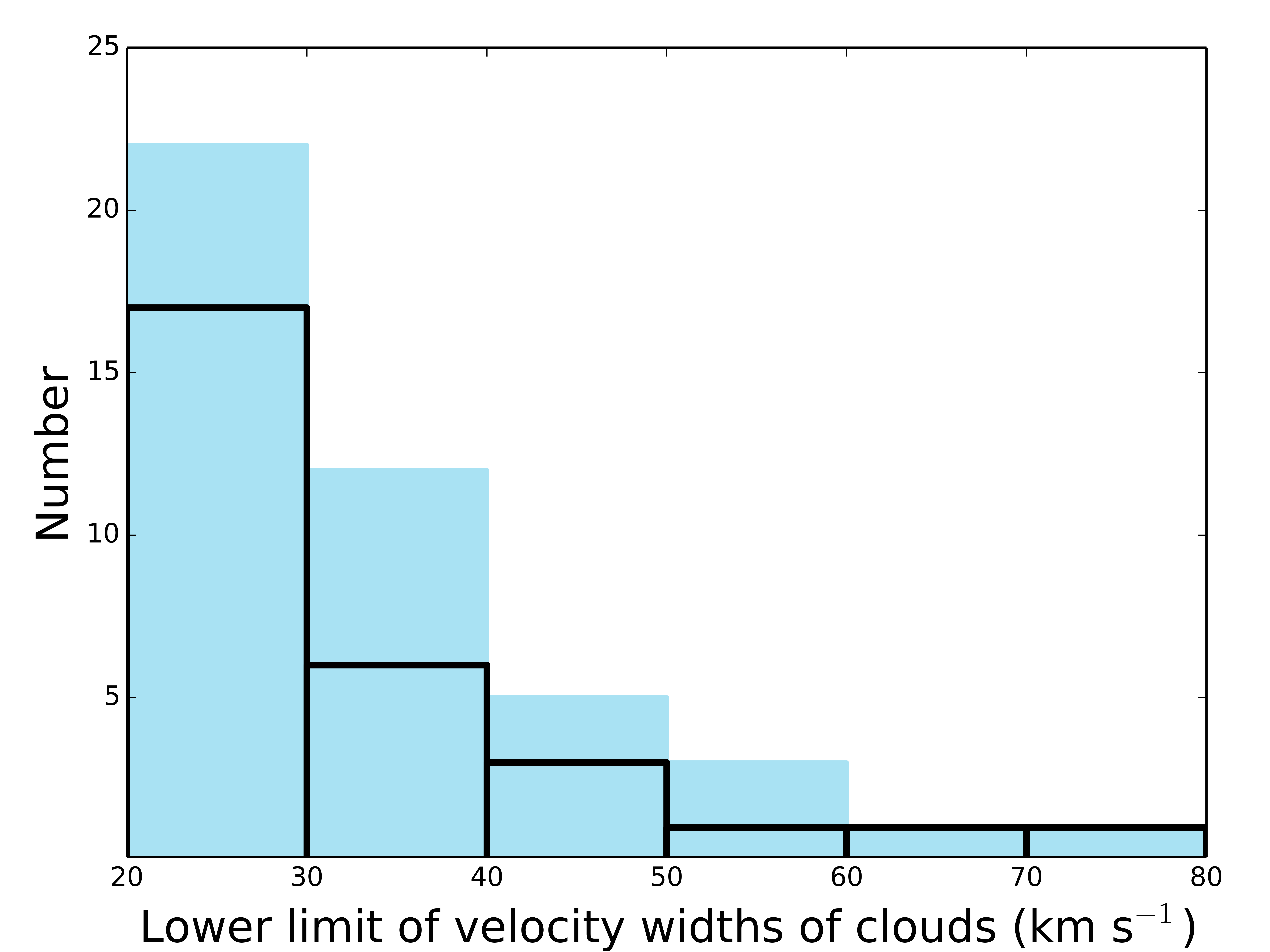}
\caption{The above shows the number of clouds with velocity widths greater than a given value. The black edged bars are the number of clouds predicted from the numerical simulation described in the text. The blue bars are for the discrete AGESM33 HI clouds we observe.}
\label{bar}
\end{figure}

As stated in the introduction it has been suggested that there may be a link between Galactic HVCs, dwarf galaxies and the missing sub-halo population (\citealt{1999ApJ...514..818B} and \citealt{2004ASSL..312..297B}). It has also been suggested that a particular class of HVC, known as compact HVC or CHVC, may actually be part of the missing sub-halo population (\citealt{1999A&A...341..437B}, \citealt{1999ApJ...514..818B}). We can compare our discrete \HI\ clouds to CHVCs detected around the Milky Way. \cite{2001A&A...369..616B}, who also used the Arecibo telescope, specifically targeted ten Milky Way CHVCs. They measured their central \HI\ column densities to be in the range $2 \times 10^{19} - 2 \times 10^{20}$ cm$^{-2}$ and typical line of sight velocity dispersions of $\sim$11 km s$^{-1}$. Sternberg et al. (2002) describe CHVCs as having average column densities of $\sim 5 \times 10^{18}$ cm$^{-2}$ and typical line of sight velocity dispersions of $\sim$14 km s$^{-1}$. Comparing these values with those given in Table 1 clearly shows that our \HI\ detections are different to Milky Way CHVCs. Our detections generally have lower peak \HI\ column densities (as measured by our 3.5 arcmin beam) and larger velocity widths, values more typical of HVCs rather than CHVCs.

\begin{figure}
\centering
\includegraphics[scale=0.43]{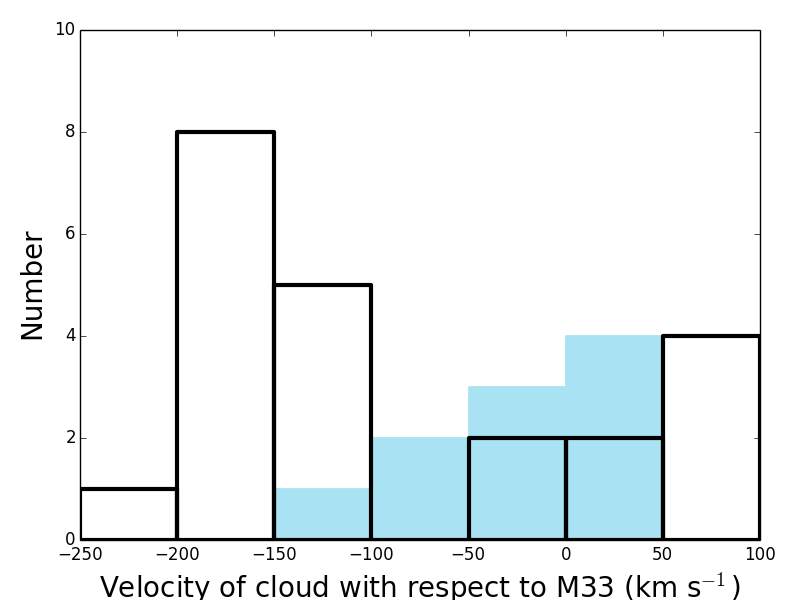}
\caption{Histogram showing the distribution of AGESM33 cloud velocities with respect to M33. The clouds we have described as 'discrete' are represented by the black edged bars and the clouds that show some low column density connection to M33 are represented by the blue bars.}
\label{veldist}
\end{figure}

Both clouds AGESM33-31 and AGESM33-32 are especially interesting as they are both extremely extended spatially compared to the other clouds detected and have comparably small velocity widths for their extent. 
This is particularly so for AGESM33-31, which has the second smallest velocity width of all of our detections.
AGESM33-31 was detected by G08 as AA21a and AA21b, however in our data it appears as a single much larger ring-like structure (see Fig. 1, 3 and 5). It was also previously detected by \cite{2002ASPC..276..370T} and designated 'M33 CHVC' but was unresolved in their survey. So, this is the first time that this object has been observed as a ring-like structure. It is extremely spatially extended, similar to Wright's cloud (seen in our data as AGESM33-32, Fig. 3), and also occupies almost the same velocity range. It has been suggested that Wright's cloud is associated with the Magellanic Stream due to an extended low N$_{HI}$ component of the cloud that extends in the direction of the stream (e.g. \citealt{2010ApJ...723.1618N}), however it is of order 50 kpc away from the main structure of the stream and $\sim$30 kpc from the S0 filament of the stream (as defined by \citealt{2010ApJ...723.1618N}).  It is possible that AGESM33-31 is also in some way related to the Magellanic Stream, rather than being a HVC of M33, though it is even further away from the main section of the Magellanic Stream stream than Wright's cloud \footnote{ As we only observe a part of Wright's cloud we will not discuss it further here.}. So, a Magellanic Stream origin is a possibility, but another interpretation is that AGESM33-31 is a Galactic HVC projected along this line of sight. At the distance of M33 AGESM33-31 has a rather large size of $\sim18$ kpc (bigger than M33) but a rather low \HI\ mass about 1\% of that in M33. This in itself may be just the characteristics of the 'failed' galaxies we have been looking for, but of course if AGESM33-31 was much closer to the Galaxy the \HI\ mass and size would reduce to something much more like a HVC - so this is another possibility that cannot be ruled out.

Being a little more speculative and accepting a location close to M33, AGESM33-31 appears to be a rather large \HI\ cloud with a big hole in it. A feature like this is plausibly tidal in origin, however, if this was \HI\ in a known galaxy its appearance could be interpreted as a void in the \HI\ vacated by a stellar wind or supernova explosion (e.g. \citealt{2002AJ....123..255O}).  Thilker et al. (2002) identify similar \HI\ 'holes' and 'shells' in the disc of M33, but with a size of $\sim$10 kpc the hole in AGESM33-31 is about an order of magnitude larger than that normally thought to have resulted from stellar winds.  Of course a hole created by a stellar wind implies stars and so a link between these \HI\ clouds and what more traditionally has been described as a galaxy\footnote{\cite{2009ApJ...698L..77H} identify a globular cluster, HM33-B, which is within the coordinate range of AGESM33-31. As no mass or distance is given for this cluster we cannot draw any further conclusions here.  \cite{Grossi2011b} have carried out follow up observations of regions around M33 searching for stellar populations. They report no evidence for a young stellar population in their clouds and hence conclude that there has been no in situ star formation.}. Further detailed investigation of the distribution of stars across this area will have to await the release of data from surveys such as PAndAS CUBS (e.g. \citealt{2009ApJ...705..758M}). However, continuing this 'galaxy' theme the very low measured velocity width (18 km s$^{-1}$) is consistent with the velocity dispersion of atomic gas seen perpendicular to the plane of a typical disc galaxy (of order 10 km s$^{-1}$, Putman et al. (2009) give a velocity dispersion for the atomic hydrogen in M33 of 19 km s$^{-1}$) - it may be possible that we are seeing a face-on disc of \HI\. In Fig. 5(b) we show the velocity field for AGESM33-31, there is clearly a velocity gradient across the ring of about 30 km s$^{-1}$, but whether this is due to rotation or some form of streaming motion is unclear. Subsequently we found that similar ideas have been reached by \cite{2002ASPC..276..370T} - they describe AGESM33-31 (M33CHVC) as a 'dark companion' to M33.

One striking feature of Table 1 is that clouds AGESM33-23 to 30 all lie at very similar velocities. This is all the more remarkable because spatially they extend some 4$^{o}$ or about 60 kpc across the sky (Fig. 6). The clouds also have velocities not too dissimilar to that of AGESM33-31 and 32 discussed above (differ by about 40 km s$^{-1}$). This raises the question as to whether they are all part of some larger structure for which we are only detecting the higher column density parts and that it is possibly debris from the Magellanic stream as discussed above. We currently list them as separate clouds, but their similar velocities are intriguing.

The possibility of this larger structure could explain a very un-isotropic distribution of clouds in the vicinity of M33 - they are concentrated in a band roughly to the south of M33. In addition the distribution of line-of-sight velocities (see below) is also highly anisotropic - 14 out of 22 clouds have velocities between -100 and -200 km/s relative to M33. Both the spatial and velocity distributions are different from what is expected for a population of CDM satellites residing in the halo of M33, unless it has been significantly disturbed in a tidal interaction with M31.

Given that we detect 22 discrete \HI\ clouds around M33 we now consider whether the number of clouds found is consistent with galaxy formation models. We assume that the detected clouds act as tracers of the dark matter halos that the models predict.  \cite{1999ApJ...522...82K} have used a numerical simulation, which implements the $\Lambda$CDM cosmology, to model the dark matter sub-halo population around individual galaxies.  \cite{2002ApJS..143..419S} have parametrised the results of  \cite{1999ApJ...522...82K} into the useful equation given below that can be used to predict the number of dark matter halos expected. \\\\
$ N(>v_{s},<d)= \\\\
1.06\times 10^3 
\biggl({M_{{\rm vir,p}} \over 10^{12} \ {\rm M_\odot}}\biggr)
\biggl({v_{s} \over 10 \ {\rm km} \ {\rm s}^{-1}}\biggr)^{-2.75}
\biggl({d \over 1 \ {\rm Mpc}}\biggr) $ \\\\\\
Where  $N(>v_{s},<d)$ is the total number of sub-halos with scale (circular) velocities greater than $v_{s}$, which we take to be $\frac{\Delta w_{50}}{2}$, contained within a distance $d$ from the centre of a parent halo, ${M_{\rm vir,p}}$ is the virial mass of the parent galaxy.  For the virial mass of M33 we use 5$\times$10$^{11}$M$_{\odot}$ (\citealt{2003MNRAS.342..199C}). We observe out to $\sim$2.25$^{o}$ from M33, which corresponds to a distance of about 33 kpc at a distance of 840 kpc. Assuming that we have detected all \HI\ clouds with velocities greater than our minimum detected velocity ($\Delta v_{50}$) we have $v_{s} \approx 9$ km s$^{-1}$. Putting these numbers into the above equation the model predicts $\sim25$ sub-halos. This compares remarkably well with the 22 \HI\ clouds detected. We can actually do a little better than this and compare the distribution of sub-halo velocities widths with that predicted by the model (Fig. 7). A Kolmogorov-Smirnov test on the data shown in Fig. 7 gives a KS statistic of 0.166
and a P-value of 0.999 indicating that we cannot reject the hypothesis that the two samples are from the
same distribution. In summary, the numbers of HI clouds and the distribution of their velocity widths is consistent with the properties of sub-halos produced in numerical cosmological simulations.

However, the main issue with the above calculation is whether the 'observed' velocities can be used as a measure of the mass of the halo that the \HI\ cloud might reside within. One distinguishing feature of dwarf galaxies is their very high mass-to-light ratios and hence the inference of large quantities of dark matter (\citealt{2012AJ....144....4M}). This inference of course also assumes that dwarf galaxies are gravitationally bound and that they are supported against collapse by either ordered rotation or velocity dispersion. Both our \HI\ clouds and other dwarf galaxies within the Local Group may have the velocities of both their gas and stars severely affected by other streaming (possibly tidal) motions, thus rendering mass-to-light calculations invalid. 

To investigate this further we have created moment 1 (velocity) maps - these maps display the velocity range across each object. The maps are shown in Appendix A along with integrated flux maps and spectra. Our conclusion after inspecting the velocity maps is that with the possible exception of cloud AGESM33-27 and the comments already made about AGESM33-31, there is little or no evidence of ordered rotation within these clouds. 

Although we find little evidence for rotation it is still possible that the clouds are supported by velocity dispersion, in which case we make a tentative estimate of their dynamical masses ($M_{Dyn}$). We have done this by simply assuming spherical virialized clouds (Blitz et al. 1999), that the FWHM of the velocity profile ($\Delta v_{50}$) is a sufficiently accurate estimate of the 3D velocity dispersion of the gas and that the geometric mean of the measured semi-minor and semi-major axis ($r_{Geo}$) is the size of the \HI\ cloud. $M_{Dyn}$ can then simply be approximated using: \\
\begin{center}$M_{Dyn}= {\Delta v^{2}_{50} r_{Geo} / G}$ \end{center}
Derived dynamical masses range between 10$^9$ and 10$^{11}$ M$_{\odot}$ and hence mass-to-light ratios would typically need to be of order $10^{4}$ for these objects to be bound. We find no correlation between \HI\ and dynamical masses within our detections, which would be expected if they had a constant ratio of dark to baryonic matter.

These very high mass-to-light ratios assume that we have detected all of the hydrogen gas present in the cloud. Sternberg et al. (2002) have modelled \HI\ clouds immersed in an ionising background as might be expected if they reside close to a star forming galaxy like M33. They conclude that the \HI\ remains neutral for column densities greater than about $10^{19}$ cm$^{-2}$ and that for typical HVCs only 10\% may remain in the neutral form i.e. there may be as much as ten times more hydrogen present than is inferred from 21cm observations. We would still require mass-to-light ratios of order $10^{3}$ for bound objects, which although high,  is not totally inconsistent with observations of Local Group dwarf galaxies (\citealt{2012AJ....144....4M}).

Figure \ref{fig:velM33clouds} shows the detected \HI\ clouds with colours to indicate their velocities relative to M33.  A test of whether the clouds we have described as 'discrete' are a different population to those that have an apparent low column density connection to M33, is their comparative distribution of velocities. Fig. 3  clearly shows that the discrete clouds are biased towards higher velocities in relation to M33 and could be described as its HVC population. On the other hand Fig. \ref{velM33cloudssep} clearly shows that those objects with a low column density connection to M33 are also connected in velocity space. Fig. 2 is a renzogram, which is a plot of contours of M33 at different velocities - the objects listed in Table 2 fall within these contours at their specific velocity - this does not happen for the discrete objects listed in Table 1. We feel that both Fig 2 and 8 indicate that we have been able to separate discrete \HI\ structures from those that have an origin within the disk of M33.

\section{Summary}
In this paper we have utilised deep 21cm Arecibo data from the AGES survey to investigate the nature of the halo around M33. Our particular interest lies in the lack of previously identified dwarf galaxy (sub-halos) companions of M33, something that at first sight is counter to expectations from galaxy formation models. We identify 22 discrete hydrogen clouds that have a distribution of internal velocity dispersions consistent with expectations from standard galaxy formation models. However, the issue remains whether the observed velocity dispersions can be used as a measure of the \HI\ clouds total mass i.e. are the velocities indicative of virialised structures or have they been influenced by tidal interactions with other structures in the Local Group. If the latter is true then they probably cannot be associated with the missing sub-halo population. We identify one particularly interesting \HI\ cloud AGESM33-31 that has many of the characteristics of \HI\ distributed in the disc of a galaxy yet there is no known optical counterpart associated with it. This object has a total \HI\ mass of $1.22 \times 10^{7}$ M$_{\odot}$ if at the distance of M33. However, we find that there are numerous other HI clouds in our observed region with similar velocities to AGESM33-31 - these may be debris associated with the Magallanic stream. We are not able to rule out the Magellanic stream interpretation  and the weight of evidence may actually support this conclusion.

\section{Acknowledgements}
This work is based on observations collected at Arecibo Observatory. The Arecibo Observatory is operated by SRI International under a cooperative agreement with the National Science Foundation (AST-1100968), and in alliance with Ana G. M\'{e}ndez-Universidad Metropolitana, and the Universities Space Research Association. \\ \\
R. Taylor is supported by the Tycho project LG14013, the project RVO:67985815 and by the Czech Science Foundation project P209/12/1795.

\bibliographystyle{mn2e}
\bibliography{references.bib}

\newpage
\begin{landscape}
\begin{table}
\begin{tabular}{|c|c|c|c|c|c|c|c|c|c|c|}
\hline
AGES ID & RA & Dec & Velocity & Systemic &Velocity & HI Mass & Flux & Size & Column & Previous \\
 & (J2000) & (J2000) & Width 50 & Velocity & Relative to M33 & (M$_{\odot}$)& (Jy km s$^{-1}$) & (a,b kpc)& Density & Detection?\\
 &  &  & (km s$^{-1}$) & (km s$^{-1}$) & (km s$^{-1}$) & & & &(cm$^{-2}$) & \\
 (1) & (2) & (3) & (4) & (5) & (6) & (7) & (8) & (9) & (10) & (11) \\
\hline 
AGESM33-1 & 01:34:51.3 & + 30:59:59 & 51.0 & -86 $\pm$ 3 & 94 $\pm$ 3 & 1.18 $\times$ 10$^{6}$ & 7.1$\pm$ 0.4 &  3.0 $\times$ 1.7 & 3.9 $\times$ 10$^{18}$ & G08-AA1\\ 
\hline 
AGESM33-2 & 01:40:13.2 & + 30:50:57 & 26.0  & -111 $\pm$ 15 & 69 $\pm$ 15 & 1.62 $\times$ 10$^{6}$ & 9.7$\pm$ 2.5 & 4.9 $\times$ 2.4 & 2.4 $\times$ 10$^{18}$ & G08-AA2\\ 
\hline 
AGESM33-3 & 01:44:03.7 & + 31:14:33 & 30.5 & -121 $\pm$ 3 & 59 $\pm$ 3 & 1.48 $\times$ 10$^{6}$ & 8.9$\pm$ 0.2 & 3.7 $\times$ 2.4 & 2.9 $\times$ 10$^{18}$ & none\\ 
\hline 
AGESM33-5 & 01:39:31.4 & + 30:41:30 & 32.8 & -126 $\pm$ 3 & 54 $\pm$ 3 & 2.13 $\times$ 10$^{6}$ & 12.8$\pm$ 0.2 & 2.4 $\times$ 2.4 & 6.2 $\times$ 10$^{18}$ & G08-AA5\\ 
\hline 
AGESM33-9 & 01:39:42.4 & + 31:27:32 & 28.5 & -180 $\pm$ 7 & 0 $\pm$ 7 &  3.00 $\times$ 10$^{5} $ & 1.8$\pm$ 0.2 & 3.2 $\times$ 1.7 & 9.5 $\times$ 10$^{17}$ & none\\ 
\hline 
AGESM33-10 & 01:41:00.8 & + 30:05:17 & 21.0 & -164 $\pm$ 3 & 16 $\pm$ 3 & 2.33 $\times$10$^{5}$ & 1.4$\pm$ 0.1 & 2.4 $\times$ 2.4  & 6.9 $\times$ 10$^{17}$ & none\\ 
\hline 
AGESM33-12 & 01:33:27.8 & + 29:14:49 & 29.2 & -186 $\pm$ 4 & -6 $\pm$ 4 & 6.66 $\times$10$^{5}$ & 4.0$\pm$ 0.3 & 3.7 $\times$ 2.4 & 1.3 $\times$ 10$^{18}$ & G08-AA9\\ 
\hline 
AGESM33-14 & 01:31:18.3 & + 30:00:54 & 22.7 & -188 $\pm$ 3 & -8 $\pm$ 3 & 1.91 $\times$ 10$^{6}$ & 11.5 $\pm$0.2 & 1.7 $\times$ 1.7 & 1.1 $\times$ 10$^{19}$ & G08-AA10\\ 
\hline 
AGESM33-19 & 01:29:52.0 & + 31:04:23 & 34.1 & -288 $\pm$ 3 & -108 $\pm$ 3 & 9.99 $\times$ 10$^{4}$ & 0.6$\pm$ 0.1 & 3.7 $\times$ 1.2 & 3.5 $\times$ 10$^{17}$ & G08-AA17\\ 
\hline 
AGESM33-20 & 01:34:27.6 & + 29:30:00 & 38.4 & -315 $\pm$ 6 & -135 $\pm$ 6 & 2.00 $\times$ 10$^{6}$ & 12.0$\pm$ 0.5 & 6.1 $\times$ 3.7 & 1.6 $\times$ 10$^{18}$ & none\\ 
\hline 
AGESM33-21 & 01:39:41.7 & + 29:58:32 & 32.6 & -307 $\pm$ 5 & -127 $\pm$ 5 & 1.67 $\times$ 10$^{5}$ & 1.0$\pm$ 0.1 & 1.7 $\times$ 1.7 & 1.0 $\times$ 10$^{18}$ & none\\ 
\hline 
AGESM33-22 & 01:33:41.3 &+ 30:57:00 & 32.8 & -338 $\pm$ 15 & -158 $\pm$ 15 & 9.99 $\times$ 10$^{5}$ & 6.0$\pm$ 0.6  & 6.6 $\times$ 3.7 & 7.2 $\times$ 10$^{17}$ & none \\ 
\hline 
AGESM33-23 & 01:37:47.2 &+ 30:23:48 & 49.5 & -352 $\pm$ 4 & -172 $\pm$ 4 & 2.53 $\times$ 10$^{6}$ & 15.2$\pm$ 0.4 & 6.6 $\times$ 3.7 & 1.8 $\times$ 10$^{18}$ & none \\ 
\hline 
AGESM33-24 & 01:34:37.0 & + 30:08:00 & 26.7 & -328 $\pm$ 3 & -148 $\pm$ 3 & 1.18 $\times$ 10$^{6}$ & 7.1$\pm$ 0.3 & 4.9 $\times$ 3.7 & 1.1 $\times$ 10$^{18}$ & G08-AA18 \\ 
\hline 
AGESM33-25 & 01:31:36.5 & + 30:13:55 & 36.9 & -324 $\pm$ 5 & -144 $\pm$ 5 & 1.78 $\times$ 10$^{6}$ & 10.7$\pm$ 0.4 & 6.1 $\times$ 4.4 & 1.2 $\times$ 10$^{18}$ & G08-AA20 \\ 
\hline 
AGESM33-26 & 01:24:32.8 & + 30:36:43 & 56.4 & -359 $\pm$ 5 & -179 $\pm$ 5 & 5.99 $\times$ 10$^{5}$ & 3.6$\pm$ 0.2 & 3.2 $\times$ 2.4 & 1.3 $\times$ 10$^{18}$ & none \\ 
\hline 
AGESM33-27 & 01:27:17.8 & + 30:05:21 & 16.9 & -377 $\pm$ 3 & -197 $\pm$ 3 & 6.66 $\times$ 10$^{5}$ & 4.0$\pm$0.2 & 7.3 $\times$ 2.4 & 6.5 $\times$ 10$^{17}$ & none \\ 
\hline 
AGESM33-28 & 01:38:04.7 & + 29:56:46 & 44.8 & -335 $\pm$ 5 & -155 $\pm$ 5 & 2.26 $\times$ 10$^{6}$ & 13.6$\pm$ 0.5 & 9.7 $\times$ 3.7 & 1.1 $\times$ 10$^{18}$ & none \\ 
\hline 
AGESM33-29 & 01:33:18.6 &+ 29:29:59 & 71.1 & -335 $\pm$ 7 & -155 $\pm$ 7 & 5.16 $\times$ 10$^{5}$ & 3.1$\pm$0.2 & 3.7 $\times$ 2.4 & 9.9 $\times$ 10$^{17}$ & G08-AA19 \\ 
\hline 
AGESM33-30 & 01:40:20.3 &+ 29:11:24 & 24.5 & -341 $\pm$ 3 & -161 $\pm$ 3 & 5.00 $\times$ 10$^{5}$ & 3.0$\pm$ 0.2 & 3.7 $\times$ 2.4 & 9.8 $\times$ 10$^{17}$  & none \\ 
\hline 
AGESM33-31 & 01:37:00:2 & + 30:01:52 & 17.7 & -374 $\pm$ 3 & -194 $\pm$ 3 & 1.22 $\times$ 10$^{7}$ & 73.5$\pm$ 0.5 & 18.2 $\times$ 14.6 & 7.9 $\times$ 10$^{17}$ & G08-AA21a, 21b\\ 
\hline 
AGESM33-32 & 01:24:06.1 & +29:38:35 & 25.4 & -383 $\pm$ 3 & -203 $\pm$ 3 & 4.54 $\times$ 10$^{7}$ & 272.6$\pm$0.7 & 10.9 $\times$ 10.9  & 6.5 $\times$ 10$^{18}$ & Wright's cloud \\ 
\hline 
\end{tabular} 
\caption{The above table gives our derived properties for all 'isolated' clouds. The columns are: 1) name, 2) and 3) co-ordinates in right ascension and declination (J2000), 4) the velocity width  at the FWHM (50$\%$), 5) the central velocity of the cloud given by {\sc mbspect} , 6) the velocity relative to M33 (the velocity of M33 is taken to be -180 km s$^{-1}$, 7) the \HI\ mass at the distance of M33, 8) the total measured flux density,  9) the measured size of the cloud along the semi-major and semi-minor axes in kpc. These were measured to the nearest pixel which corresponds to $\sim$0.2 kpc 10) the peak column density of the cloud as measured by our 3.5 arcmin beam and 11) whether the cloud has been previously detected by another survey. N.B. values given for AGESM33-32 (Wright's cloud) only measure the amount of the cloud visible in our cube and not the entirety. Where necessary a distance to the clouds of 840 kpc is assumed (\citealt{1991ApJ...372..455F}). The errors on velocities and fluxes were calculated with parameters from {\sc mbspect} using equations given in \citet{2006MNRAS.371.1617A}. Where the error on the velocity was calculated to be $<$3 km s$^{-1}$ the error is given as 3 km s$^{-1}$. This is because the central velocity is accurate to with half a channel.}
\end{table}
\end{landscape}
\pagebreak

\begin{landscape}
\begin{table}
\begin{tabular}{|c|c|c|c|c|c|c|c|c|c|c|c|c|}
\hline
AGES & RA & Dec & Velocity & Systemic & Velocity & HI Mass & Flux & Size & Column & G08  \\
ID & (J2000) & (J2000) & Width 50  & Velocity & Relative to M33 & (M$_{\odot}$)& (Jy km s$^{-1}$) & (x,y kpc) & Density& ID \\
& & & (km s$^{-1}$)  & (km s$^{-1}$) & (km s$^{-1}$) & & & & (cm$^{-2}$) & \\
(1) & (2) & (3) & (4) & (5) & (6)  & (7) & (8) & (9) & (10) & (11) \\
\hline 
AGESM33-4 & 01:37:56.4 & + 30:10:15 & 29.4 & -136 $\pm$ 3 & 44 $\pm$ 3 & 5.08 $\times$ 10$^{6}$ & 30.5$\pm$0.3 & 6.1 $\times$ 2.2 & 6.60 $\times$ 10$^{18}$ & AA4 \\ 
\hline 
AGESM33-6 & 01:32:30.6 & + 29:26:56 & 22.7 & -149 $\pm$ 3 & 31 $\pm$ 3 & 5.48 $\times$ 10$^{6}$ & 32.9$\pm$0.4 & 3.2 $\times$ 3.1 & 9.4 $\times$ 10$^{18}$ & AA6 \\ 
\hline 
AGESM33-7 & 01:38:11.8 & + 29:47:11 & 57.7 & -137 $\pm$ 3 & 43 $\pm$ 3 & 1.08 $\times$ 10$^{6}$ & 6.5$\pm$0.3 & 2.2 $\times$ 2.4 & 3.5 $\times$ 10$^{18}$ & AA7\\ 
\hline 
AGESM33-8 & 01:36:15.0 & + 29:58:10 & 42.3 & -158 $\pm$ 3 & 22 $\pm$ 3 & 5.33 $\times$ 10$^{5}$ & 3.2$\pm$ 0.1  & 1.5 $\times$ 1.2 & 5.2 $\times$ 10$^{18}$ & AA8 \\ 
\hline 
AGESM33-11 & 01:34:18.8 & + 31:32:00 & 43.8 & -191 $\pm$ 3 & -11 $\pm$ 3 & 7.49 $\times$ 10$^{5}$ & 4.5$\pm$ 0.1 & 1.9 $\times$ 1.2 & 5.50 $\times$ 10$^{17}$ & AA13 \\ 
\hline 
AGESM33-13 & 01:38:26.0 & + 30:50:37 & 36.6 & -191 $\pm$ 6 & -11 $\pm$ 6 & 2.33 $\times$ 10$^{5}$ & 1.4$\pm$0.1 & 1.2 $\times$ 1.2 & 2.7 $\times$ 10$^{18}$ & AA11 \\ 
\hline 
AGESM33-15 & 01:29:56.6 & + 31:16:46 & 35.6 & -199 $\pm$ 3 & -19 $\pm$ 3 & 2.61 $\times$ 10$^{6}$ & 15.7$\pm$ 0.2 & 2.4 $\times$ 2.4 & 7.6 $\times$ 10$^{18}$ & AA12\\ 
\hline 
AGESM33-16 & 01:27:51.3 & + 31:31:21 & 34.1 & -245 $\pm$ 3 & -65 $\pm$ 3 & 3.73 $\times$ 10$^{6}$ & 22.4$\pm$0.2 & 3.7 $\times$ 3.1 & 5.6 $\times$ 10$^{18}$ & AA14 \\ 
\hline 
AGESM33-17 & 01:28:49.8 & + 31:40:37 & 27.9 & -263 $\pm$ 3 & -83 $\pm$ 3 & 1.15 $\times$ 10$^{6}$ & 6.9$\pm$0.1 & 3.7 $\times$ 2.4 & 2.2 $\times$ 10$^{18}$ & AA15\\ 
\hline 
AGESM33-18 & 01:35:28.9 & + 30:43:17 & 34.6 & -292 $\pm$ 6 & -112 $\pm$ 6 & 9.66 $\times$ 10$^{5}$ & 5.8$\pm$ 0.5 & 2.4 $\times$ 2.4 & 2.8 $\times$ 10$^{18}$ & AA16 \\ 
\hline 
\end{tabular} 
\caption{The above table gives our derived properties of all clouds listed by Grossi et al 2008 which are either part of the disc of M33 or joined to M33 by low column density \HI\ . All columns are the same as in table 1, with the exception of column 11 which is the cloud ID given in G08.}
\end{table}
\end{landscape}

\section{Appendix A: Column density maps, velocity maps and spectra of the clouds around M33}

For all figures below image (a) is an integrated flux map, (b) is a velocity map and (c) is the spectrum taken from {\sc MBSPECT} with the cloud marked. In all instances except AGESM33-8, AGESM33-15 and AGESM33-18 the peak can be clearly distinguished. These are G08 clouds AA8, AA12 and AA16. The integrated flux maps show clouds AGESM33-8 and AGESM33-15 to be joined to the disk of M33 by low column density hydrogen. On inspection of the integrated flux map AGESM33-18 is indistinguishable from the disk of M33. It has been included in this paper as it can been seen as a very small peak in the spectra of M33.

\begin{figure}
\subfloat[]{\includegraphics[width = 3in]{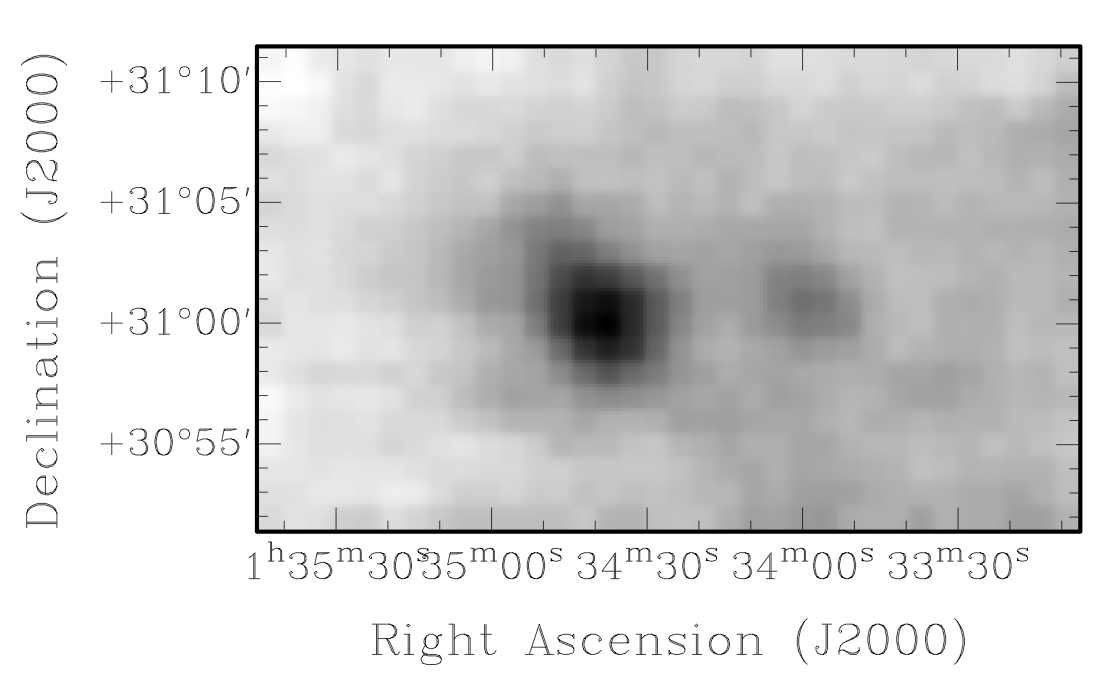}}\
\subfloat[]{\includegraphics[width = 3.4in]{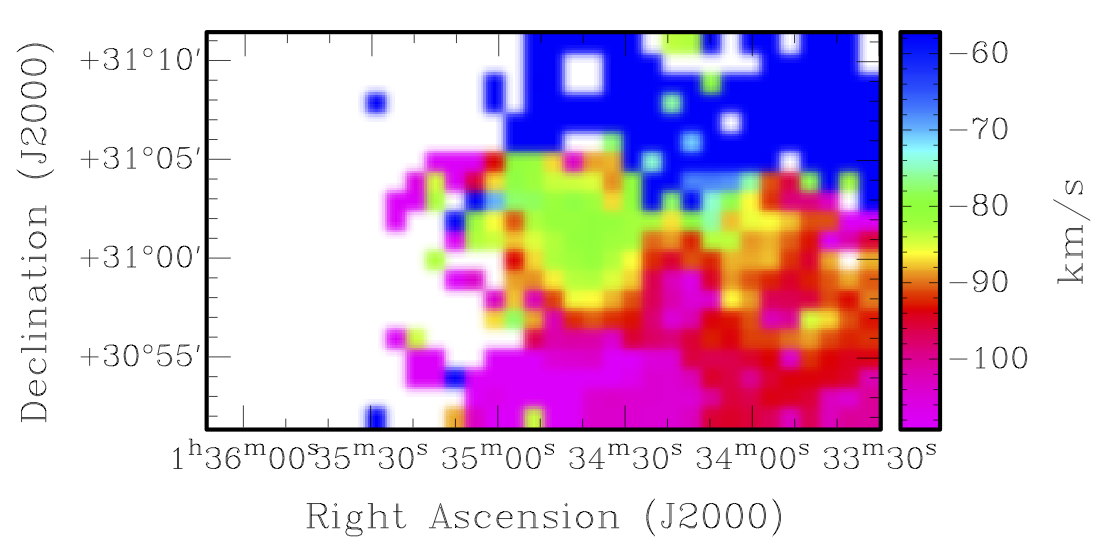}}
\newline
\centering
\subfloat[]{\includegraphics[width = 2.5in]{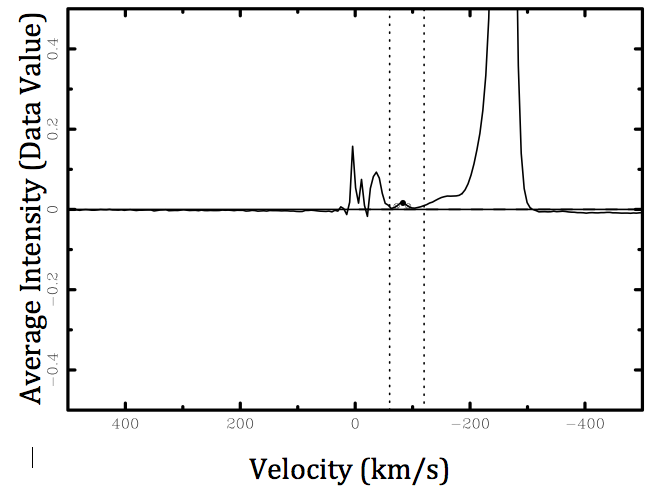}}
\caption{AGESM33-1}
\label{ok1_all}
\end{figure}

\begin{figure}
\subfloat[]{\includegraphics[width = 3in]{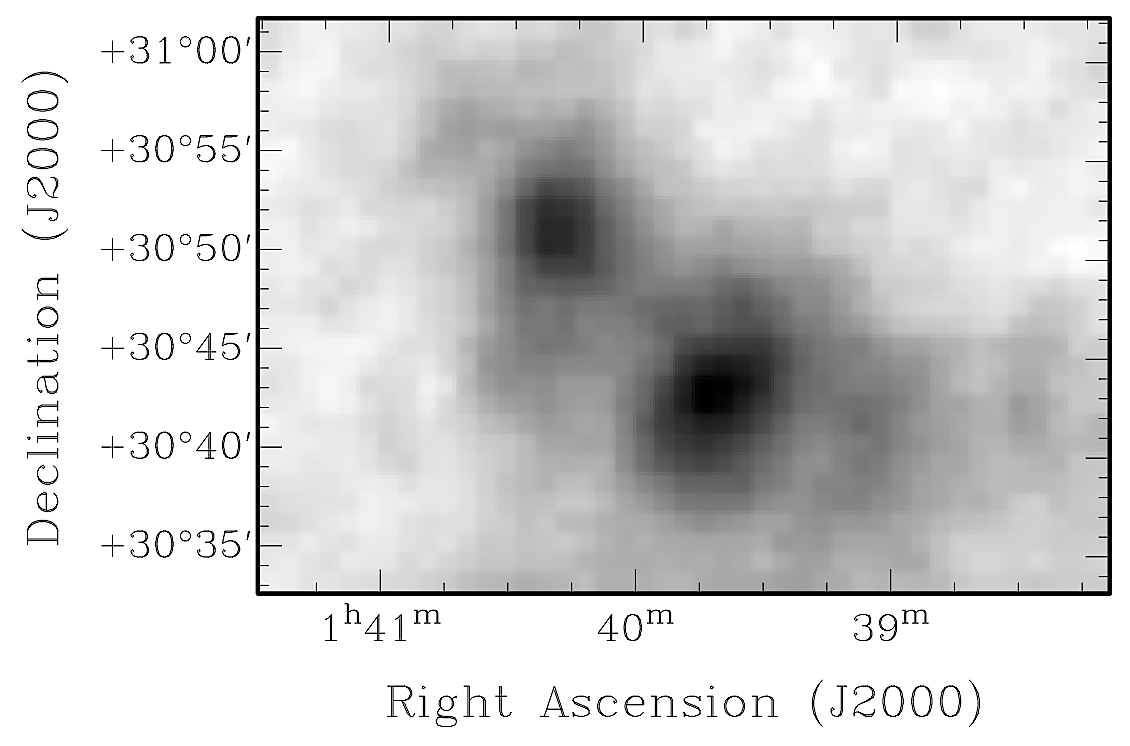}}\
\subfloat[]{\includegraphics[width = 3.4in]{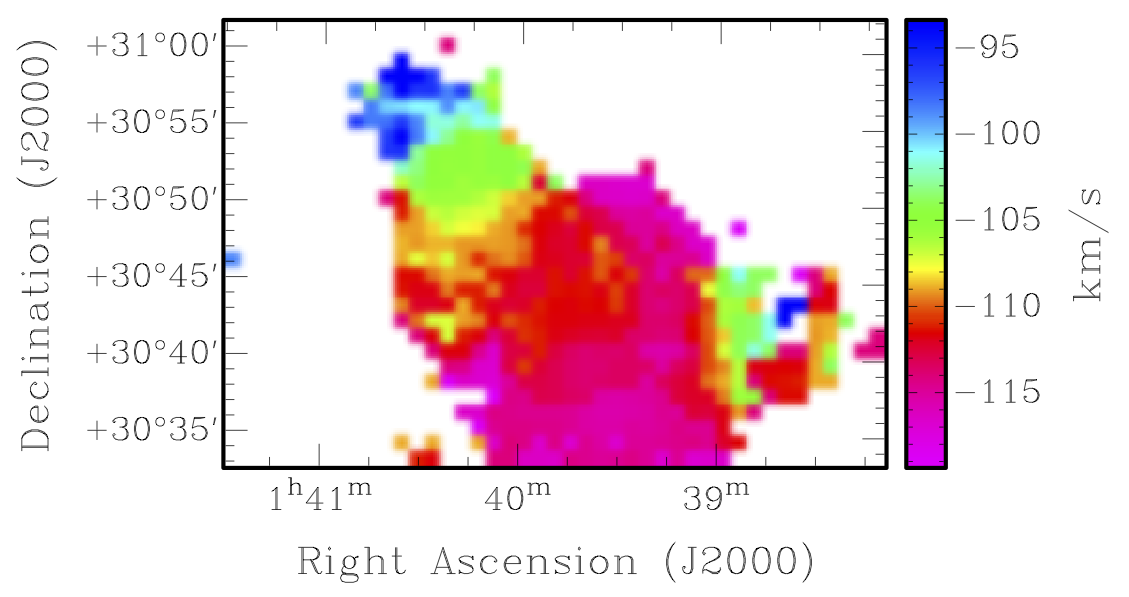}}
\newline
\centering
\subfloat[]{\includegraphics[width = 2.5in]{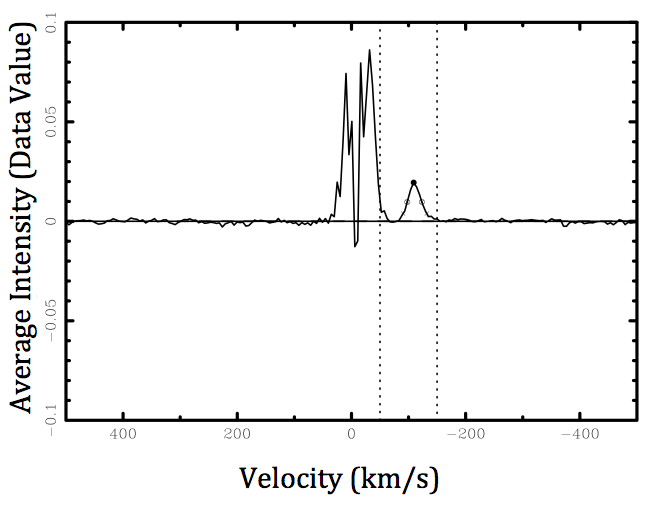}}
\caption{AGESM33-2}
\label{ok2_all}
\end{figure}

\begin{figure}
\subfloat[]{\includegraphics[width = 2.5in]{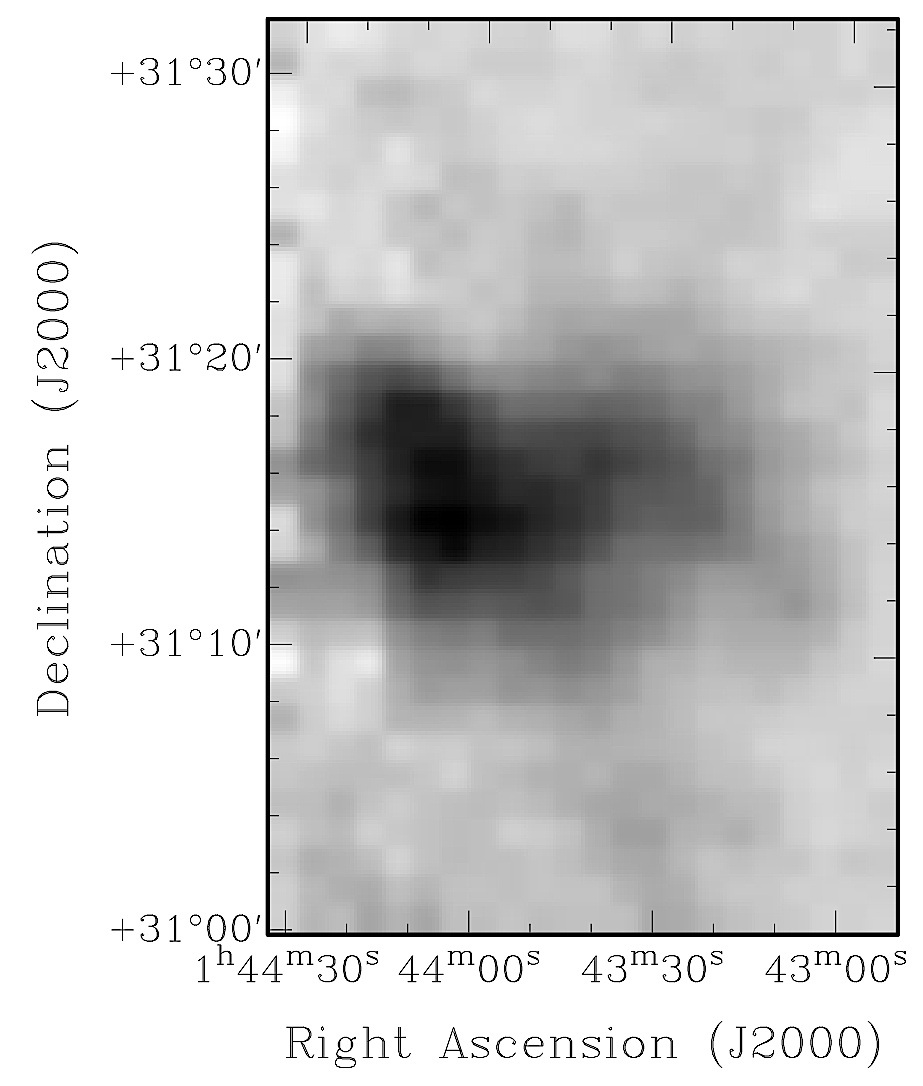}}\
\subfloat[]{\includegraphics[width = 3in]{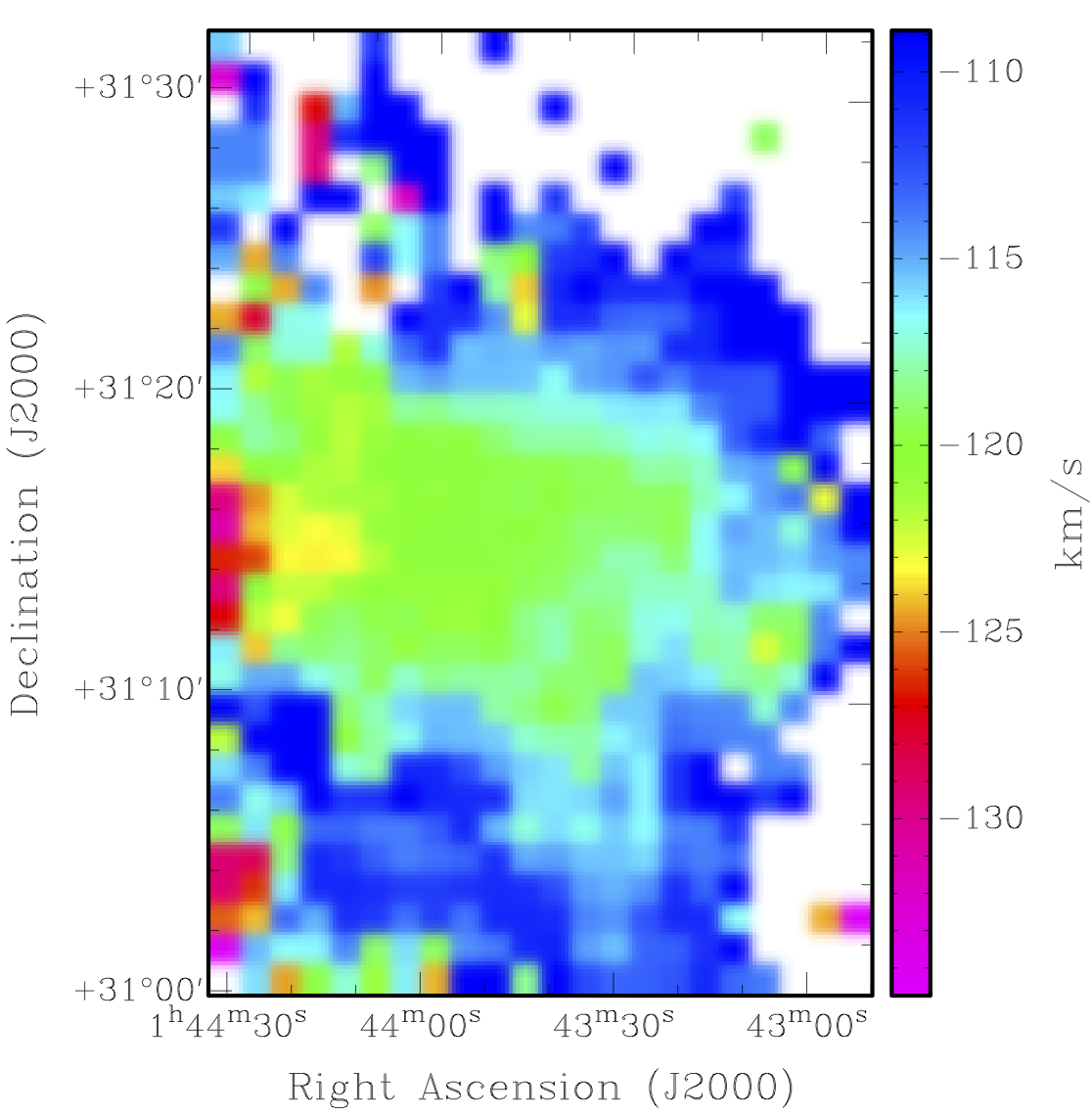}}
\newline
\centering
\subfloat[]{\includegraphics[width = 2.5in]{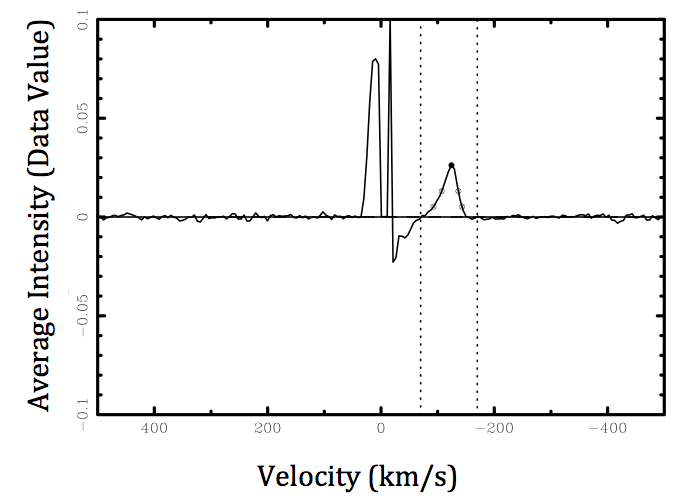}}
\caption{AGESM33-3}
\label{ok3_all}
\end{figure}

\begin{figure}
\subfloat[]{\includegraphics[width = 2.5in]{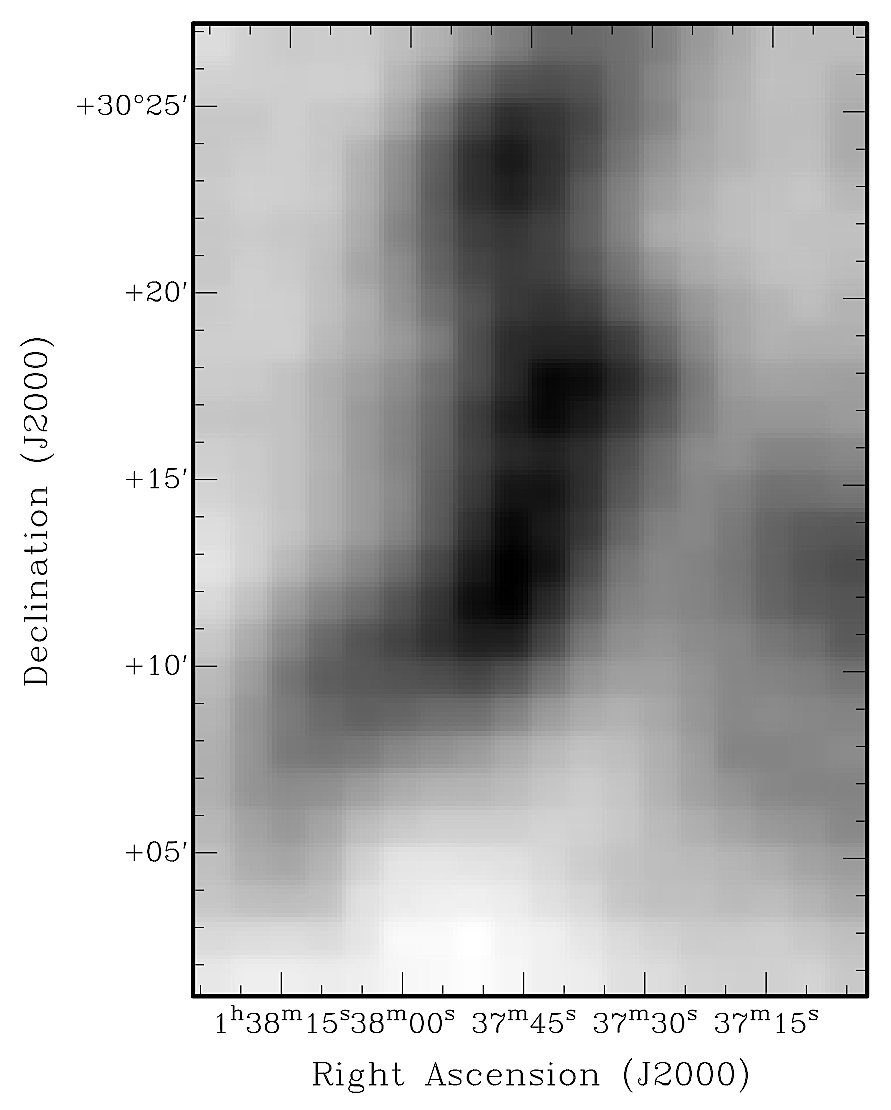}}\
\subfloat[]{\includegraphics[width = 3in]{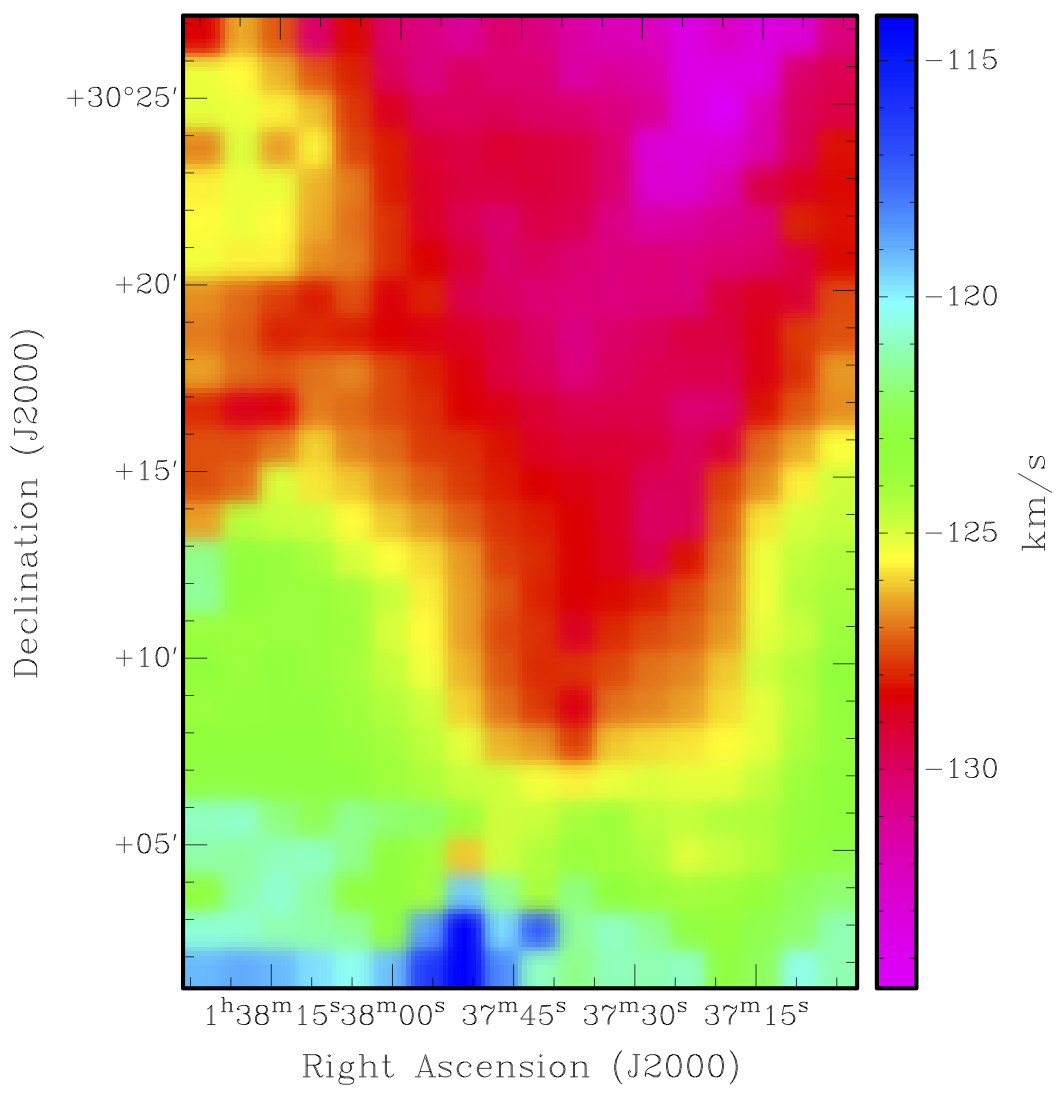}}
\newline
\centering
\subfloat[]{\includegraphics[width = 2.5in]{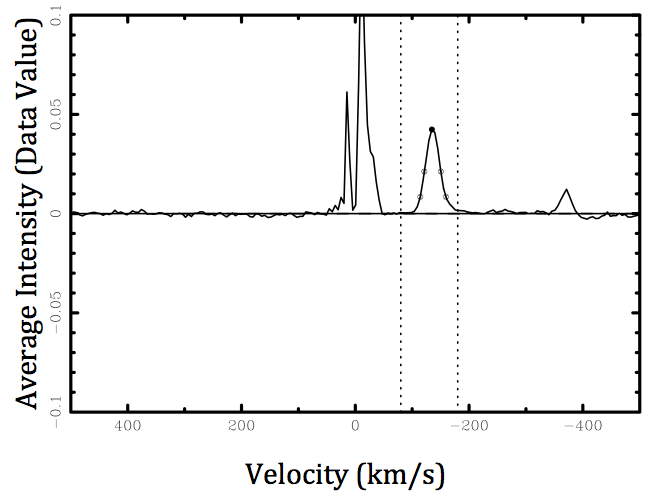}}
\caption{AGESM33-4 }
\label{ok4_all}
\end{figure}

\begin{figure}
\subfloat[]{\includegraphics[width = 3in]{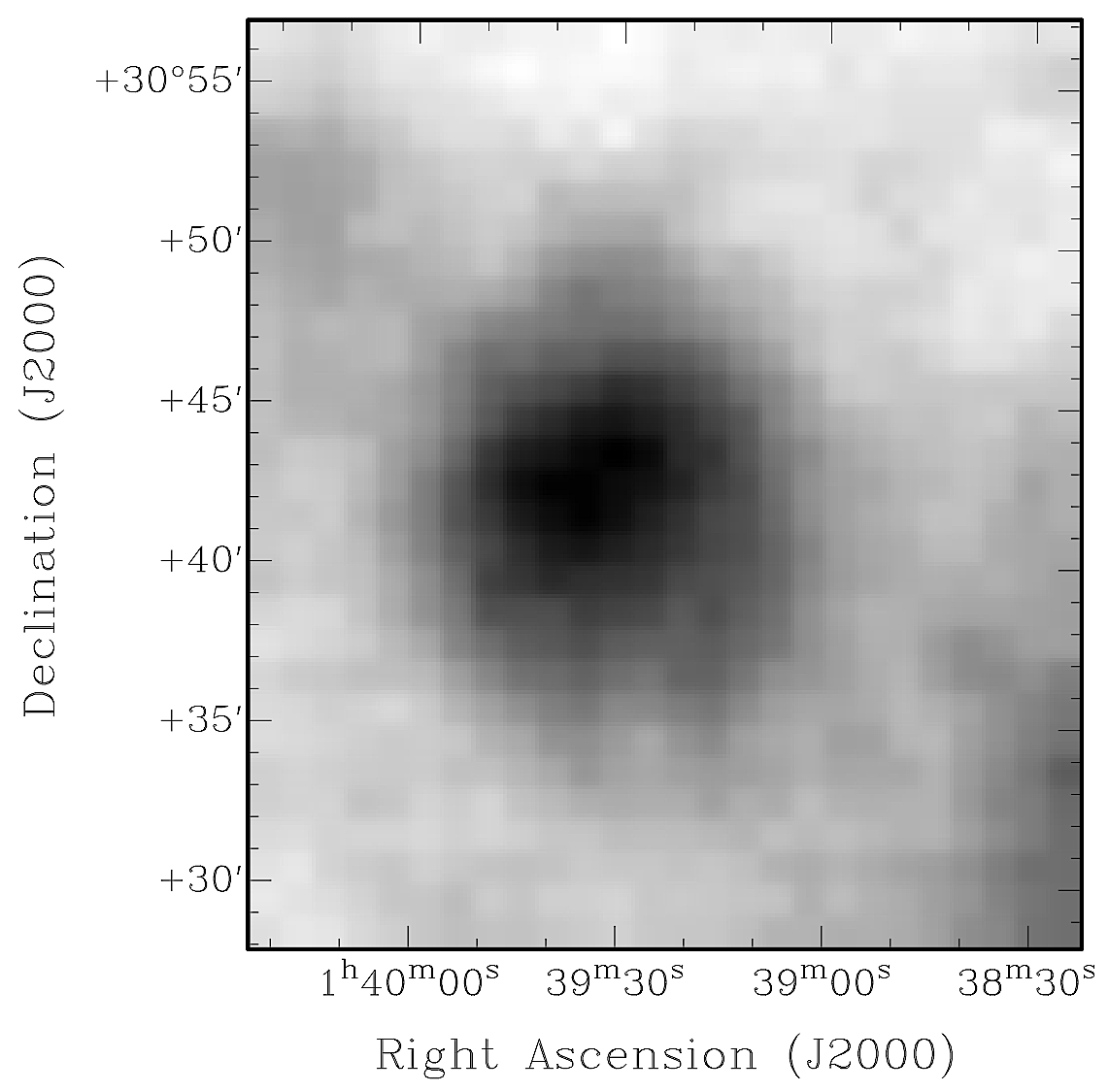}}\
\subfloat[]{\includegraphics[width = 3.4in]{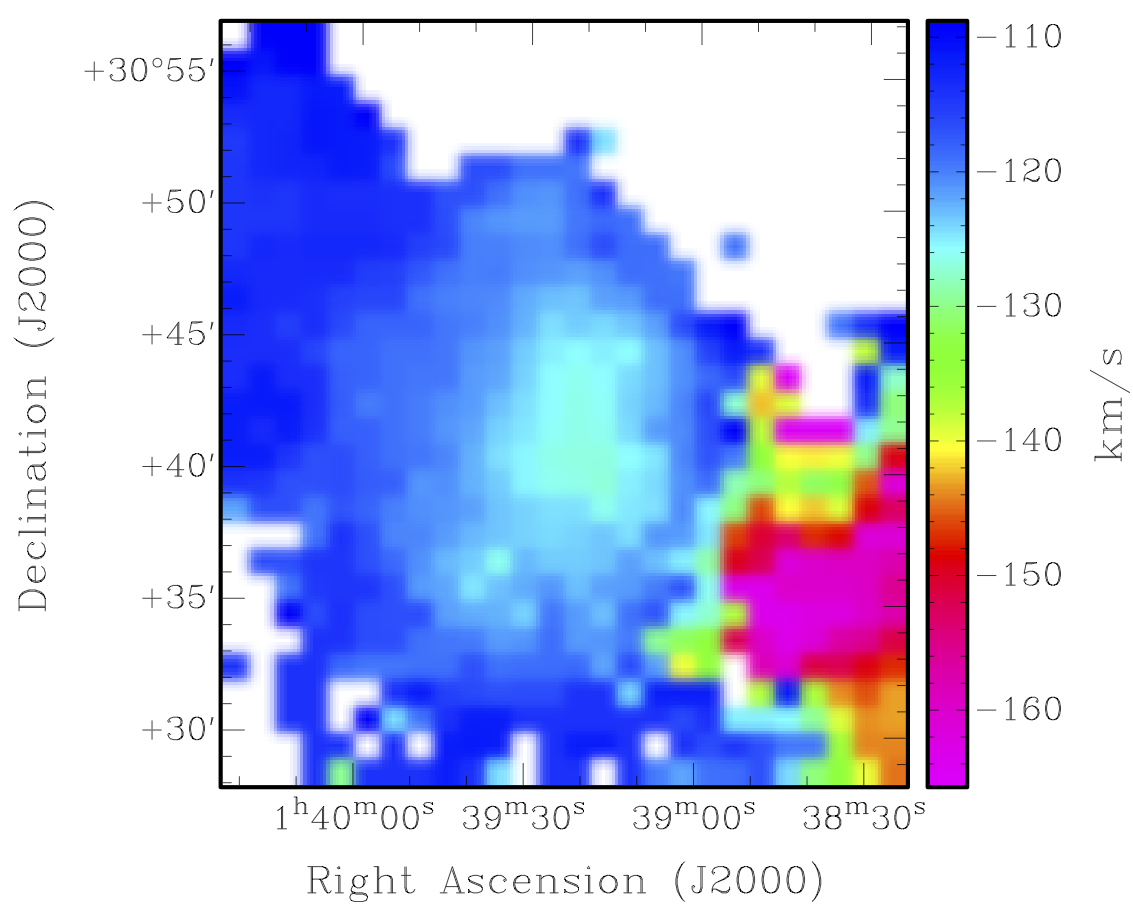}}
\newline
\centering
\subfloat[]{\includegraphics[width = 2.5in]{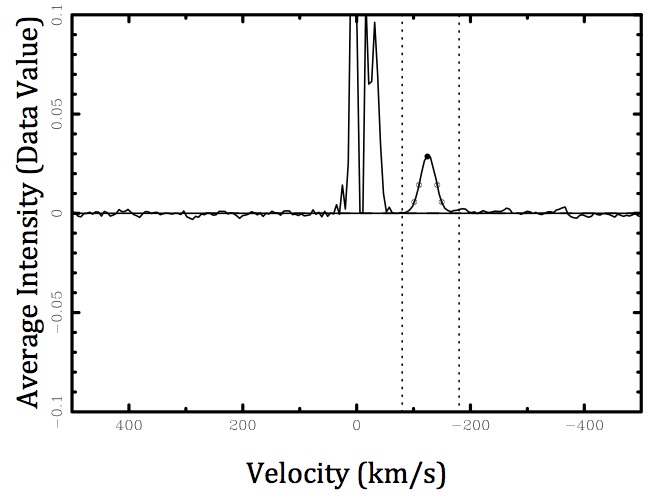}}
\caption{AGESM33-5 }
\label{ok5_all}
\end{figure}

\begin{figure}
\subfloat[]{\includegraphics[width = 3in]{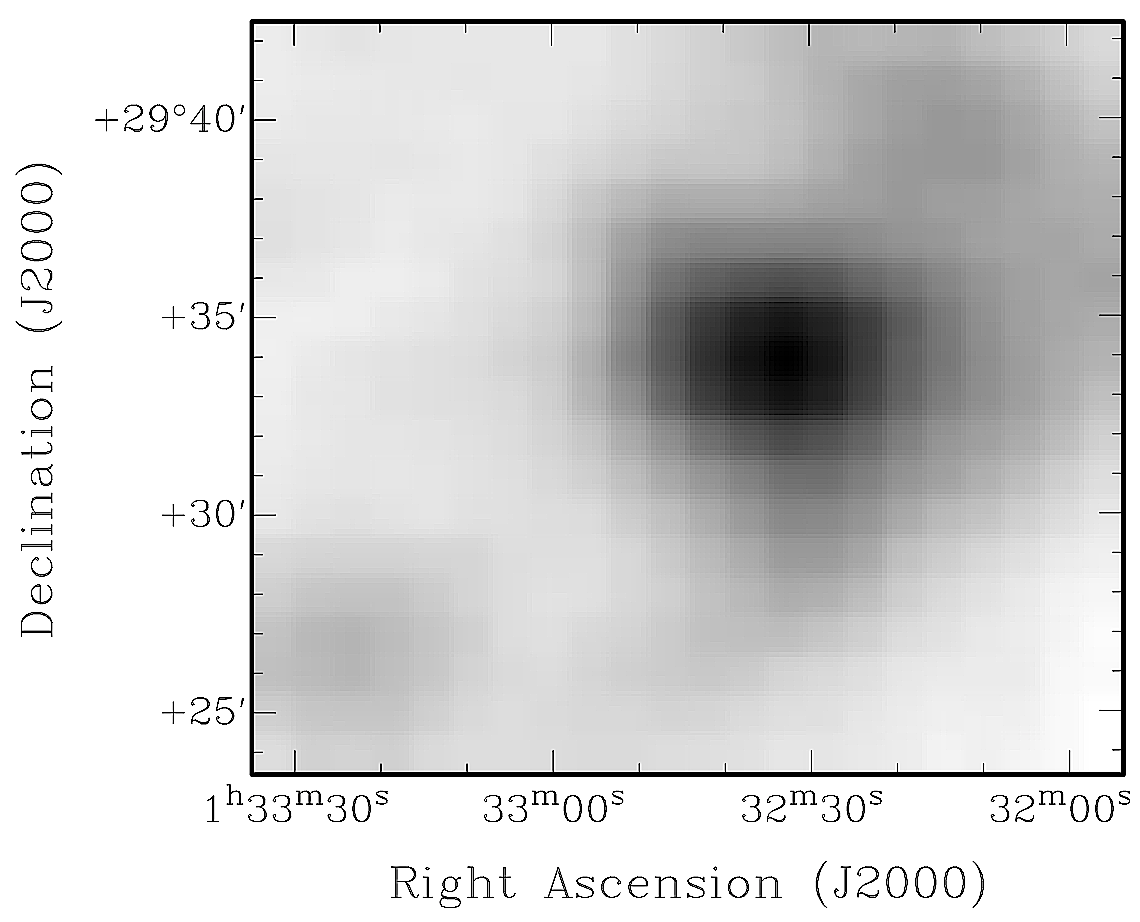}}\
\subfloat[]{\includegraphics[width = 3.4in]{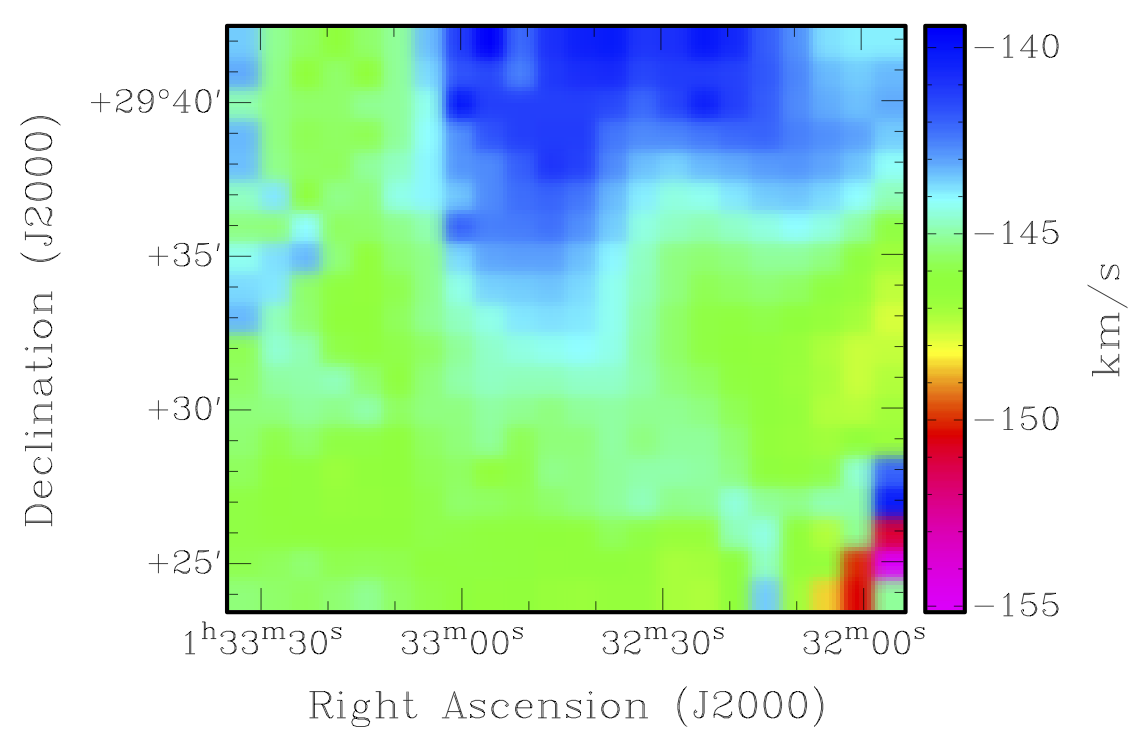}}
\newline
\centering
\subfloat[]{\includegraphics[width = 2.5in]{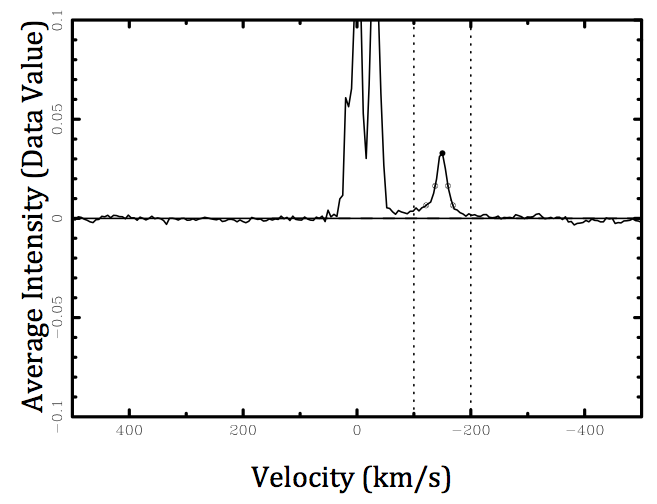}}
\caption{AGESM33-6 }
\label{ok6_all}
\end{figure}

\begin{figure}
\subfloat[]{\includegraphics[width = 3in]{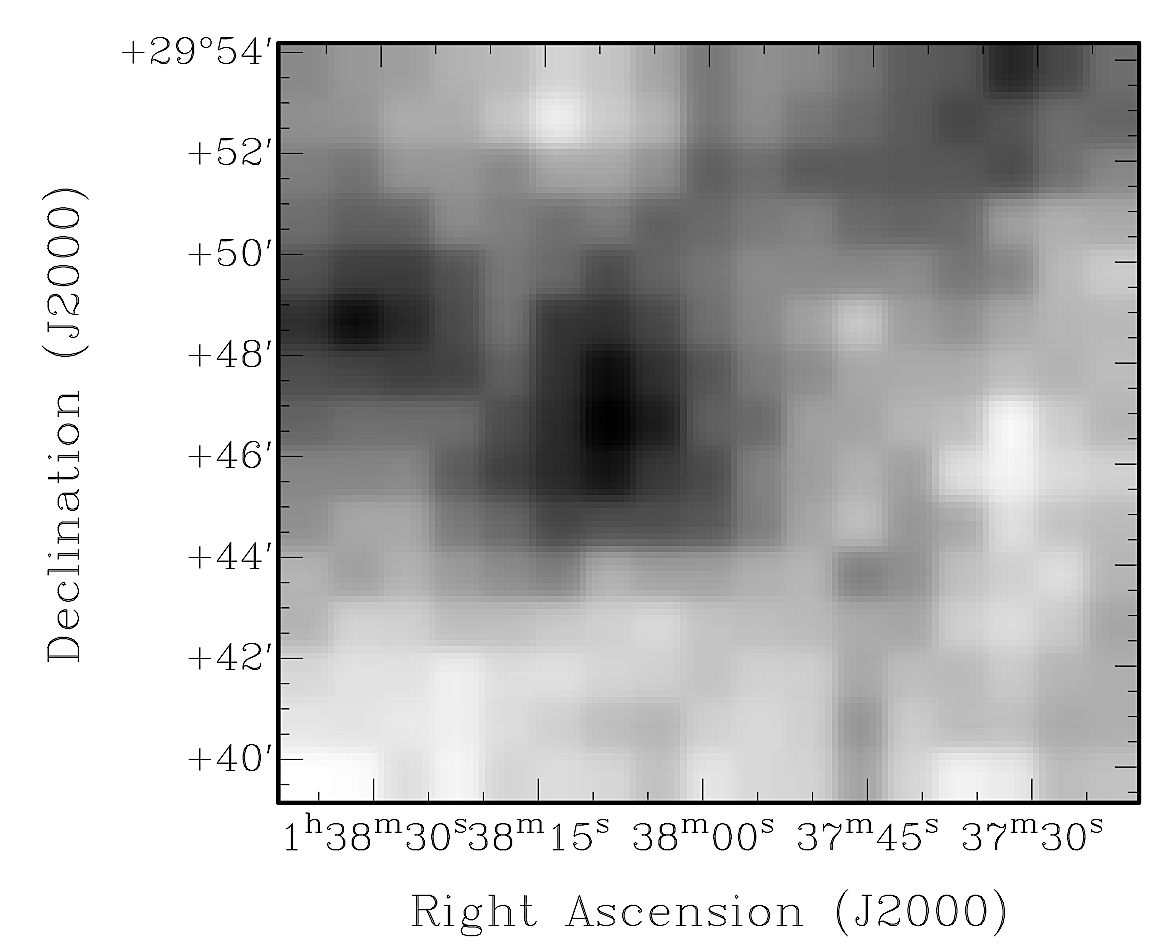}}\
\subfloat[]{\includegraphics[width = 3.4in]{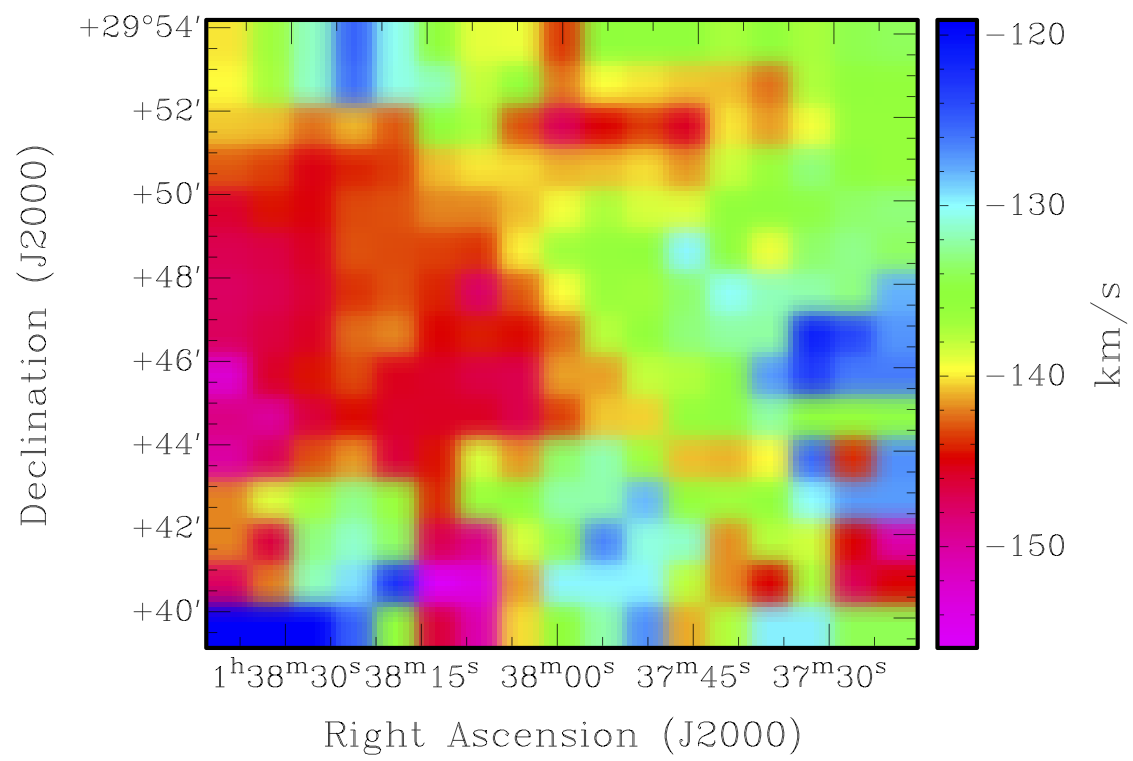}}
\newline
\centering
\subfloat[]{\includegraphics[width = 2.5in]{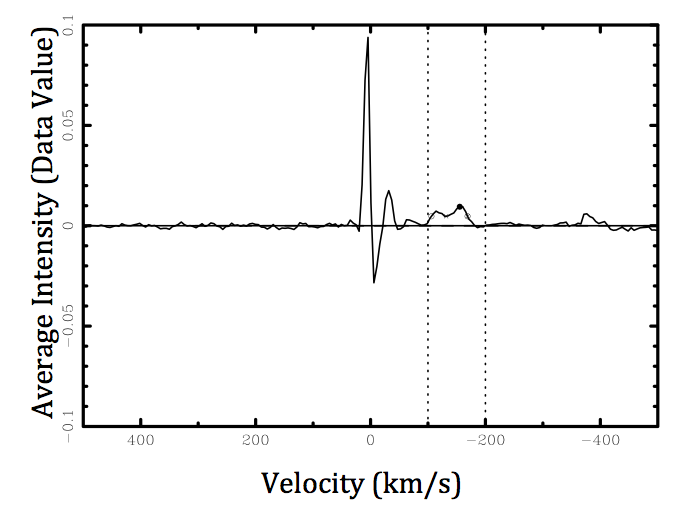}}
\caption{AGESM33-7 }
\label{ok7_all}
\end{figure}

\begin{figure}
\subfloat[]{\includegraphics[width = 3in]{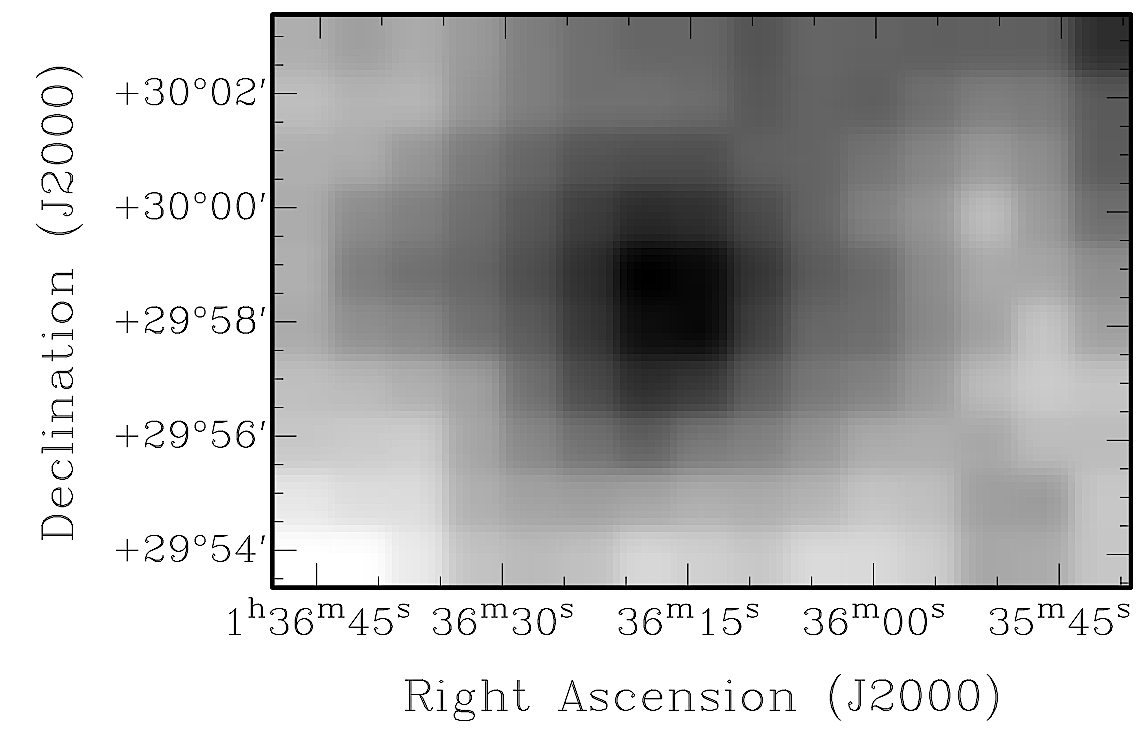}}\
\subfloat[]{\includegraphics[width = 3.4in]{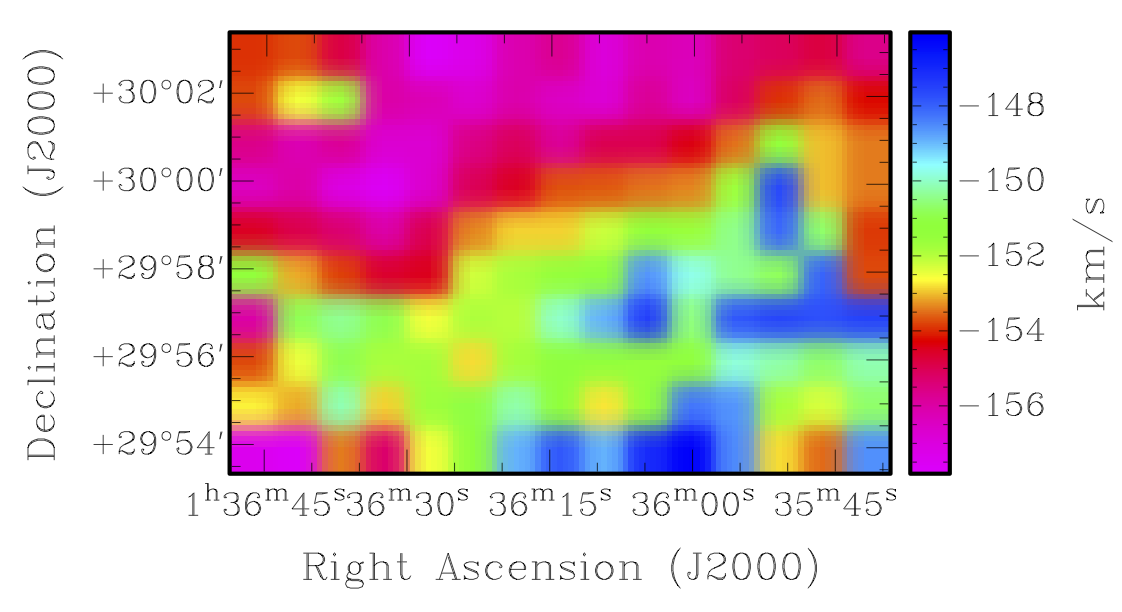}}
\newline
\centering
\subfloat[]{\includegraphics[width = 2.5in]{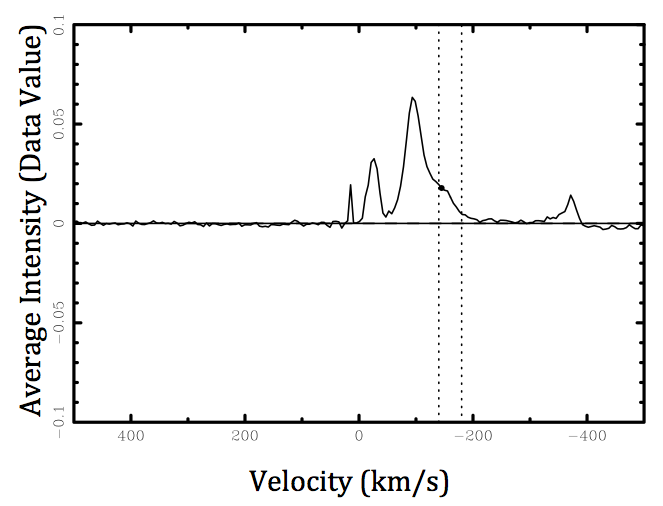}}
\caption{AGESM33-8 }
\label{ok8_all}
\end{figure}

\begin{figure}
\subfloat[]{\includegraphics[width = 3in]{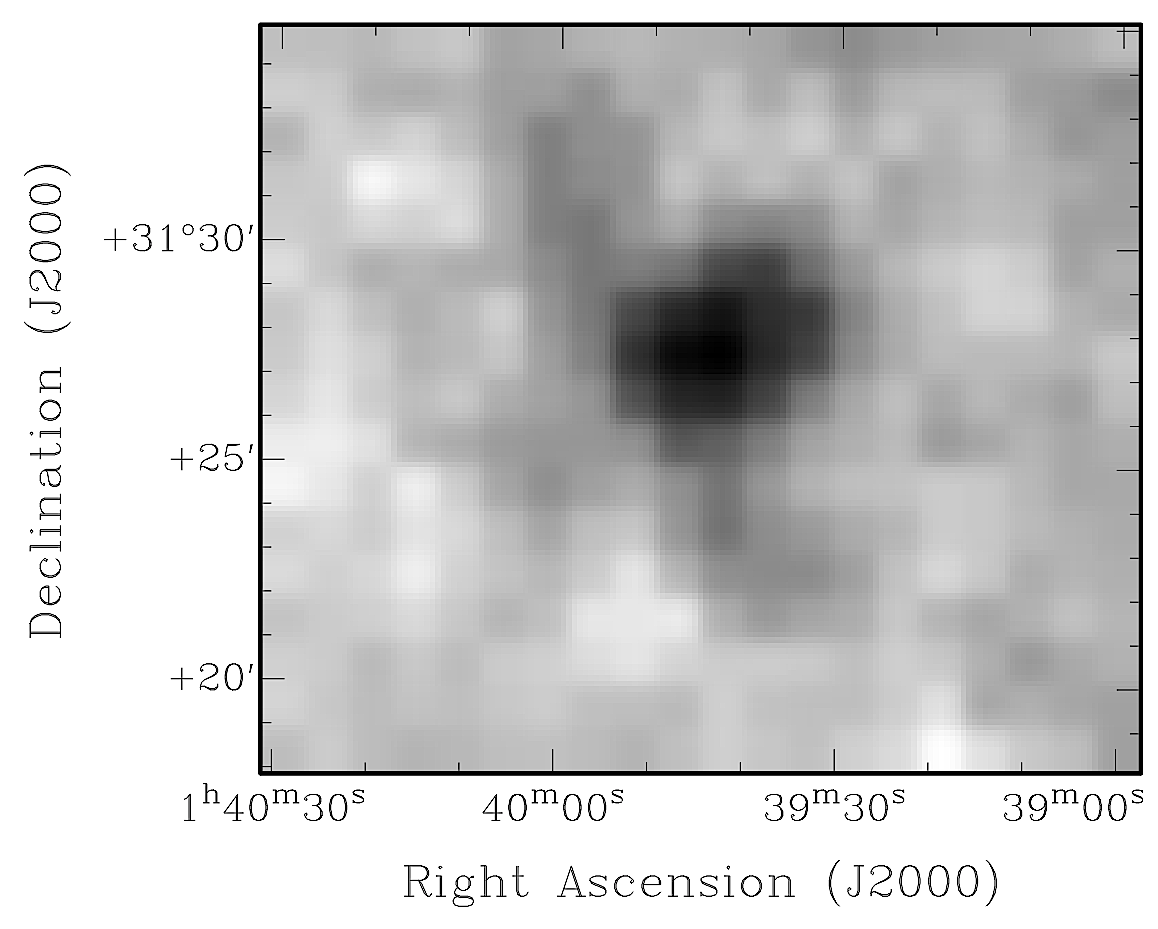}}\
\subfloat[]{\includegraphics[width = 3.4in]{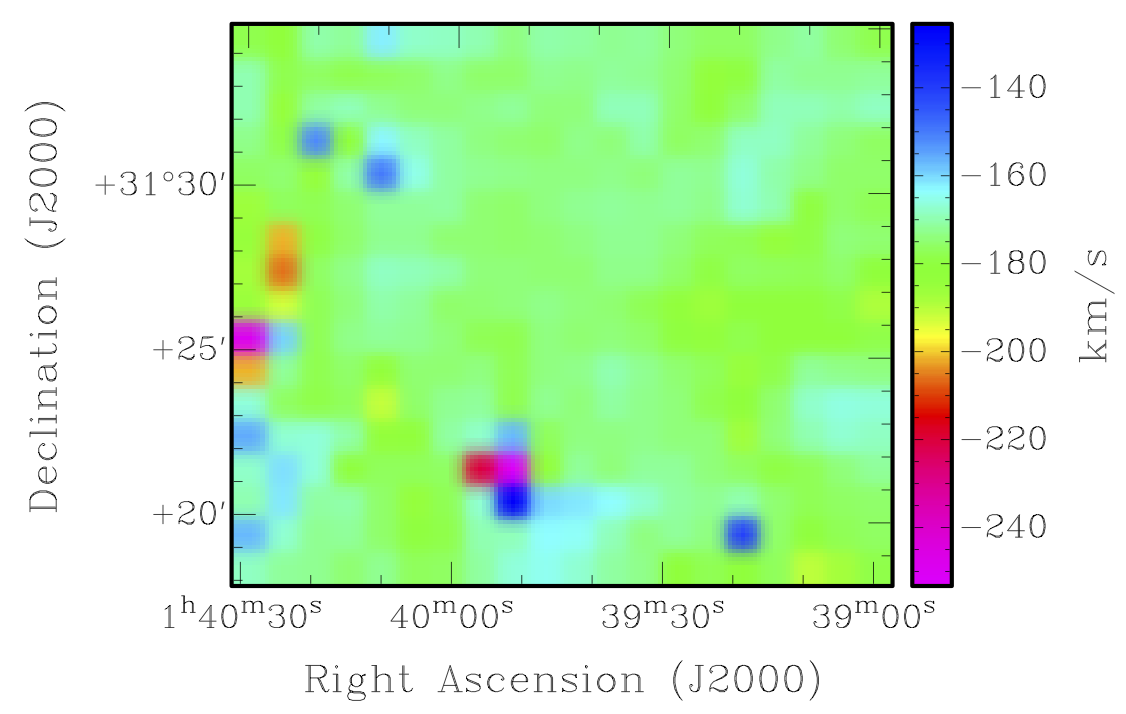}}
\newline
\centering
\subfloat[]{\includegraphics[width = 2.5in]{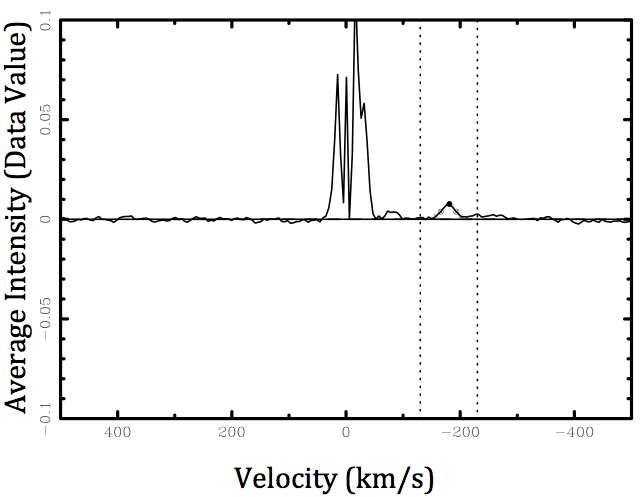}}
\caption{AGESM33-9}
\label{ok9_all}
\end{figure}

\begin{figure}
\subfloat[]{\includegraphics[width = 3in]{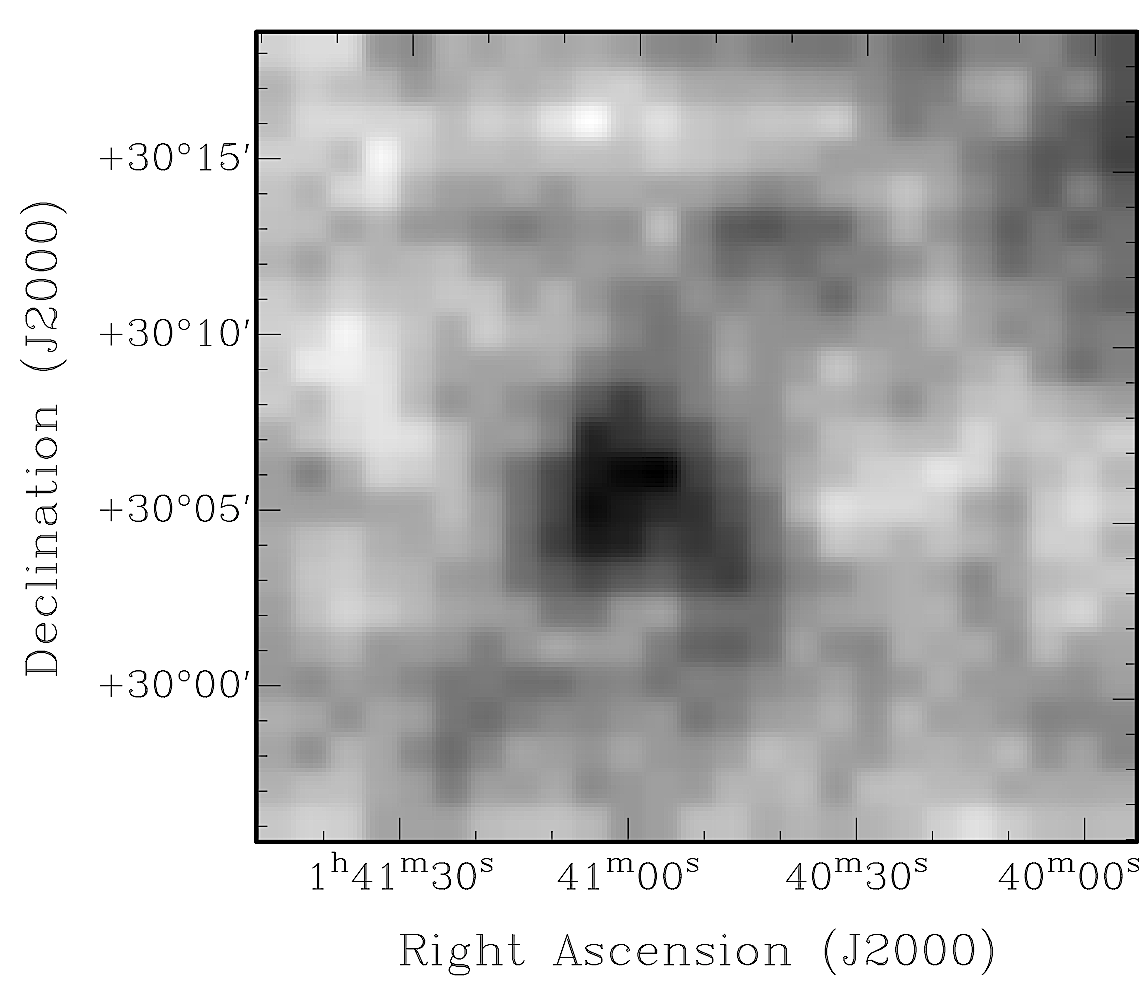}}\
\subfloat[]{\includegraphics[width = 3.4in]{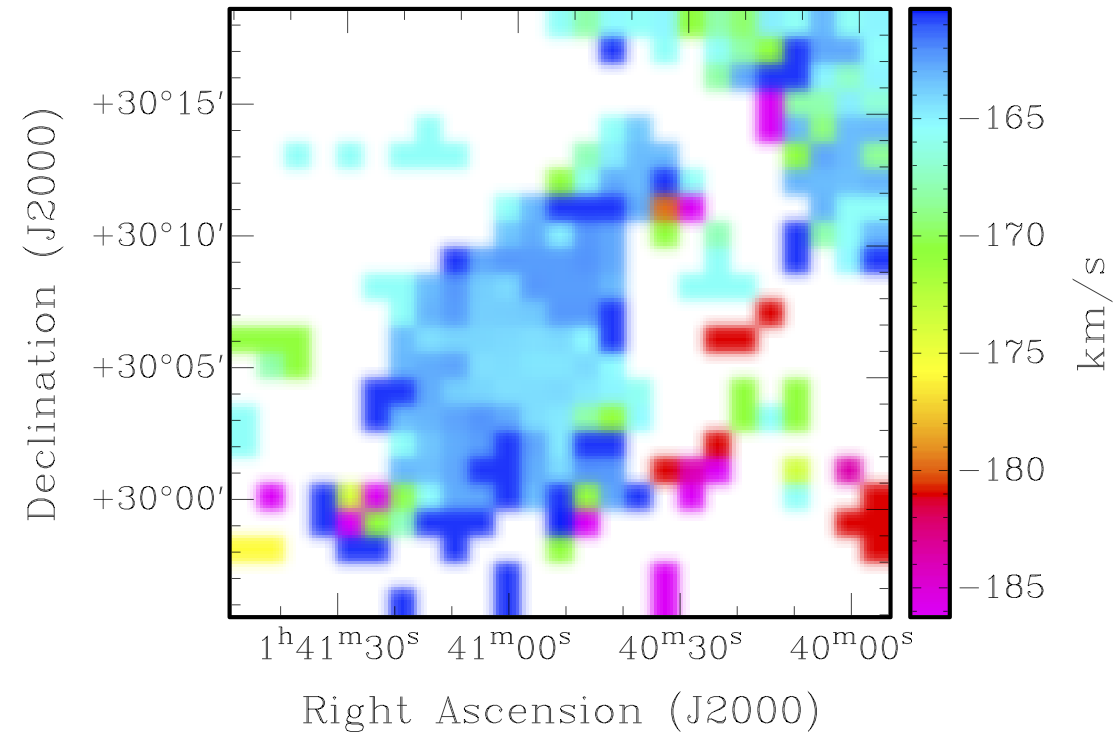}}
\newline
\centering
\subfloat[]{\includegraphics[width = 2.5in]{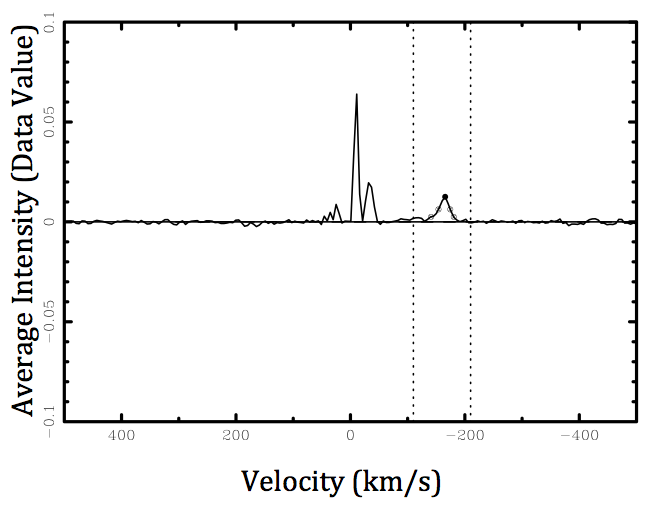}}
\caption{AGESM33-10 }
\label{ok10_all}
\end{figure}

\begin{figure}
\subfloat[]{\includegraphics[width = 3in]{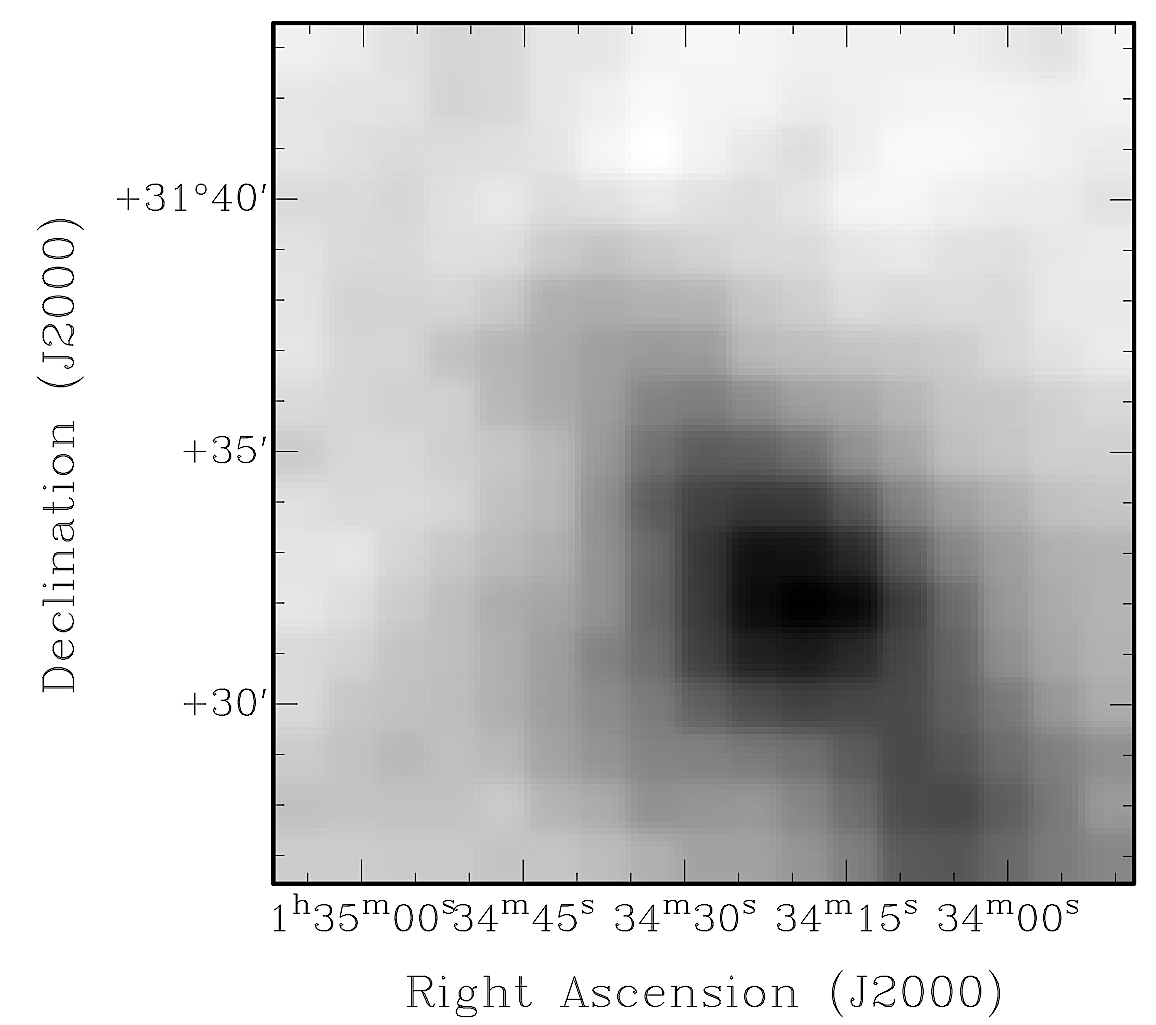}}\
\subfloat[]{\includegraphics[width = 3.4in]{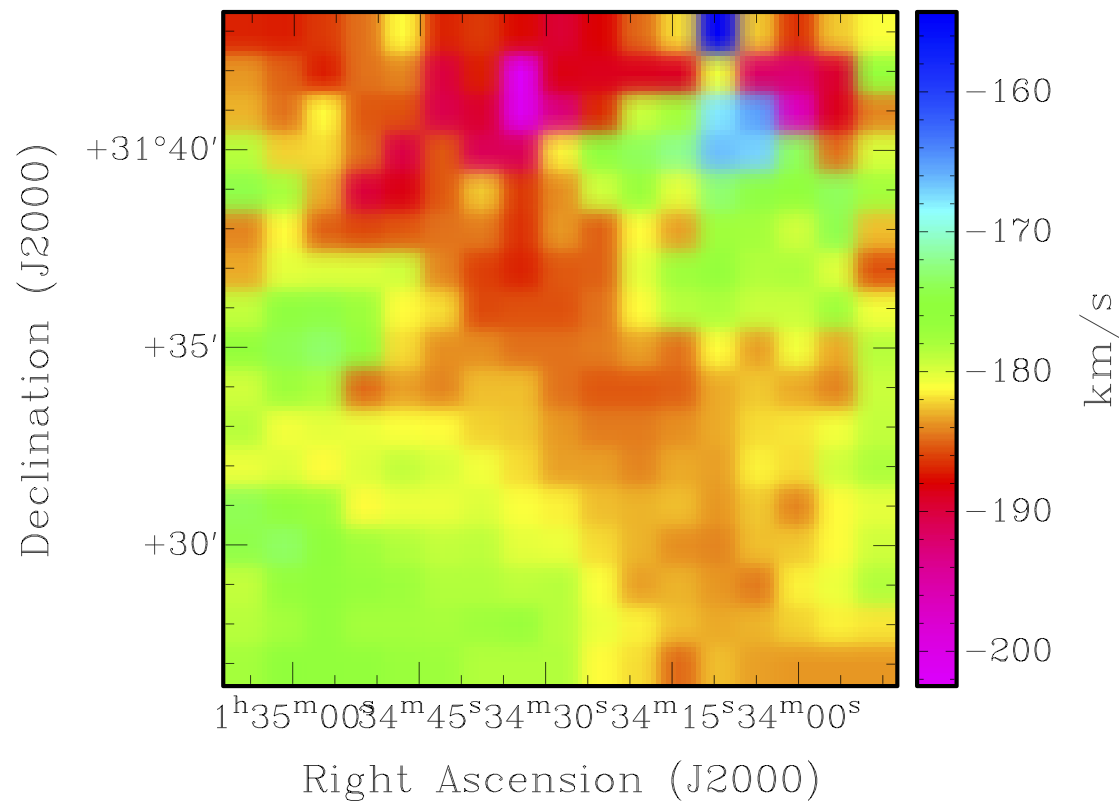}}
\newline
\centering
\subfloat[]{\includegraphics[width = 2.5in]{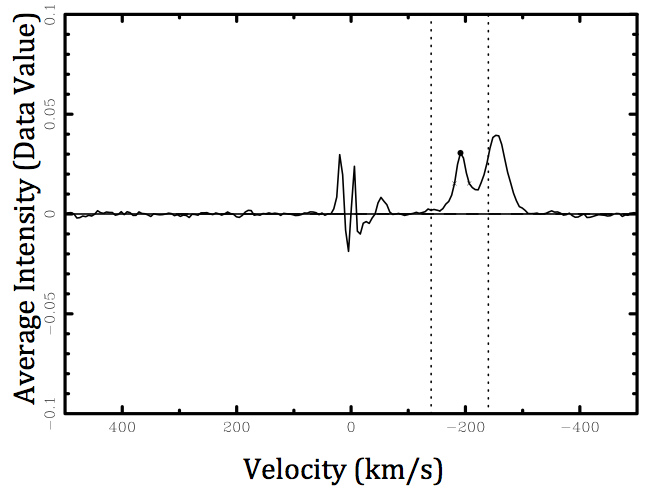}}
\caption{AGESM33-11 }
\label{ok11_all}
\end{figure}

\begin{figure}
\subfloat[]{\includegraphics[width = 3in]{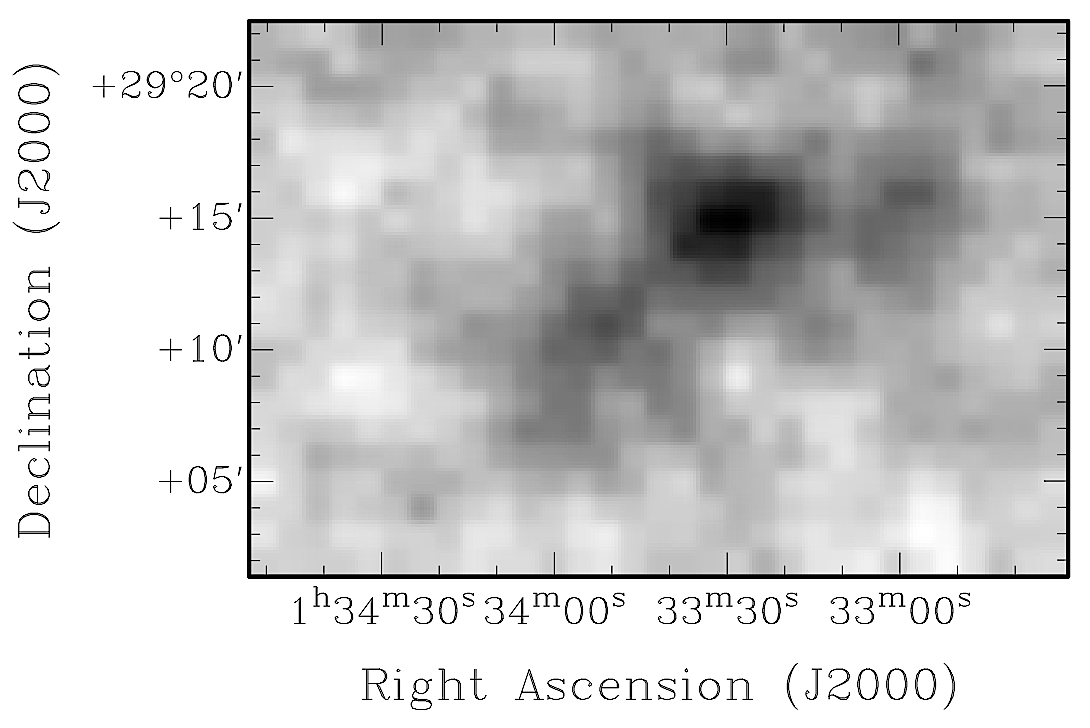}}\
\subfloat[]{\includegraphics[width = 3.4in]{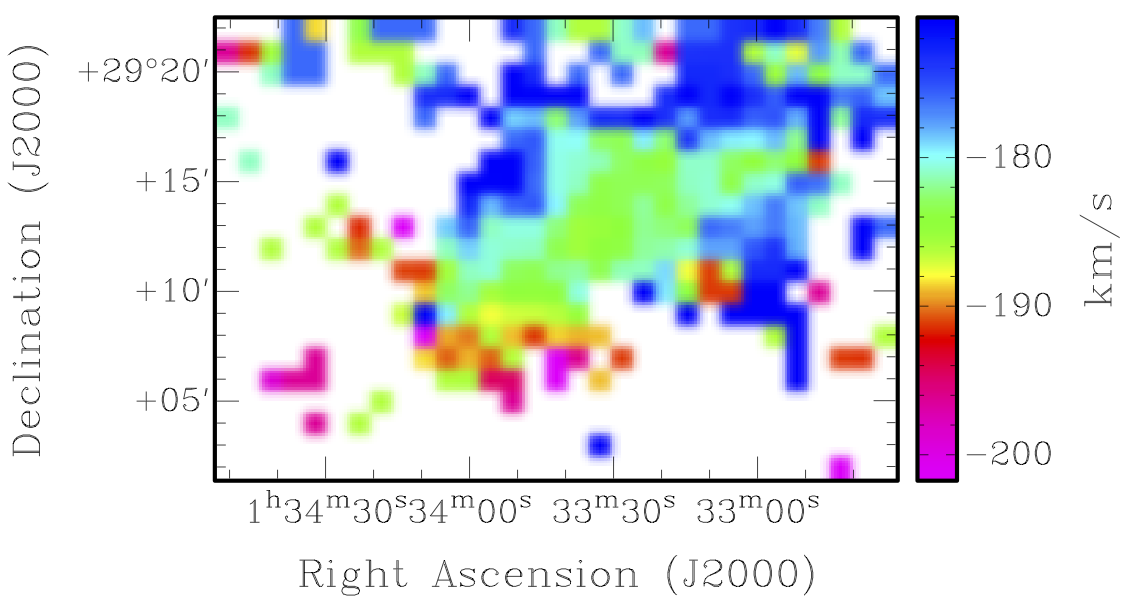}}
\newline
\centering
\subfloat[]{\includegraphics[width = 2.5in]{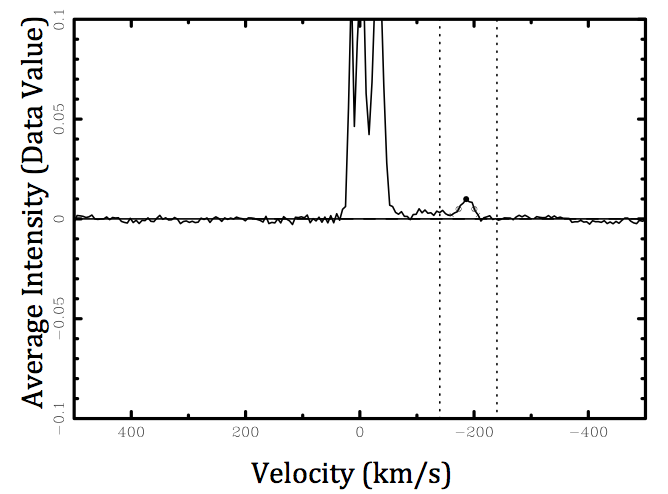}}
\caption{AGESM33-12 }
\label{ok12_all}
\end{figure}

\begin{figure}
\subfloat[]{\includegraphics[width = 3in]{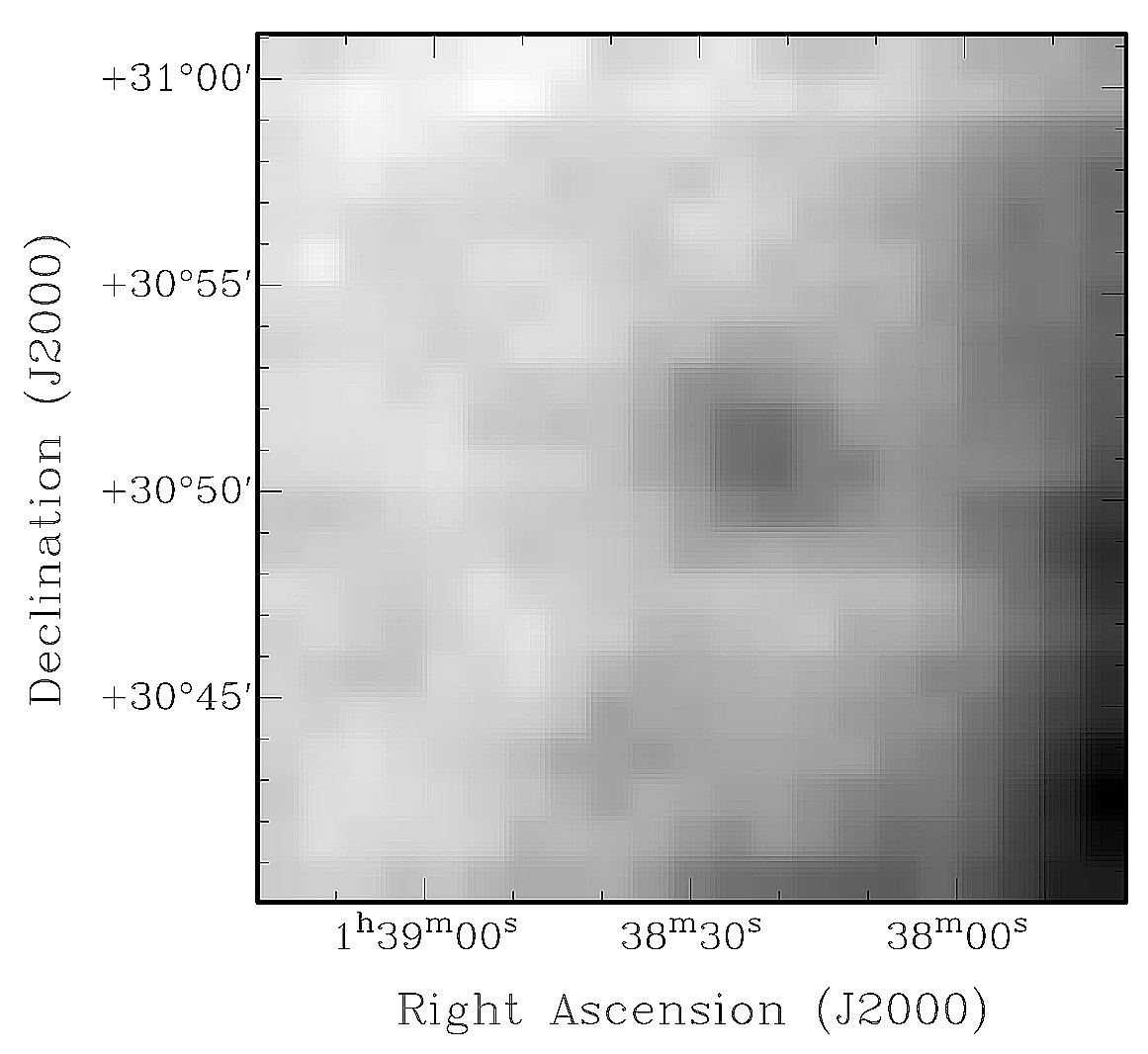}}\
\subfloat[]{\includegraphics[width = 3.4in]{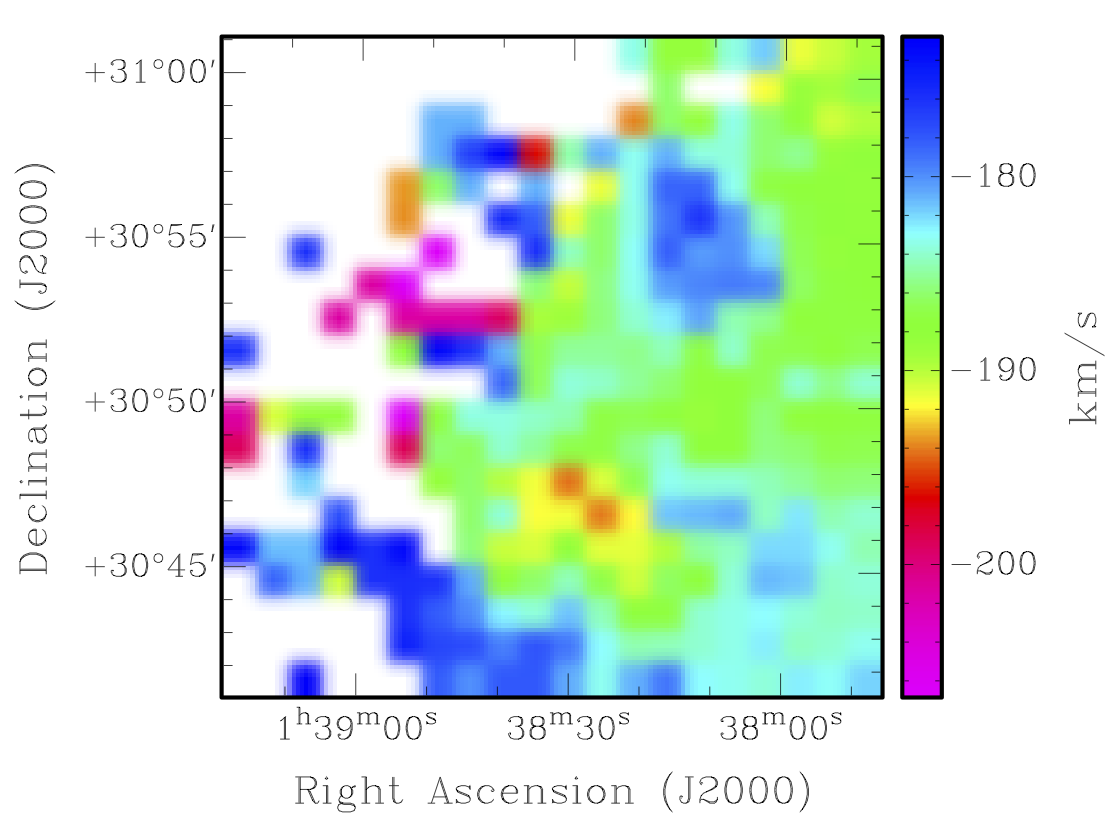}}
\newline
\centering
\subfloat[]{\includegraphics[width = 2.5in]{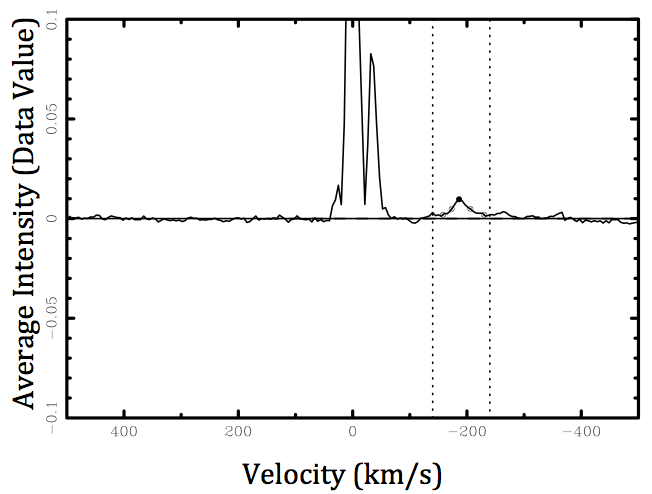}}
\caption{AGESM33-13 }
\label{ok13_all}
\end{figure}

\begin{figure}
\subfloat[]{\includegraphics[width = 3in]{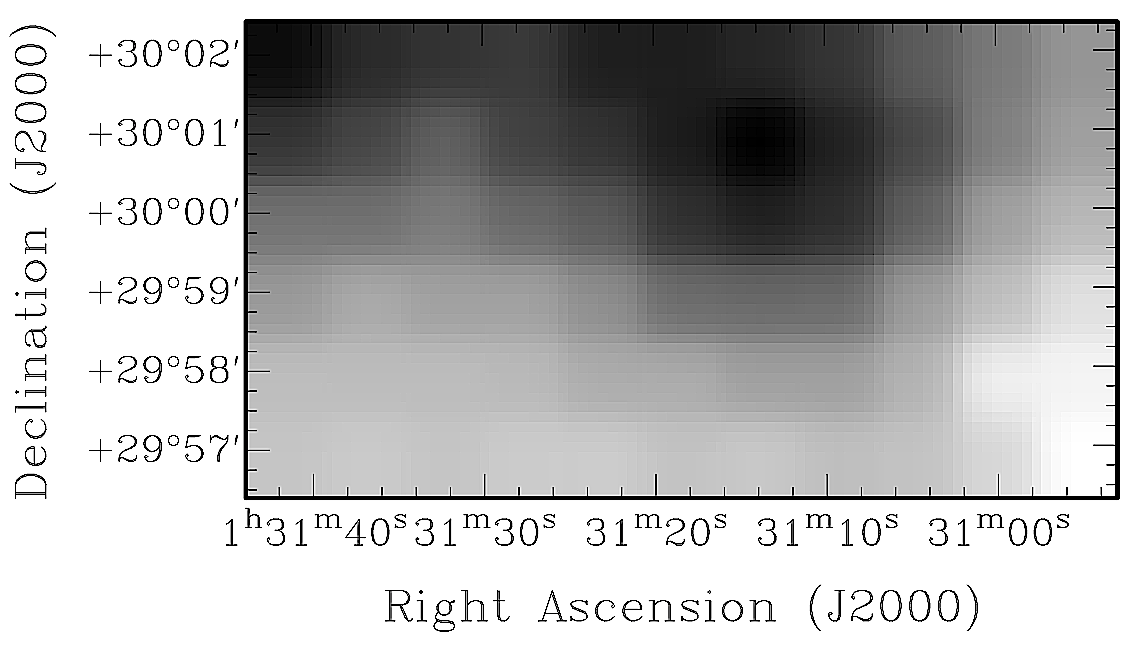}}\
\subfloat[]{\includegraphics[width = 3.4in]{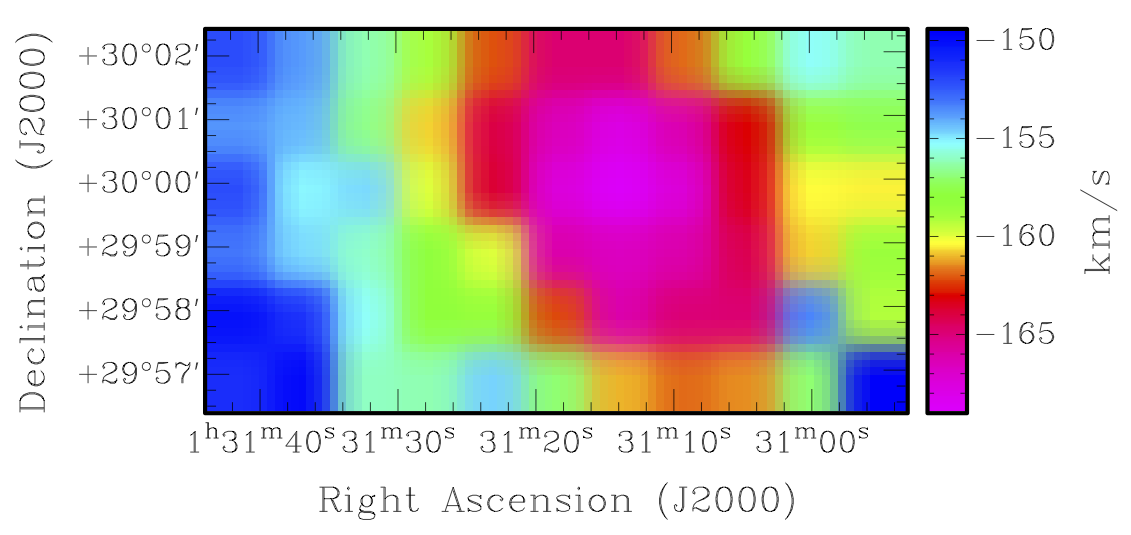}}
\newline
\centering
\subfloat[]{\includegraphics[width = 2.5in]{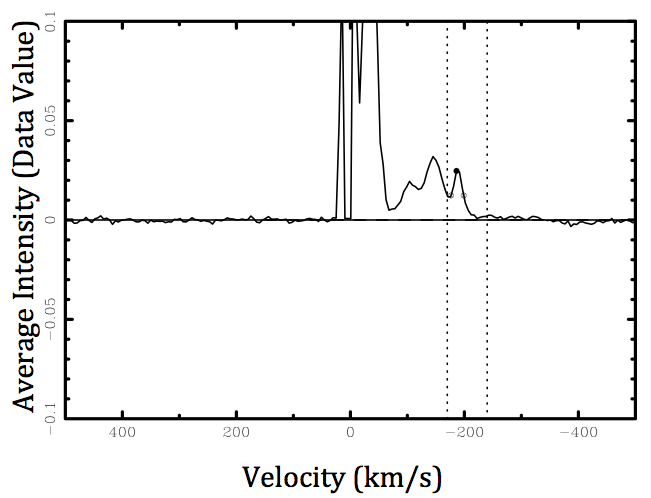}}
\caption{AGESM33-14 }
\label{ok14_all}
\end{figure}

\begin{figure}
\subfloat[]{\includegraphics[width = 3in]{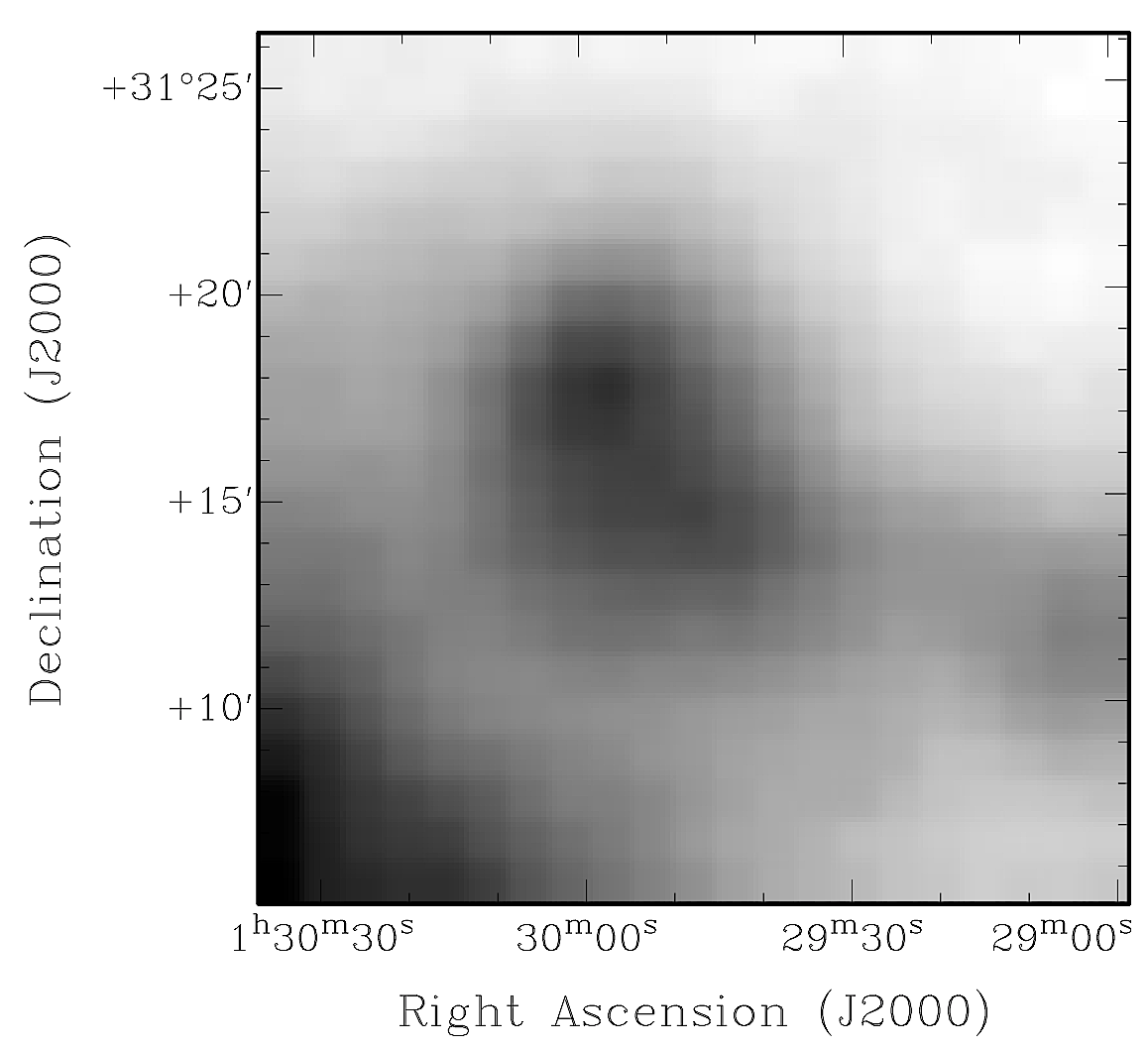}}\
\subfloat[]{\includegraphics[width = 3.4in]{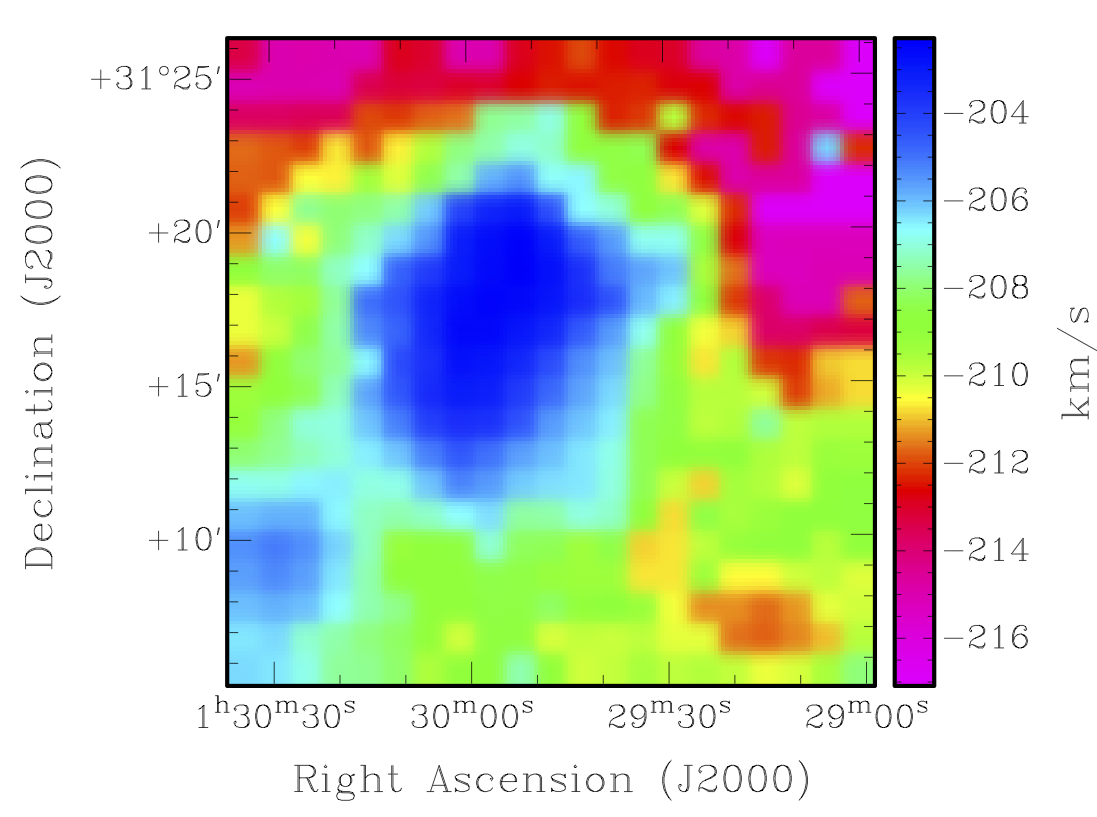}}
\newline
\centering
\subfloat[]{\includegraphics[width = 2.5in]{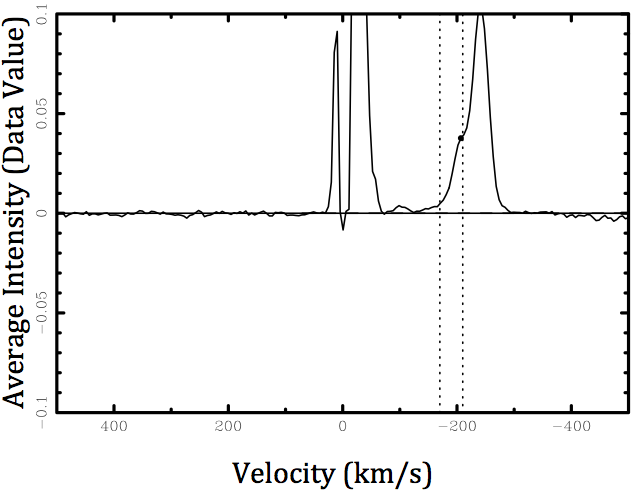}}
\caption{AGESM33-15 }
\label{ok15_all}
\end{figure}

\begin{figure}
\subfloat[]{\includegraphics[width = 3in]{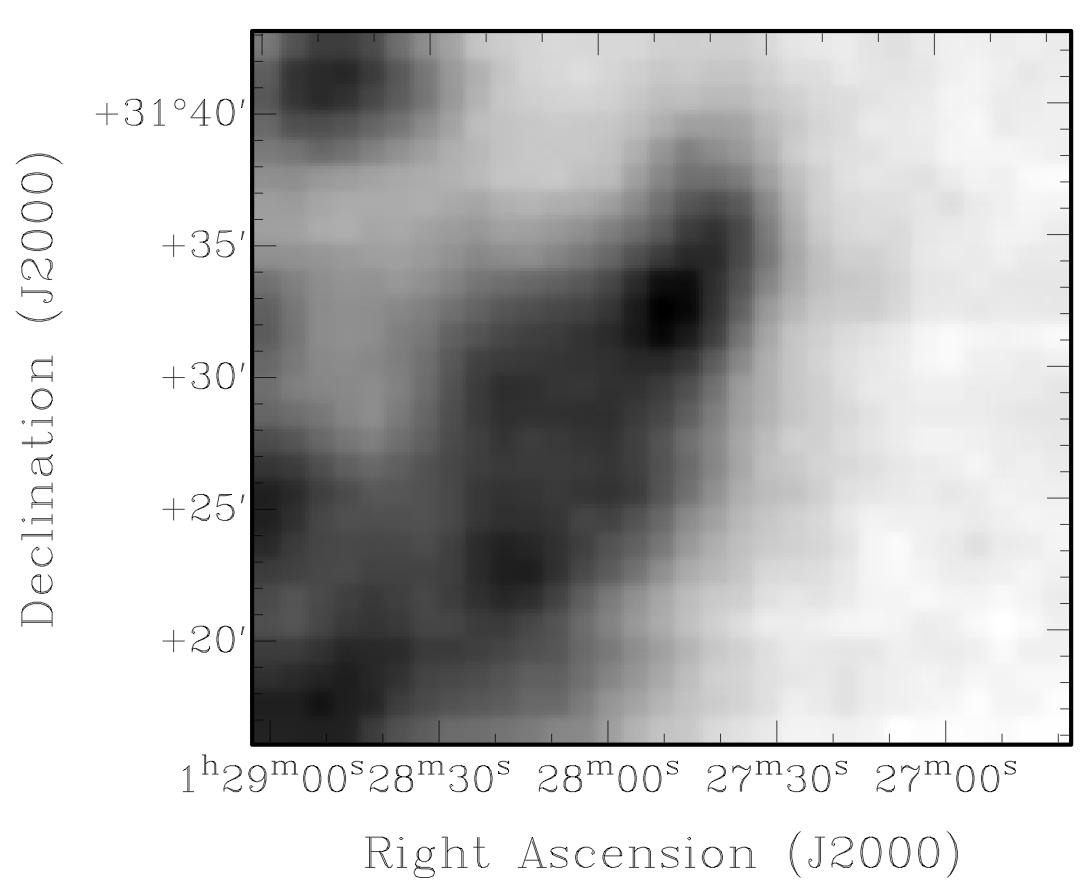}}\
\subfloat[]{\includegraphics[width = 3.4in]{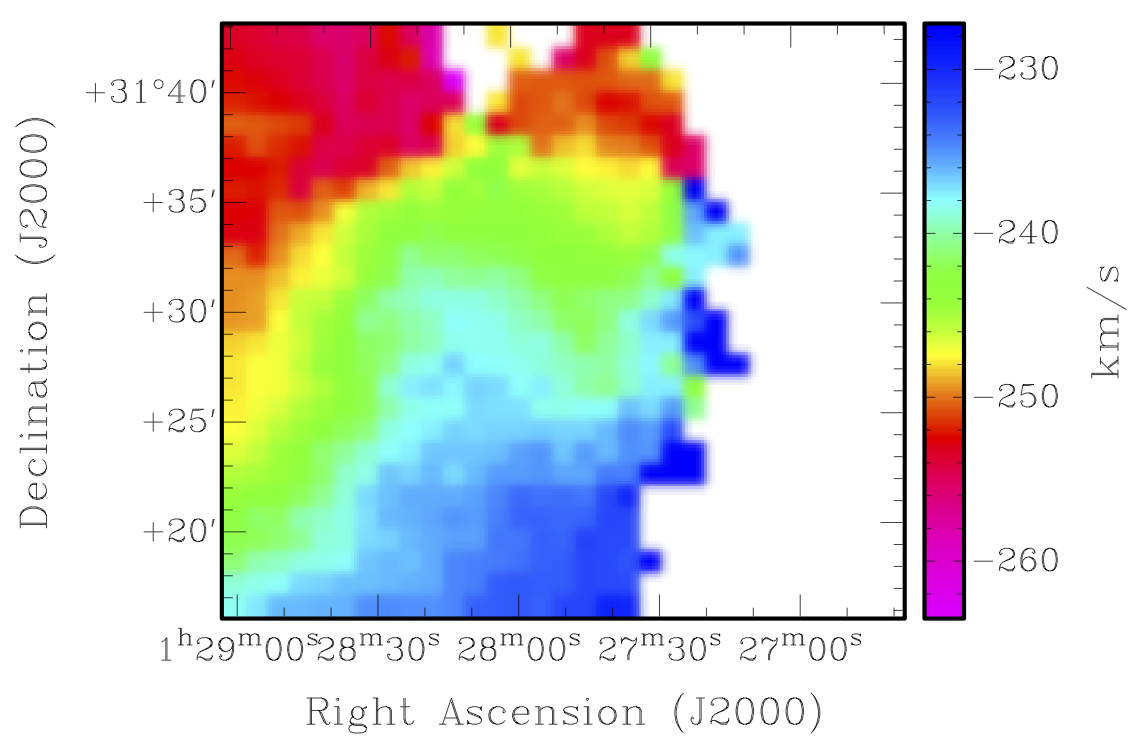}}
\newline
\centering
\subfloat[]{\includegraphics[width = 2.5in]{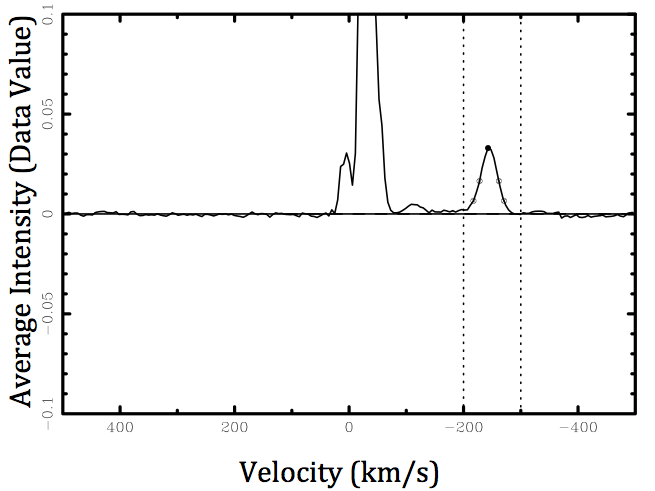}}
\caption{AGESM33-16 }
\label{ok16_all}
\end{figure}

\begin{figure}
\subfloat[]{\includegraphics[width = 3in]{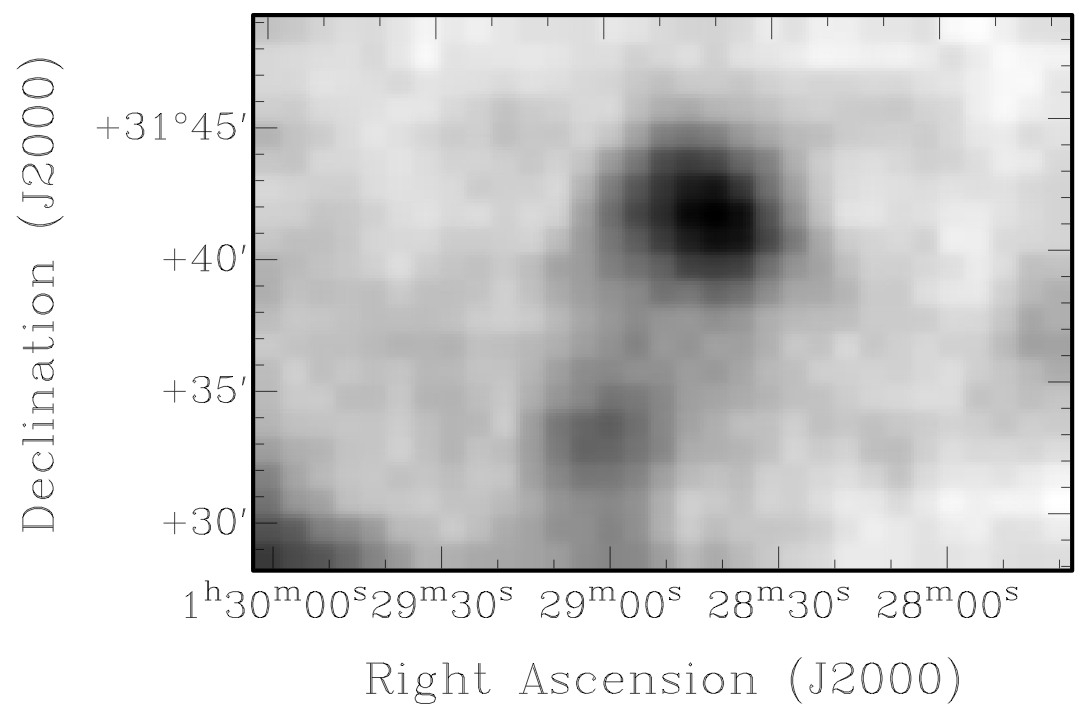}}\
\subfloat[]{\includegraphics[width = 3.4in]{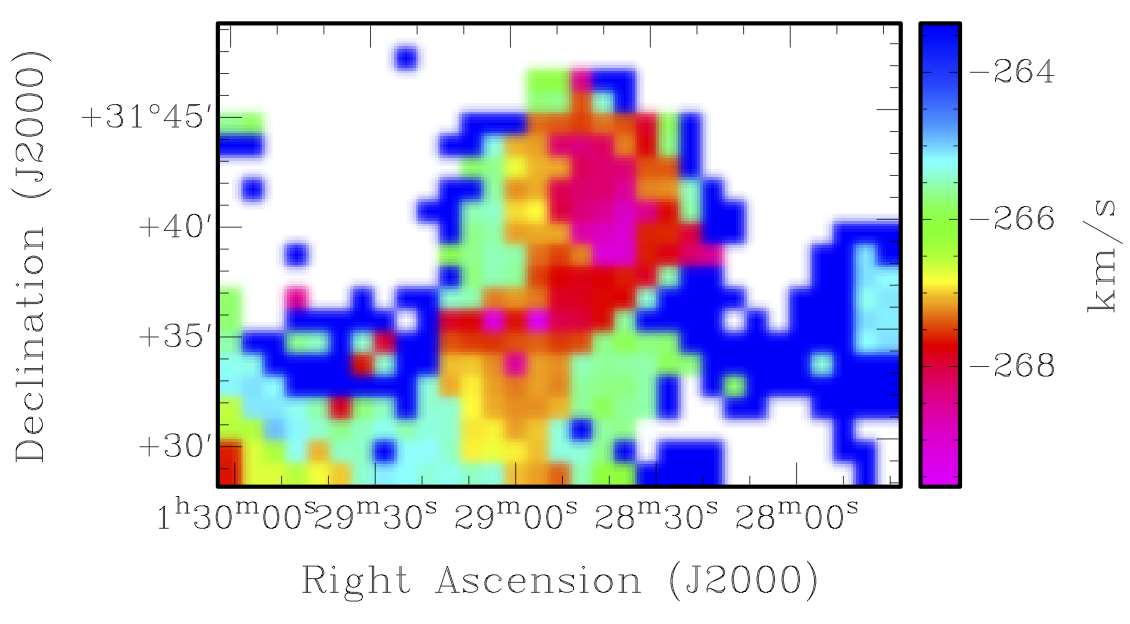}}
\newline
\centering
\subfloat[]{\includegraphics[width = 2.5in]{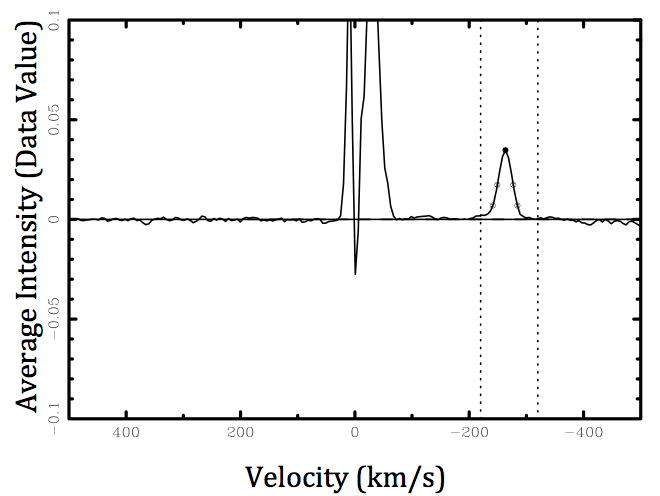}}
\caption{AGESM33-17 }
\label{ok17_all}
\end{figure}

\clearpage

\begin{figure}
\subfloat[]{\includegraphics[width = 3in]{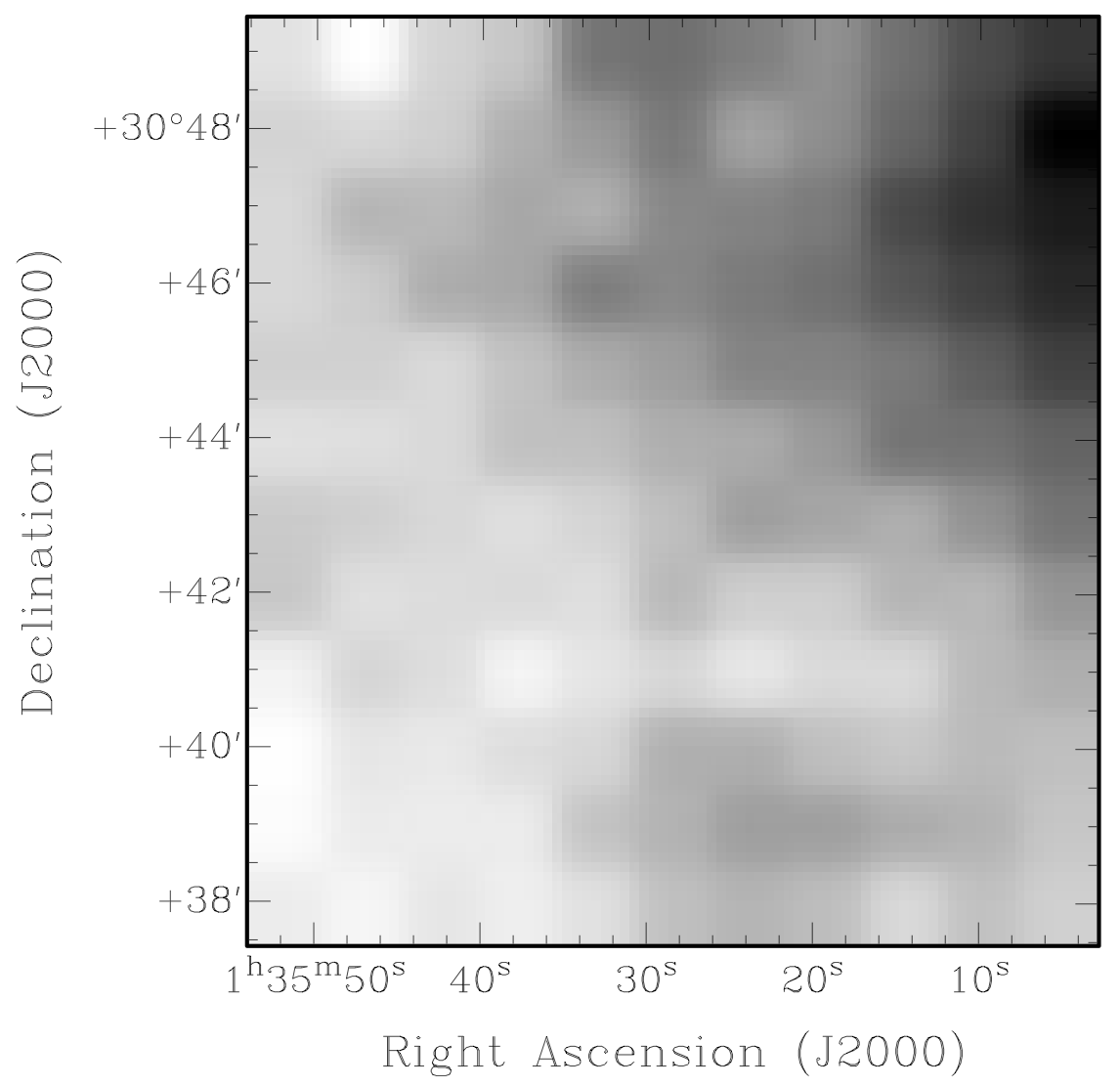}}\
\subfloat[]{\includegraphics[width = 3.4in]{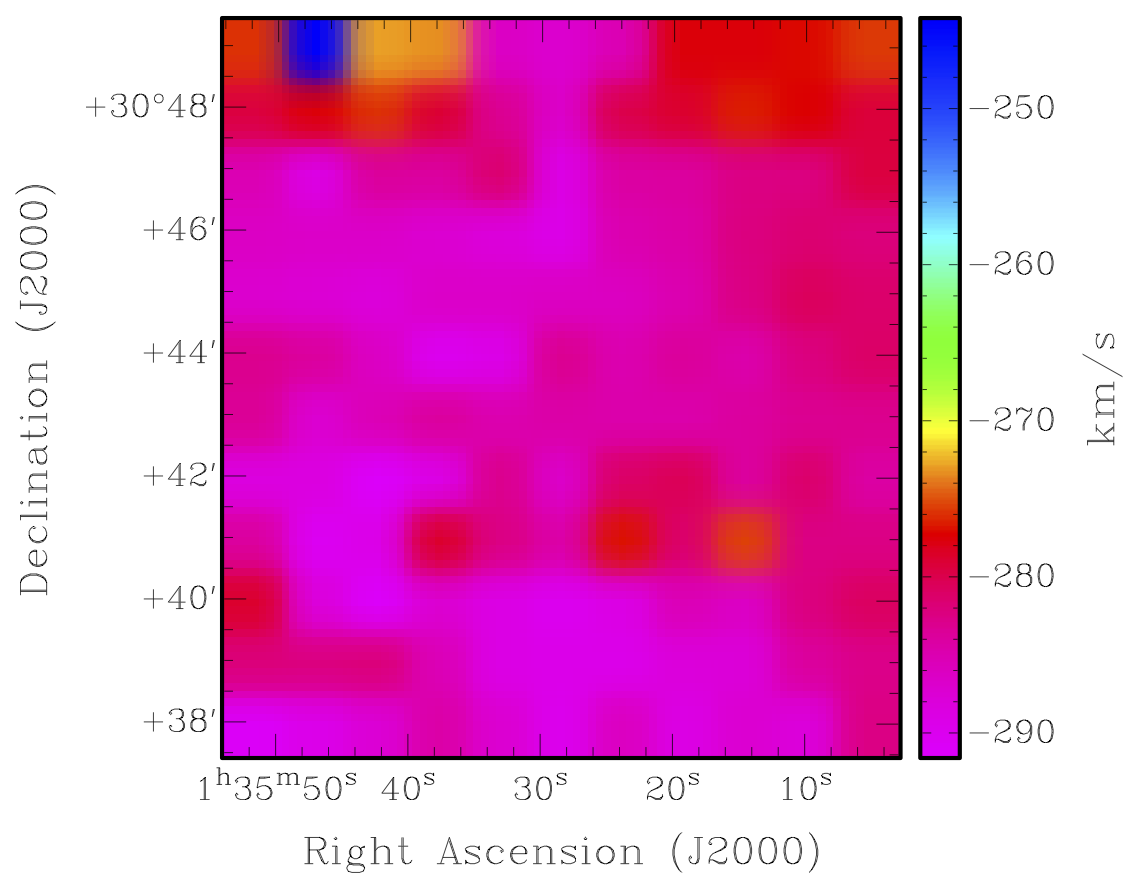}}
\newline
\centering
\subfloat[]{\includegraphics[width = 2.5in]{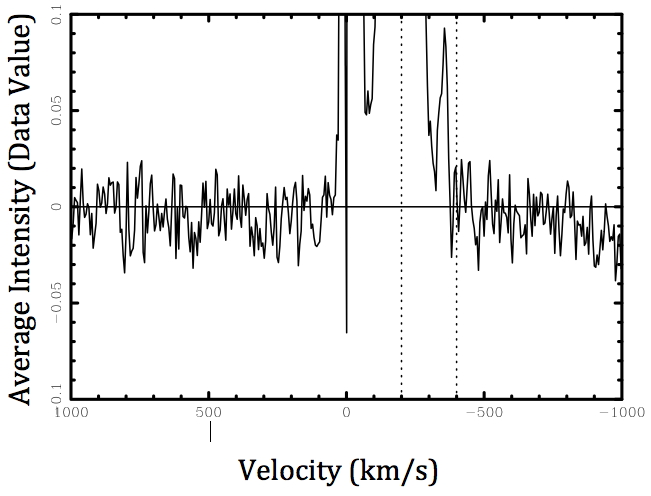}}
\caption{AGESM33-18 }
\label{ok18_all}
\end{figure}

\begin{figure}
\subfloat[]{\includegraphics[width = 2.5in]{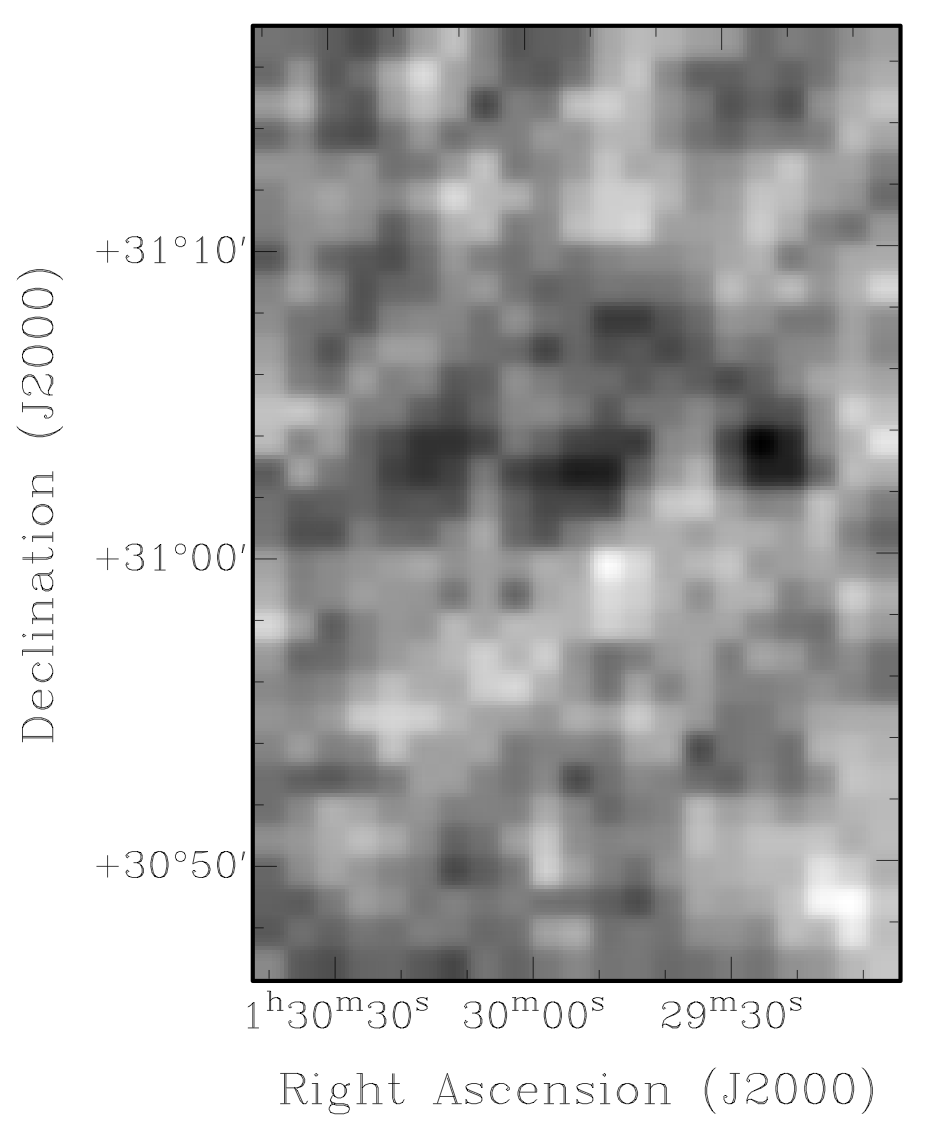}}\
\subfloat[]{\includegraphics[width = 3in]{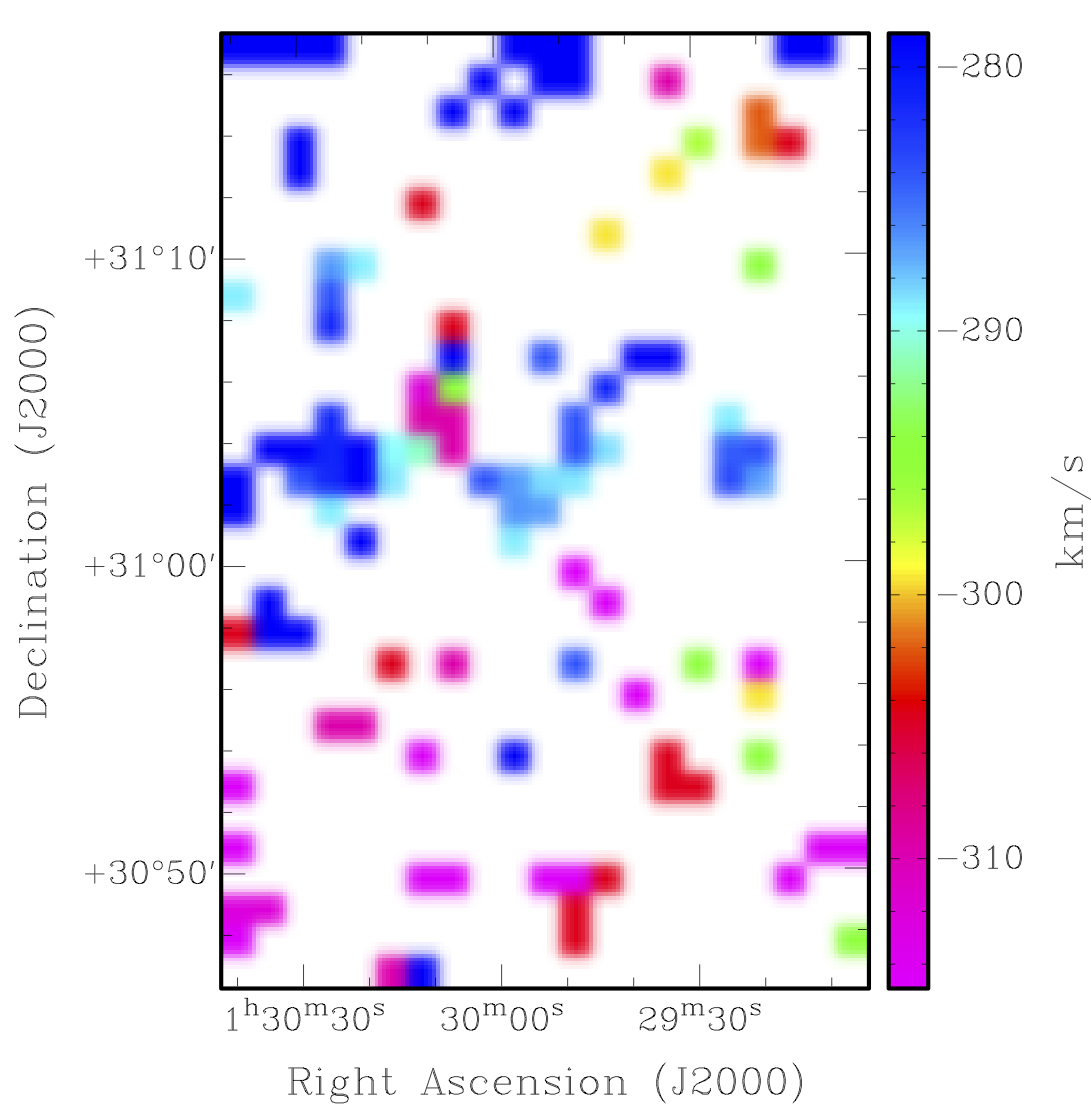}}
\newline
\centering
\subfloat[]{\includegraphics[width = 2.5in]{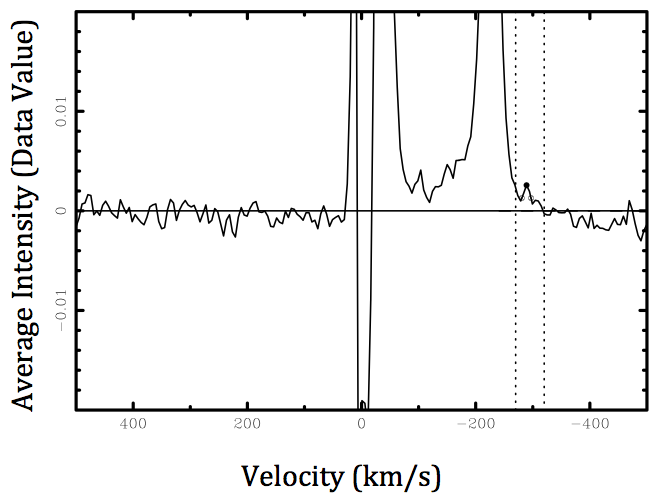}}
\caption{AGESM33-19 }
\label{ok19_all}
\end{figure}

\begin{figure}
\subfloat[]{\includegraphics[width = 2.5in]{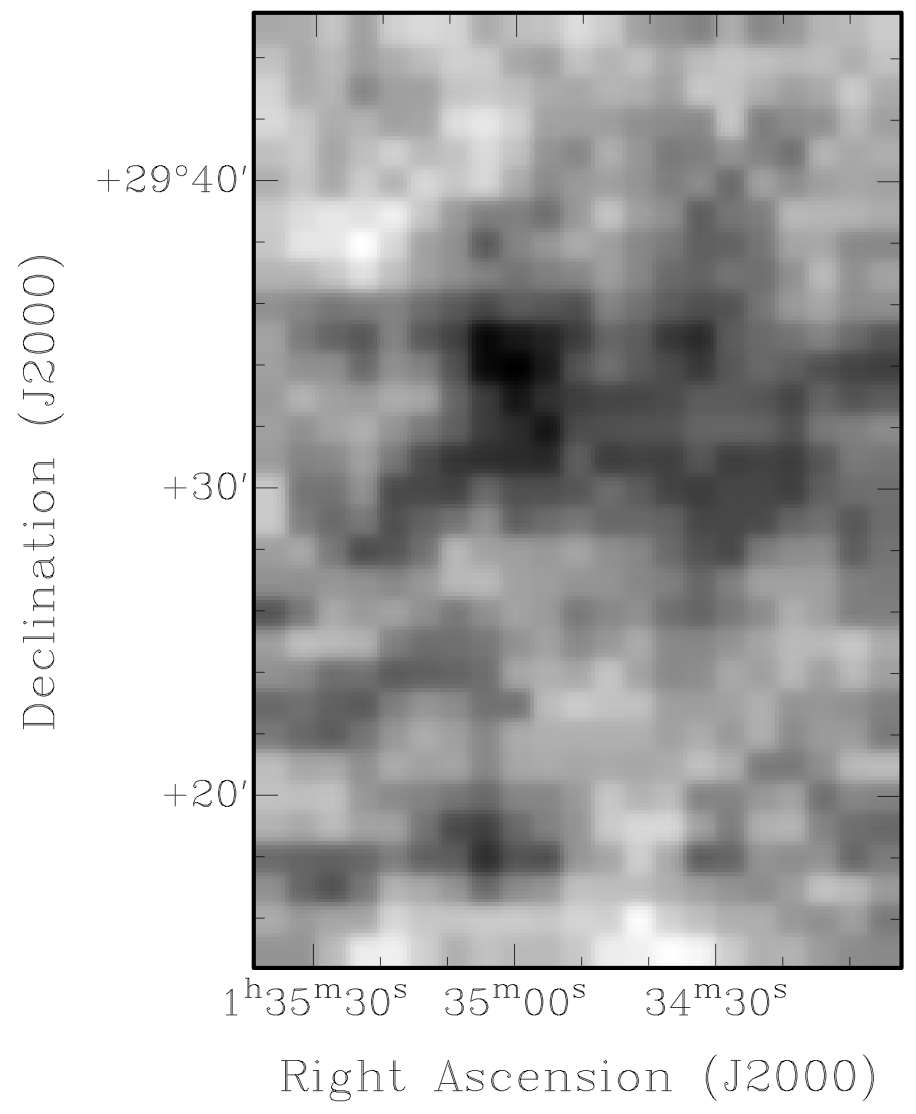}}\
\subfloat[]{\includegraphics[width = 3in]{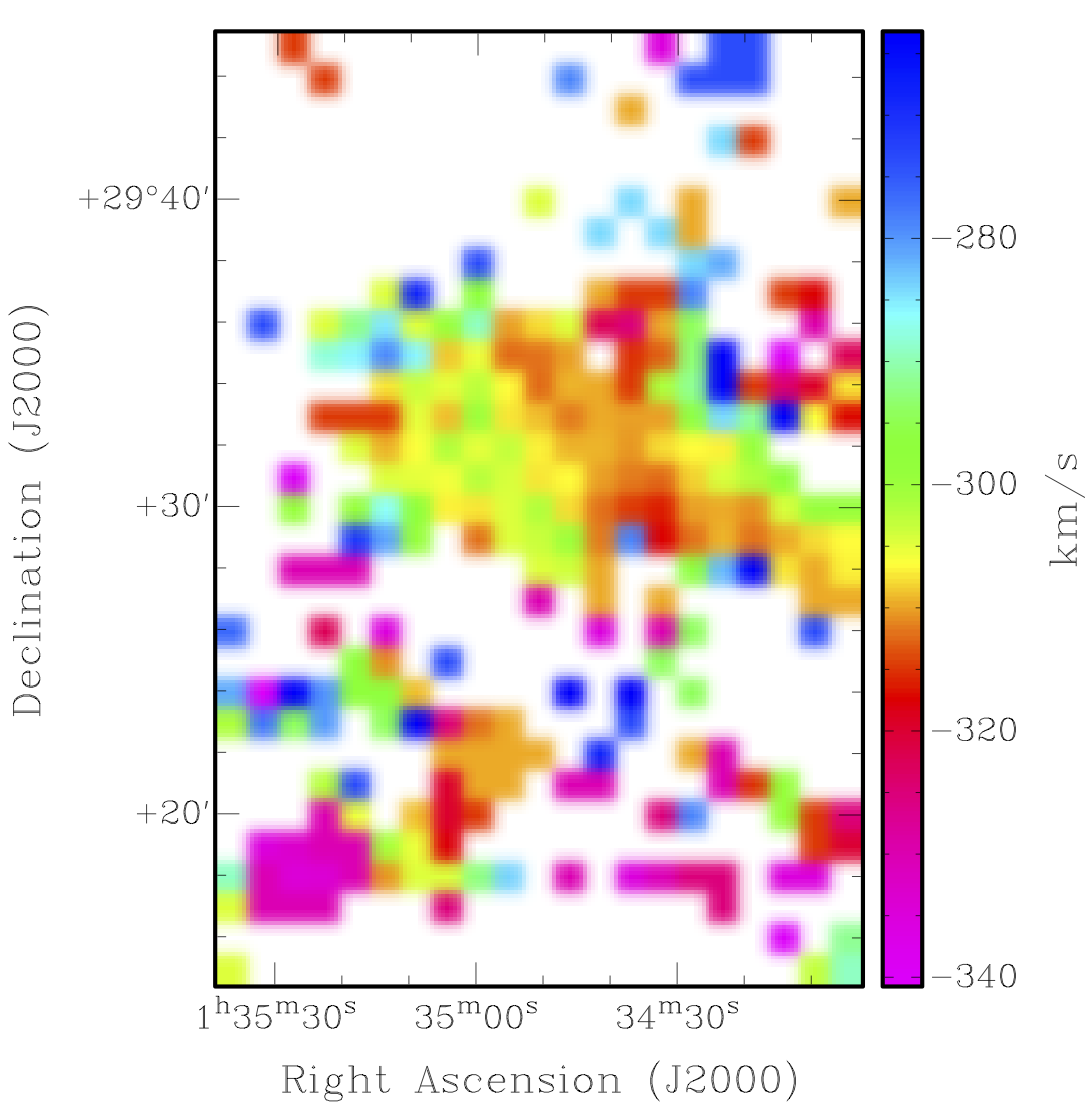}}
\newline
\centering
\subfloat[]{\includegraphics[width = 2.5in]{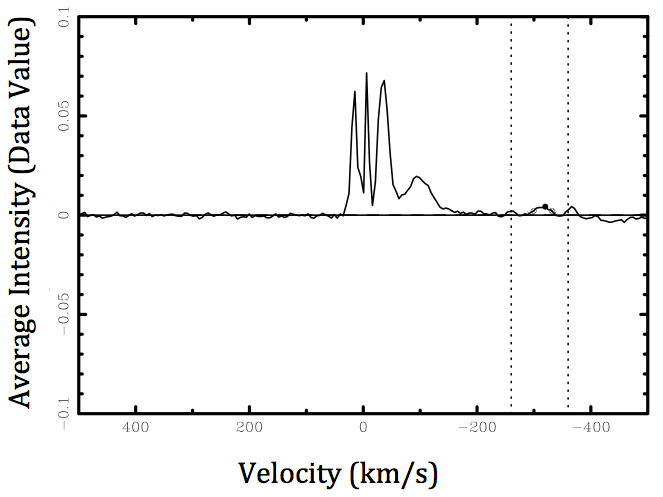}}
\caption{AGESM33-20 }
\label{ok20_all}
\end{figure}

\begin{figure}
\subfloat[]{\includegraphics[width = 3in]{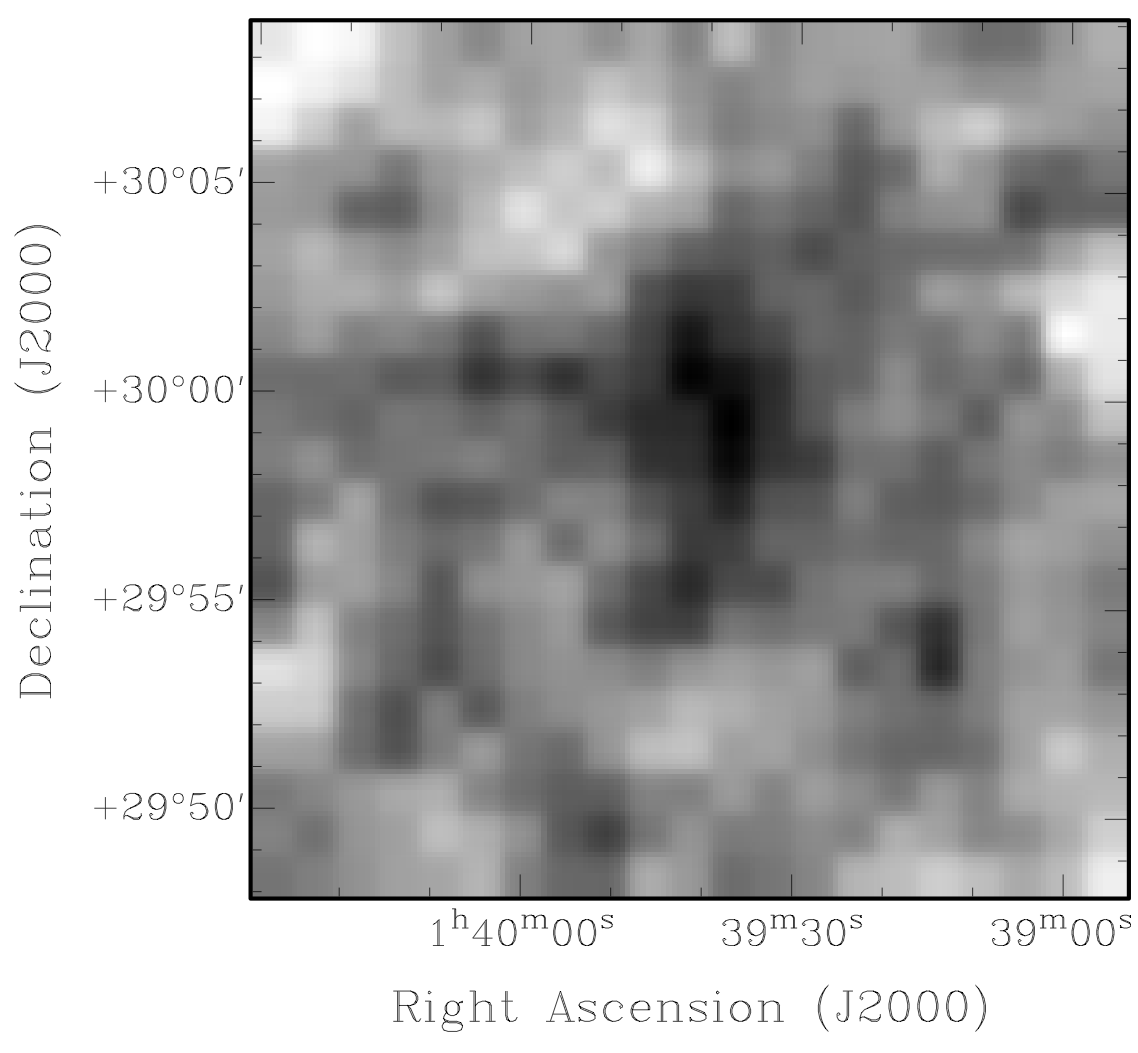}}\
\subfloat[]{\includegraphics[width = 3.4in]{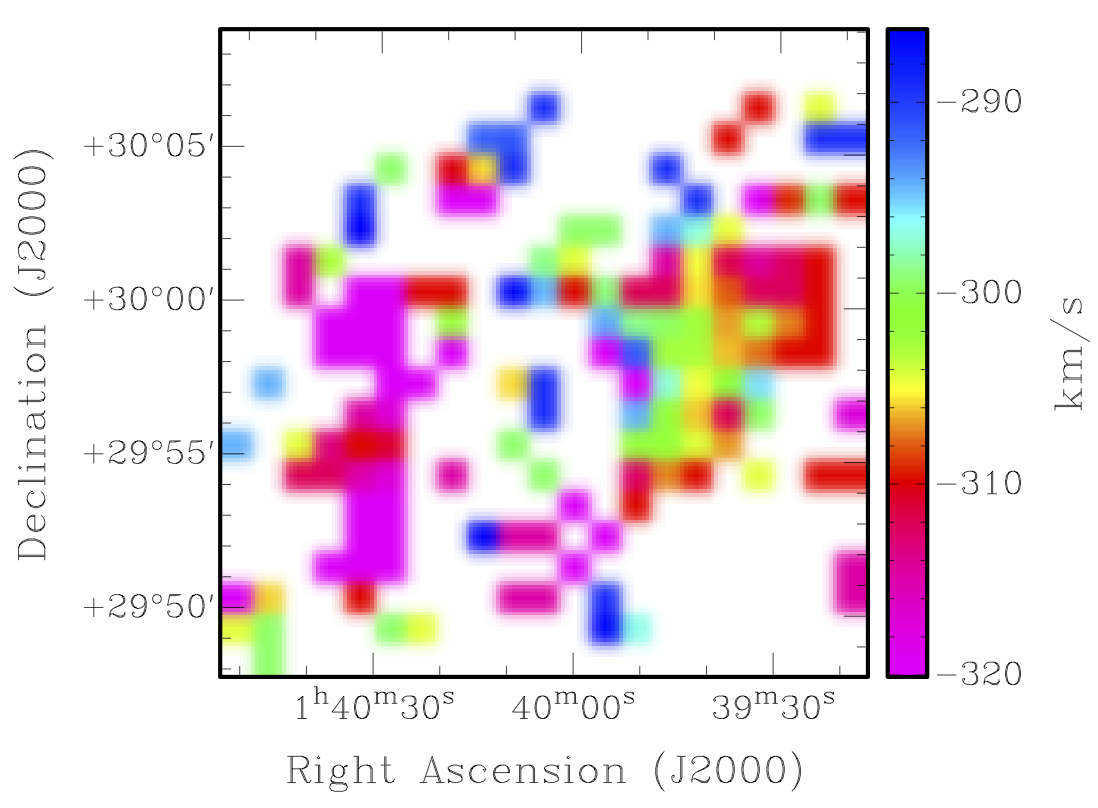}}
\newline
\centering
\subfloat[]{\includegraphics[width = 2.5in]{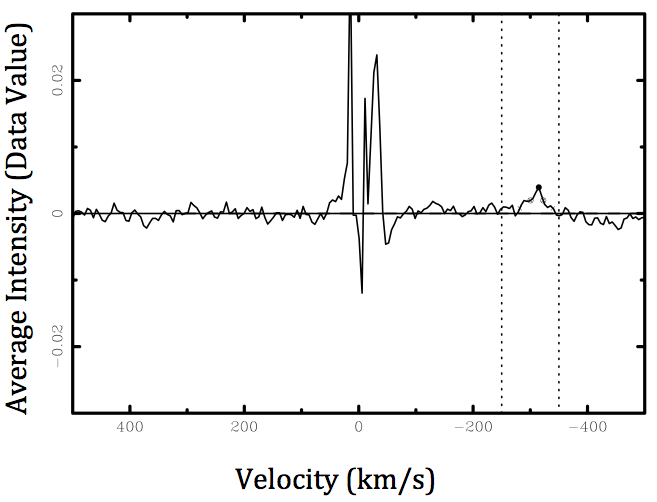}}
\caption{AGESM33-21 }
\label{ok21_all}
\end{figure}

\begin{figure}
\subfloat[]{\includegraphics[width = 3in]{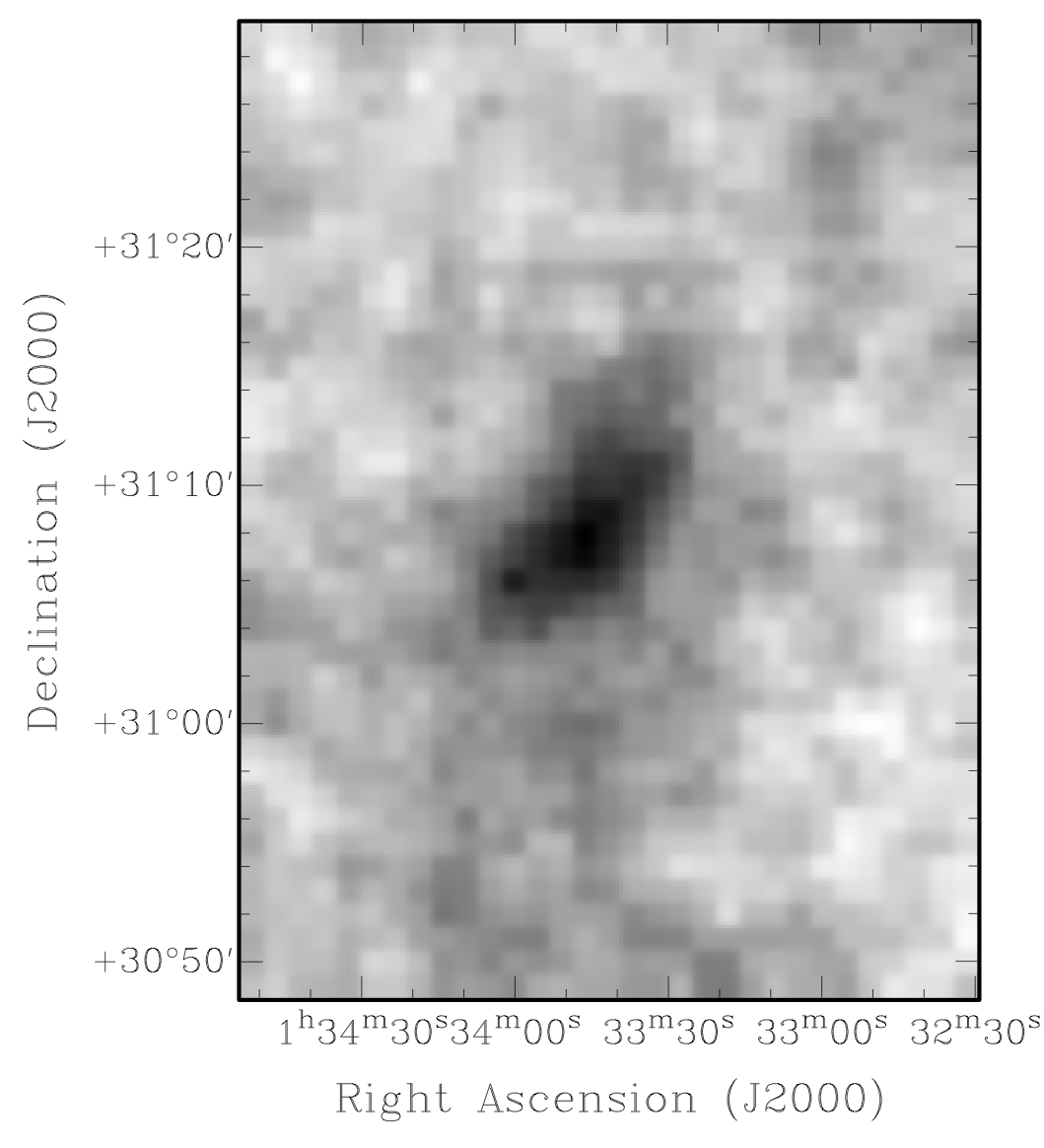}}\
\subfloat[]{\includegraphics[width = 3.4in]{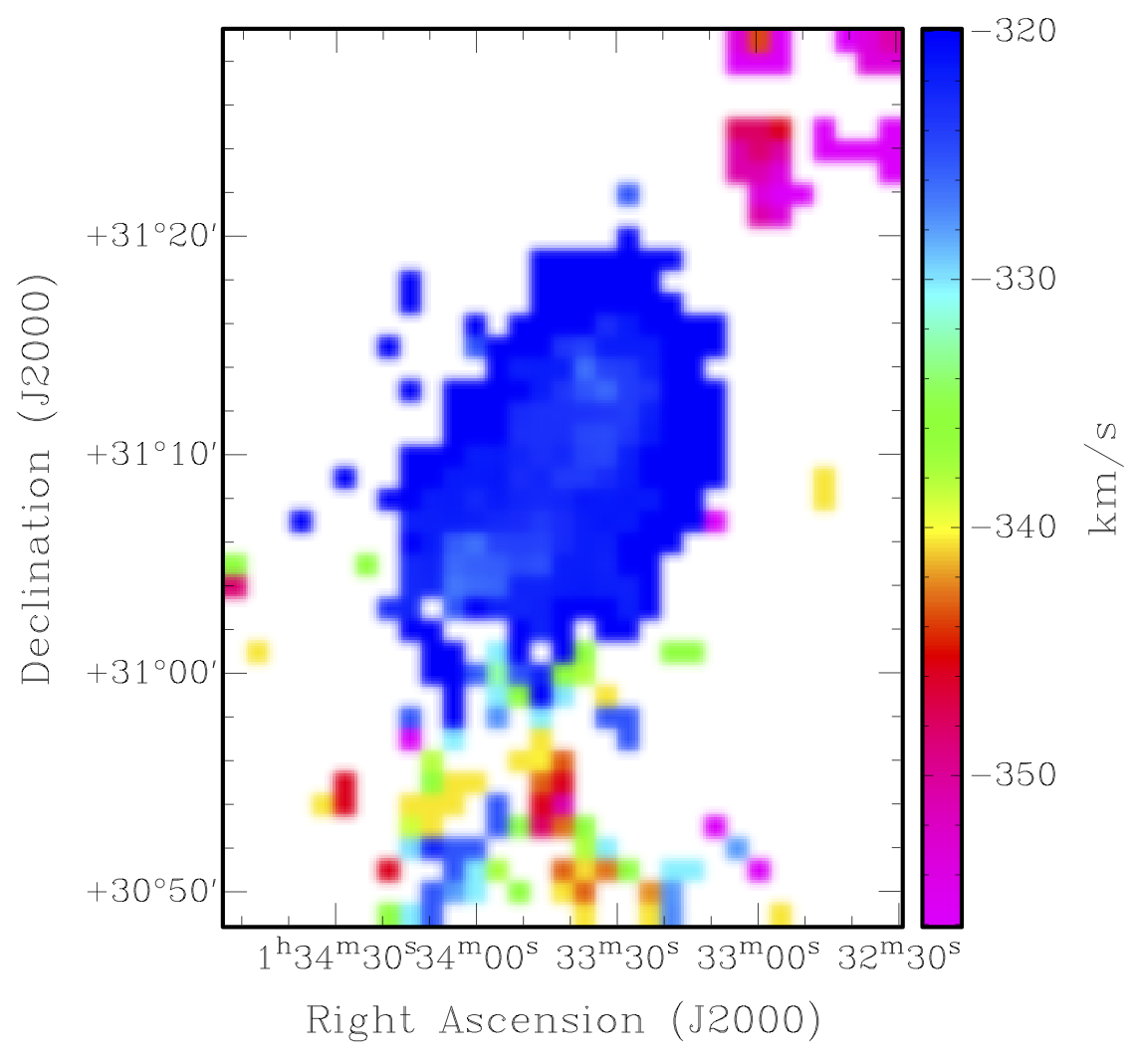}}
\newline
\centering
\subfloat[]{\includegraphics[width = 2.5in]{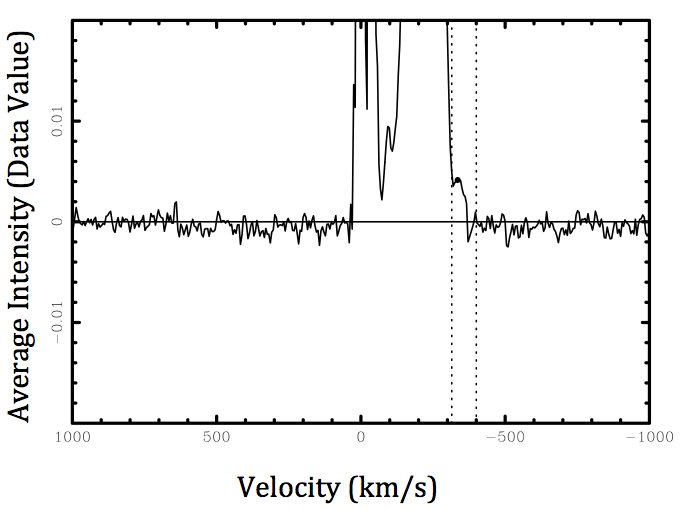}}
\caption{AGESM33-22 }
\label{ok22_all}
\end{figure}

\begin{figure}
\subfloat[]{\includegraphics[width = 3in]{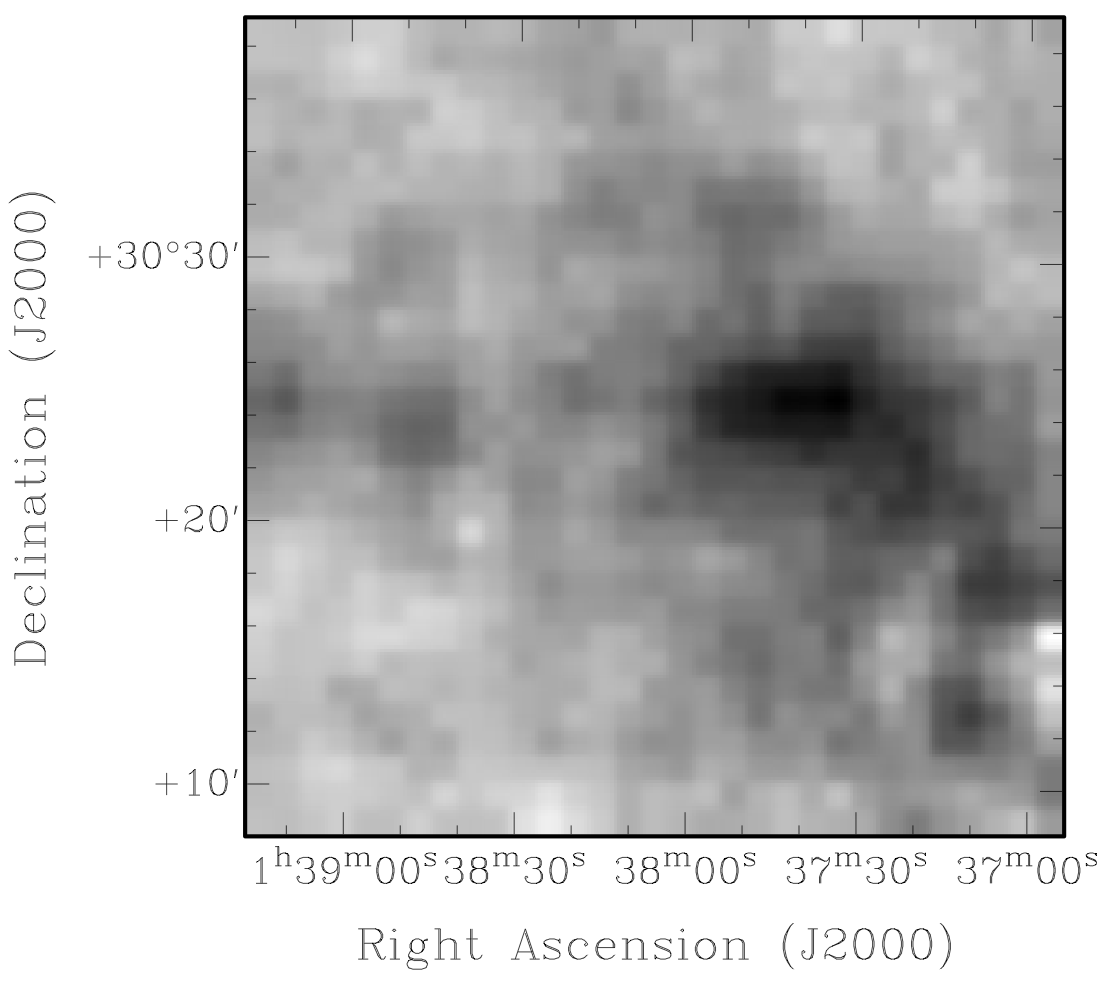}}\
\subfloat[]{\includegraphics[width = 3.4in]{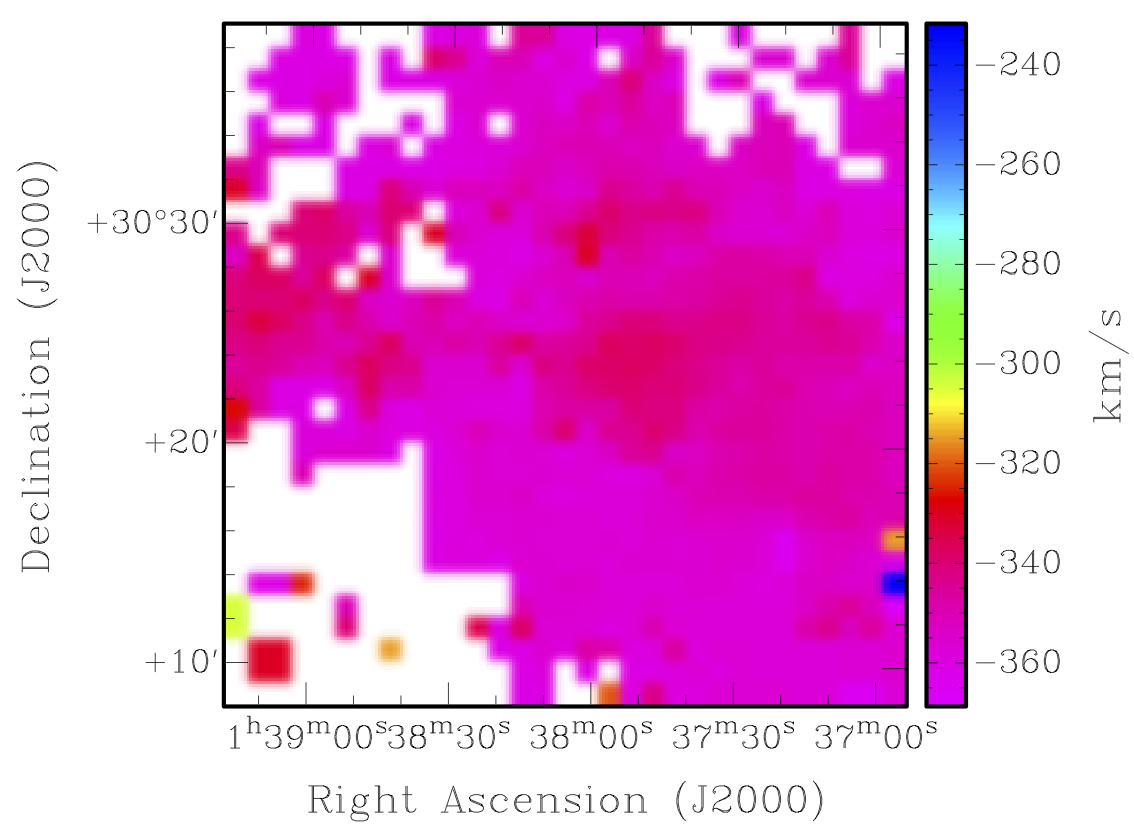}}
\newline
\centering
\subfloat[]{\includegraphics[width = 2.5in]{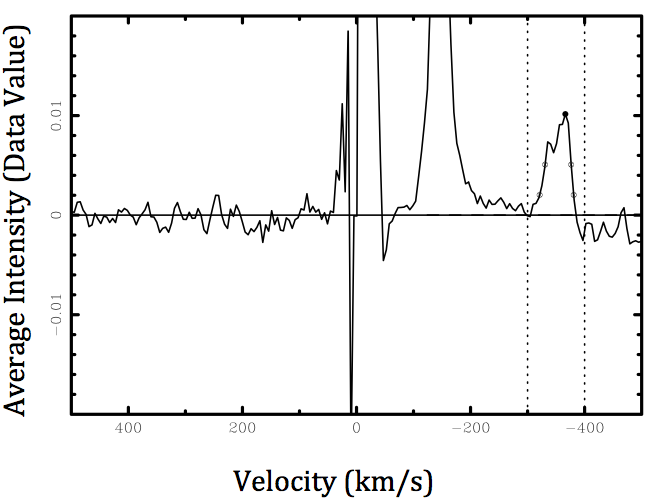}}
\caption{AGESM33-23 }
\label{ok23_all}
\end{figure}

\begin{figure}
\subfloat[]{\includegraphics[width = 3in]{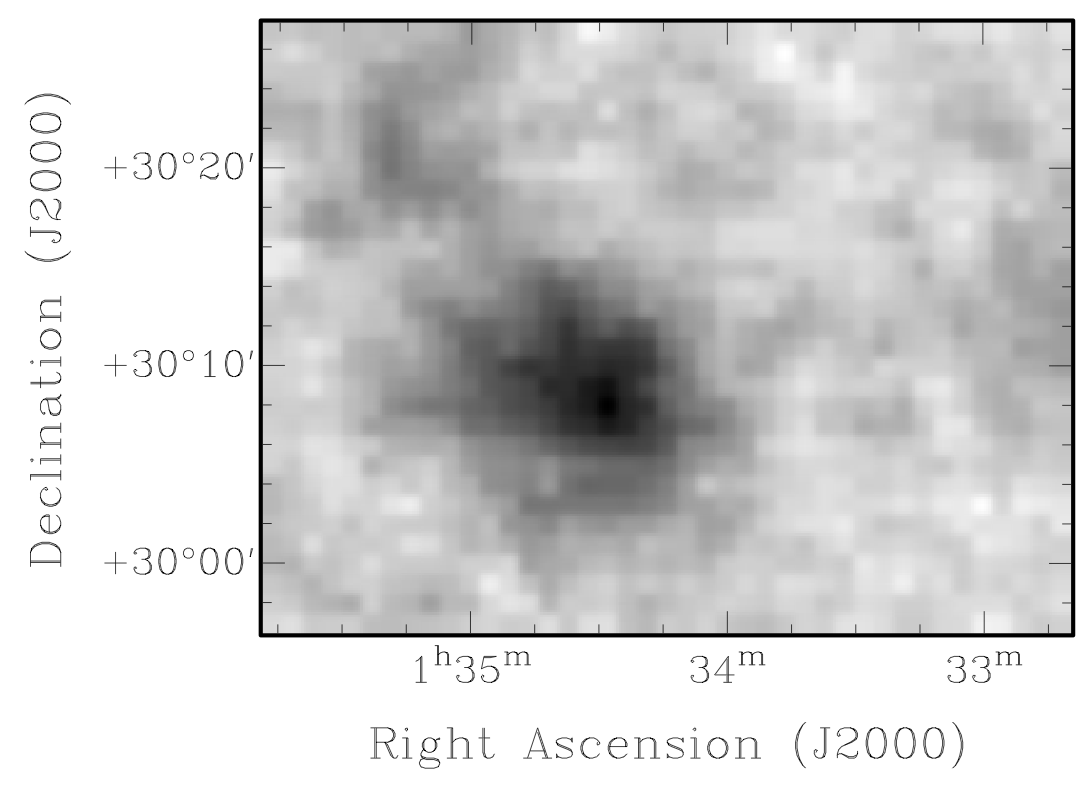}}\
\subfloat[]{\includegraphics[width = 3.4in]{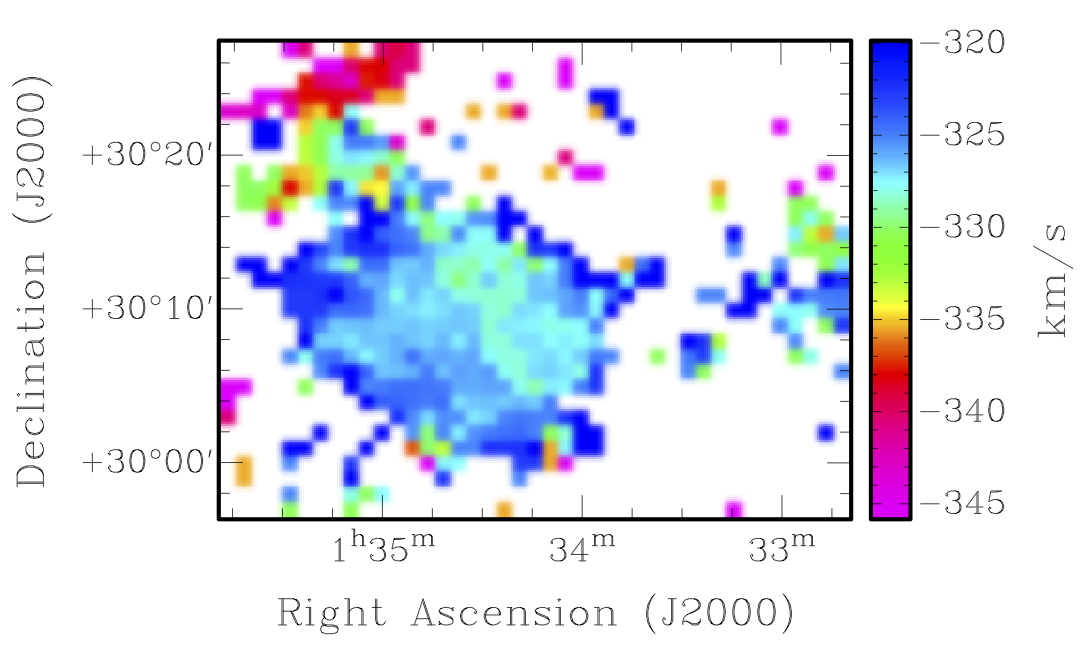}}
\newline
\centering
\subfloat[]{\includegraphics[width = 2.5in]{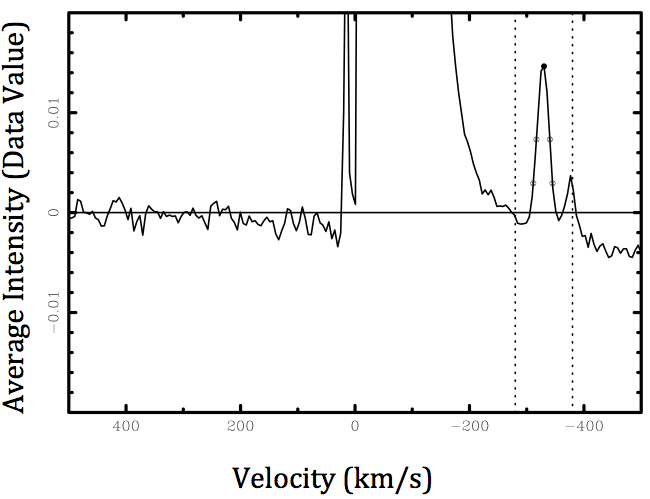}}
\caption{AGESM33-24}
\label{ok24_all}
\end{figure}

\begin{figure}
\subfloat[]{\includegraphics[width = 3in]{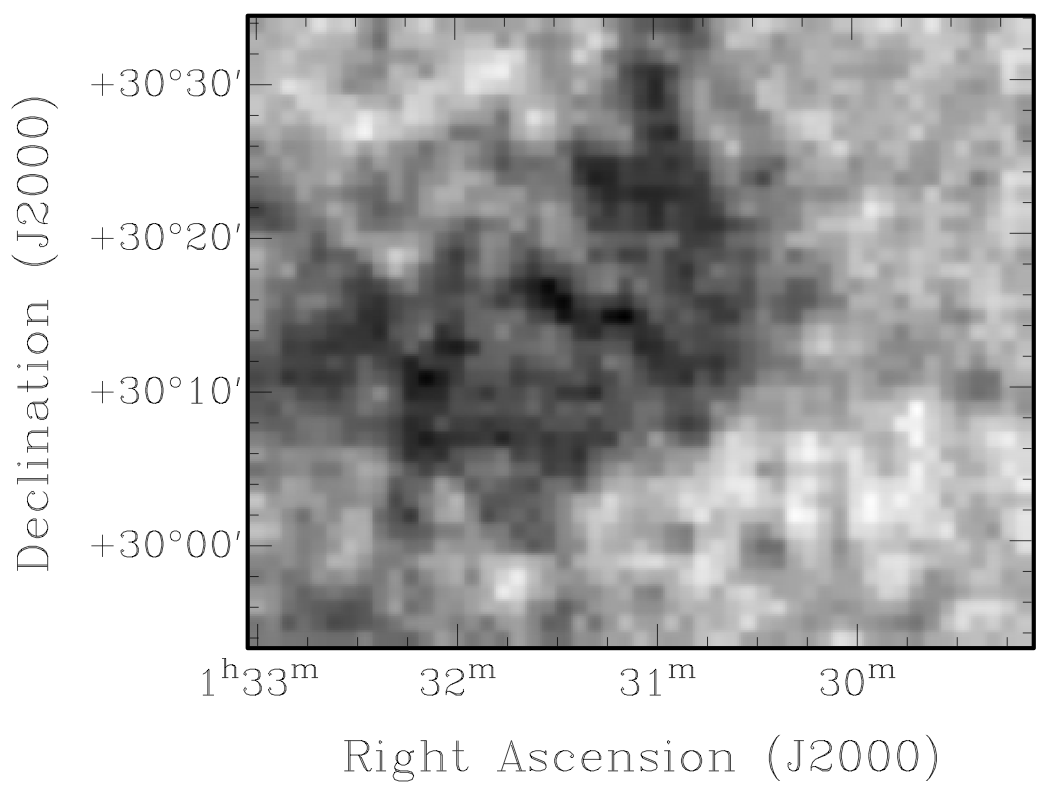}}\
\subfloat[]{\includegraphics[width = 3.4in]{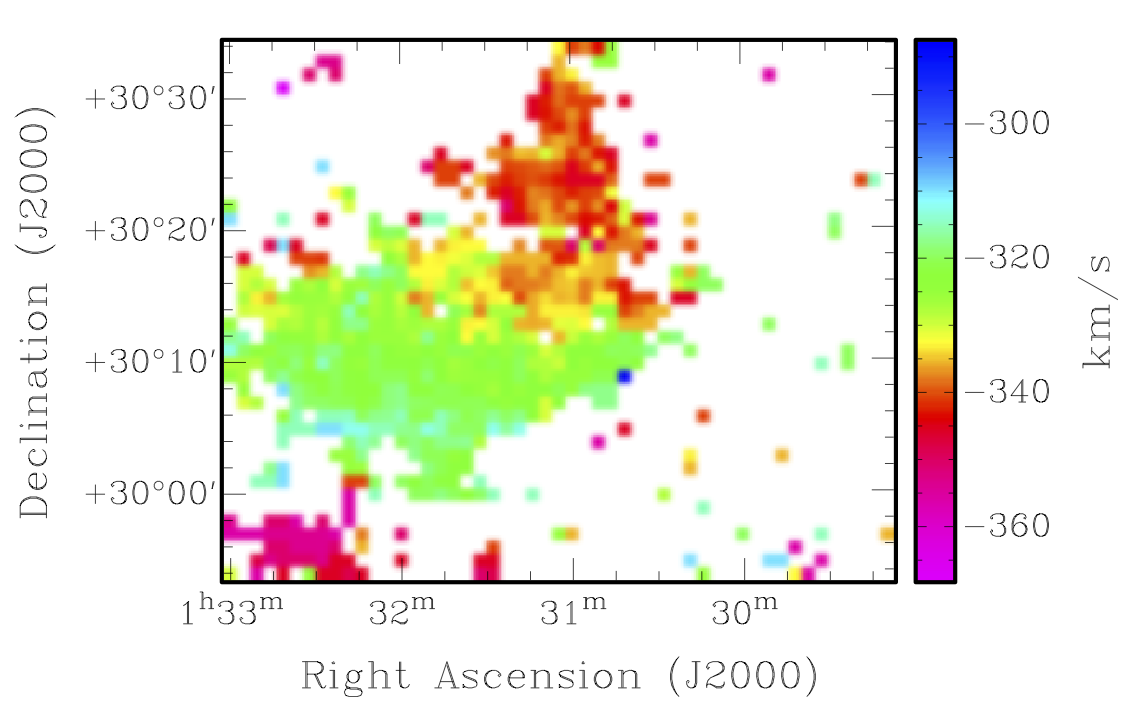}}
\newline
\centering
\subfloat[]{\includegraphics[width = 2.5in]{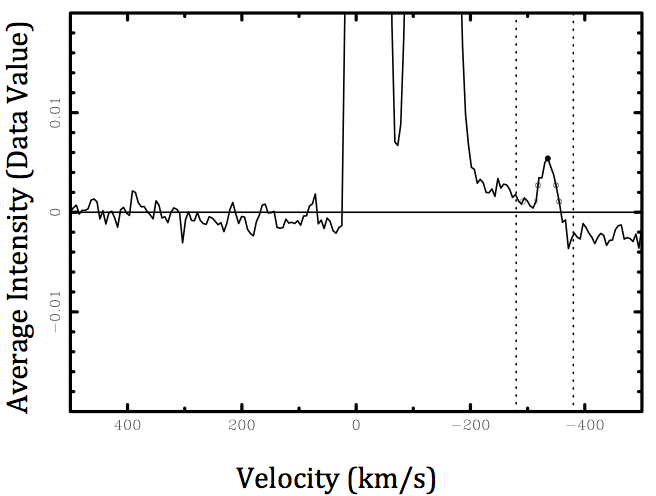}}
\caption{AGESM33-25 }
\label{ok25_all}
\end{figure}

\begin{figure}
\subfloat[]{\includegraphics[width = 3in]{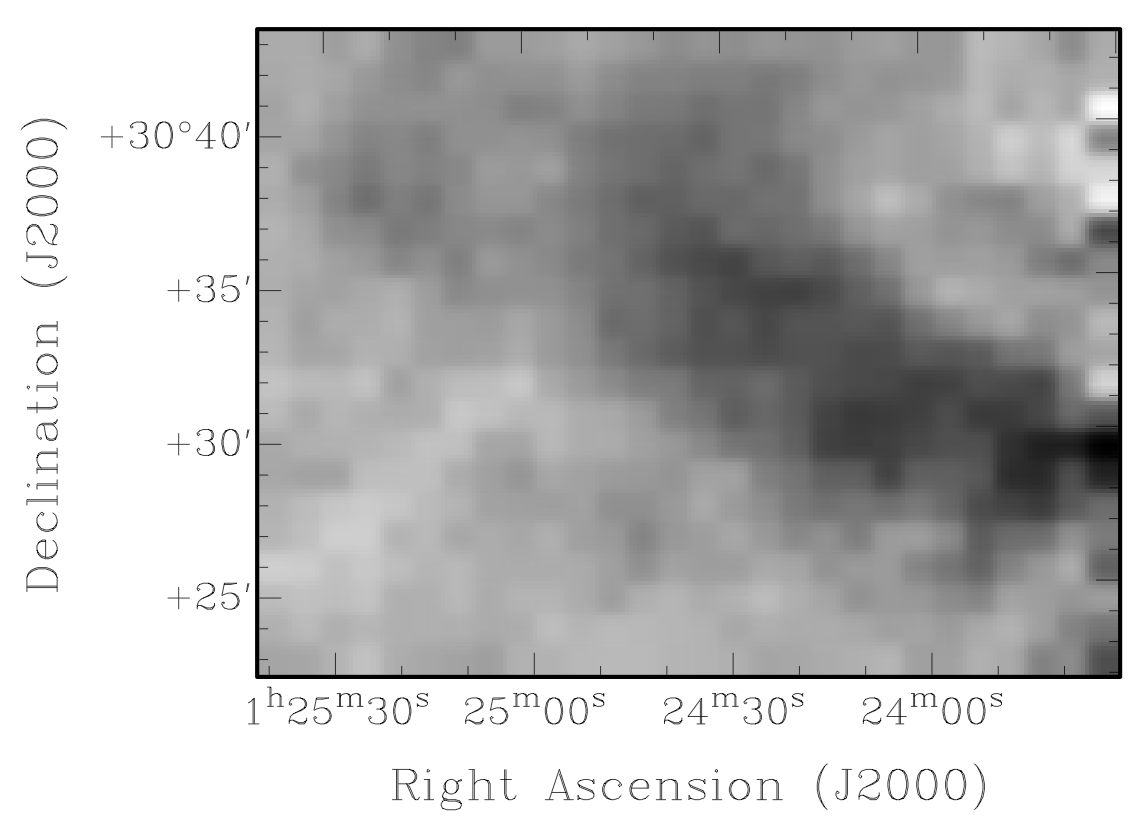}}\
\subfloat[]{\includegraphics[width = 3.4in]{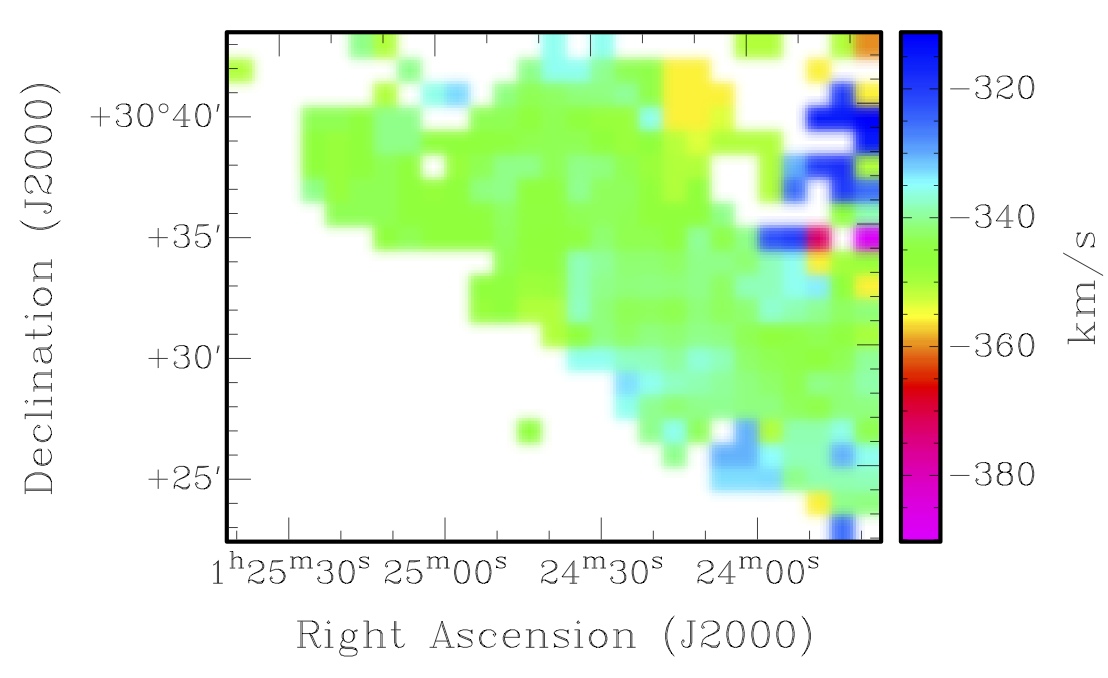}}
\newline
\centering
\subfloat[]{\includegraphics[width = 2.5in]{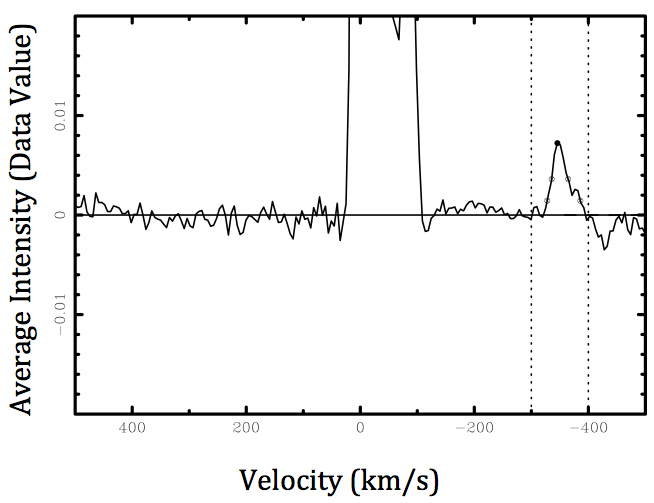}}
\caption{AGESM33-26 }
\label{ok26_all}
\end{figure}

\begin{figure}
\subfloat[]{\includegraphics[width = 3in]{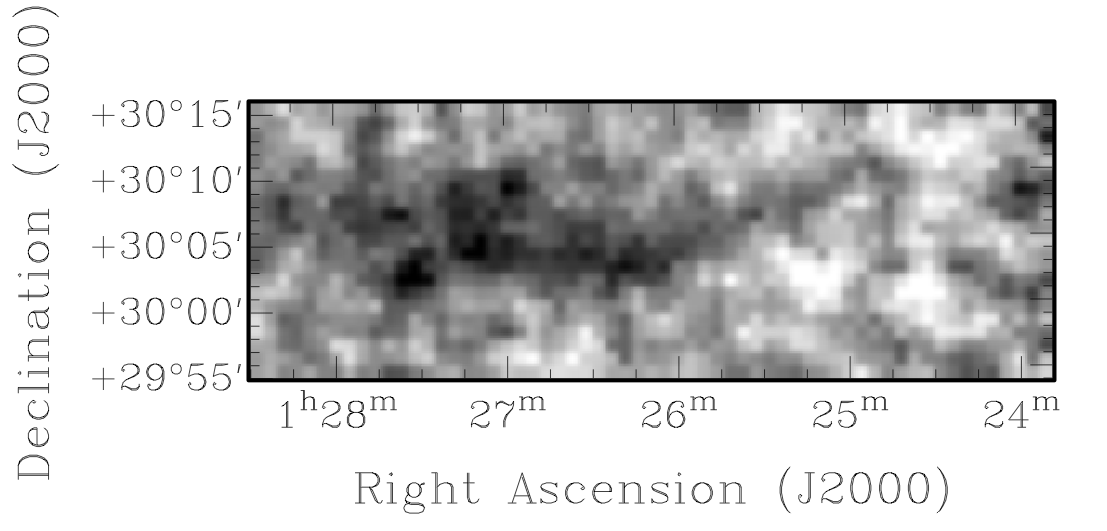}}\
\subfloat[]{\includegraphics[width = 3.4in]{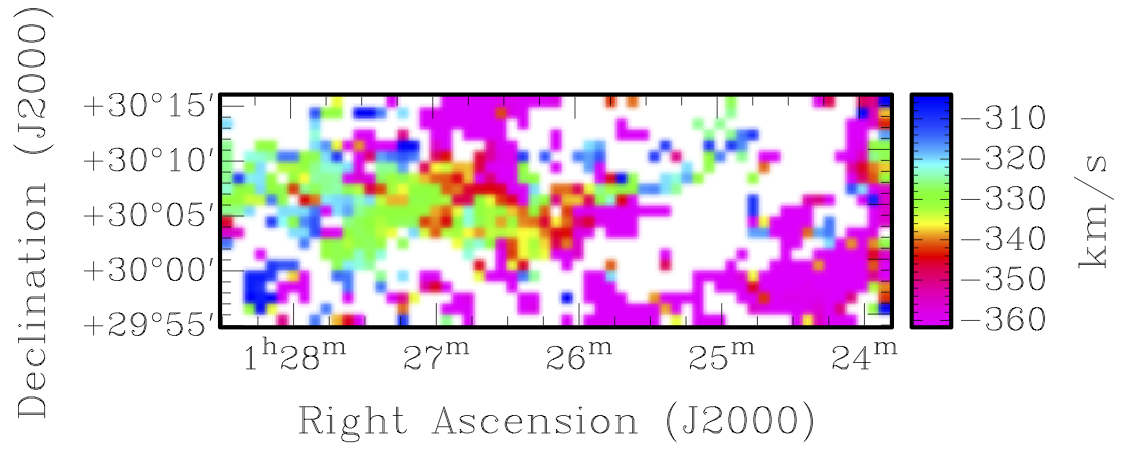}}
\newline
\centering
\subfloat[]{\includegraphics[width = 2.5in]{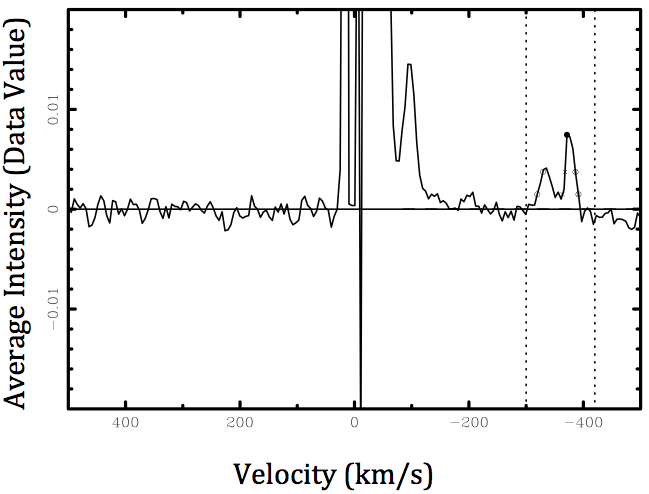}}
\caption{AGESM33-27 }
\label{ok27_all}
\end{figure}

\begin{figure}
\subfloat[]{\includegraphics[width = 3in]{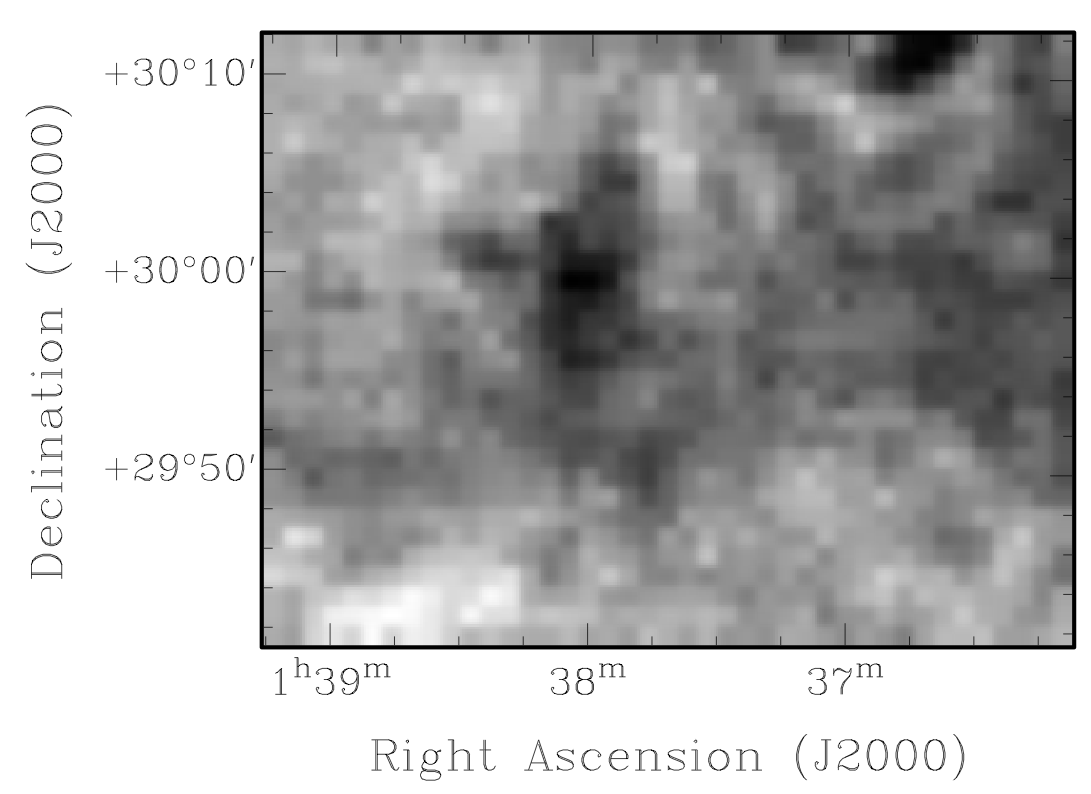}}\
\subfloat[]{\includegraphics[width = 3.4in]{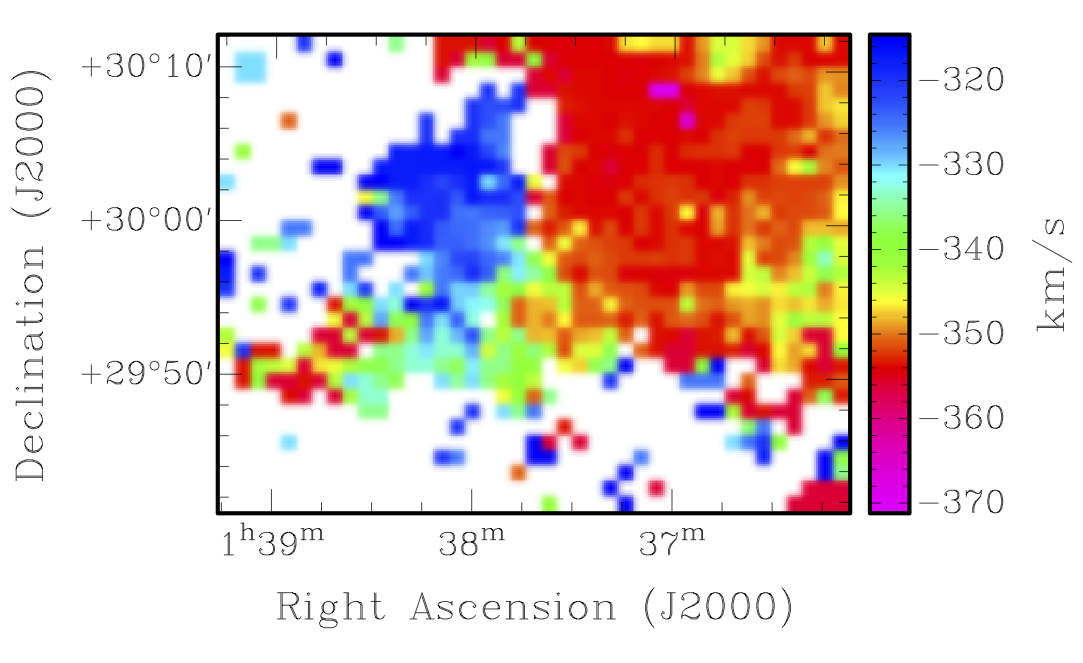}}
\newline
\centering
\subfloat[]{\includegraphics[width = 2.5in]{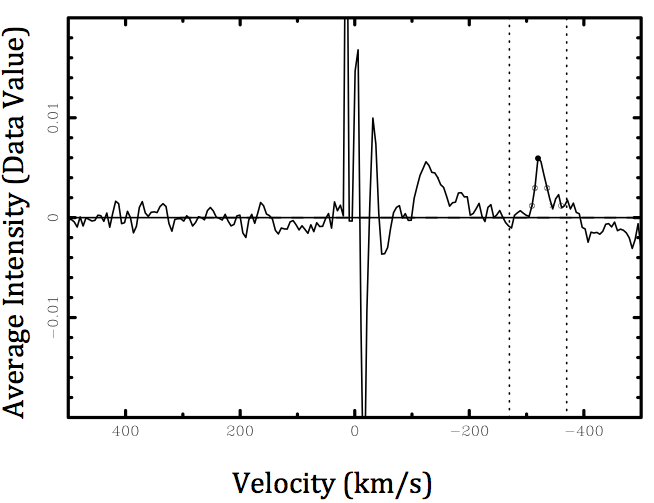}}
\caption{AGESM33-28 }
\label{ok28_all}
\end{figure}

\begin{figure}
\subfloat[]{\includegraphics[width = 3in]{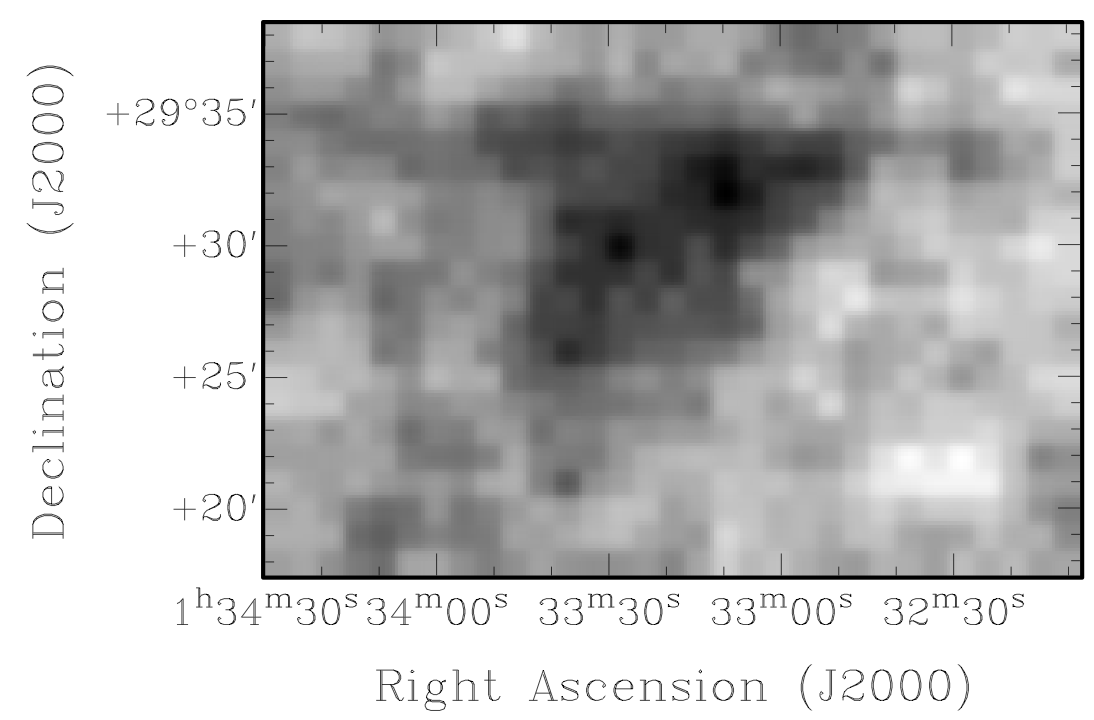}}\
\subfloat[]{\includegraphics[width = 3.4in]{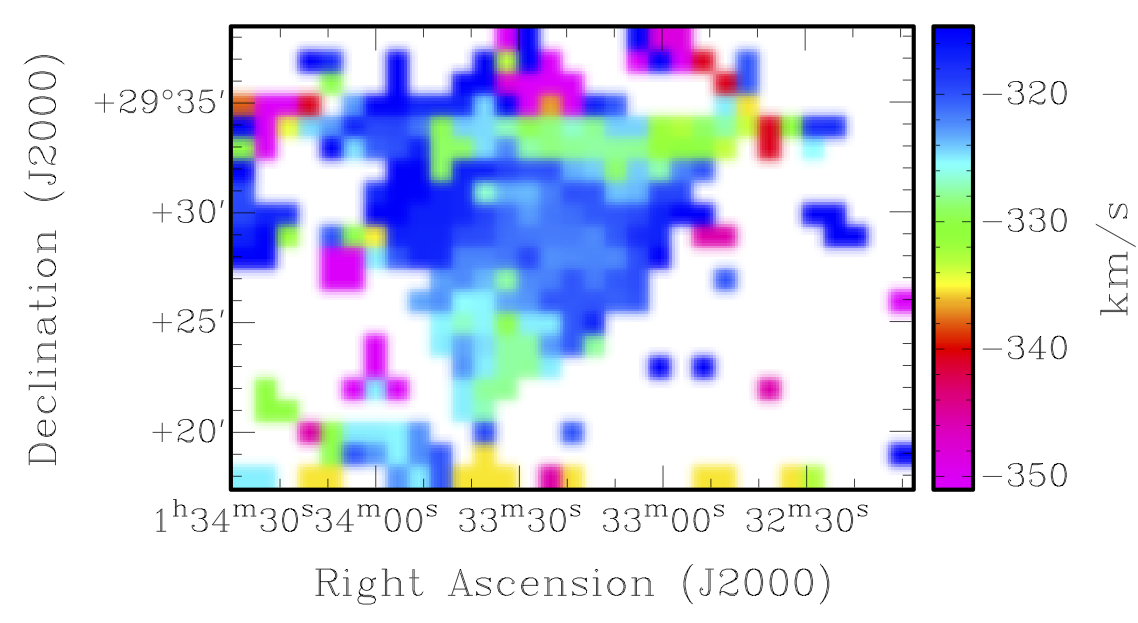}}
\newline
\centering
\subfloat[]{\includegraphics[width = 2.5in]{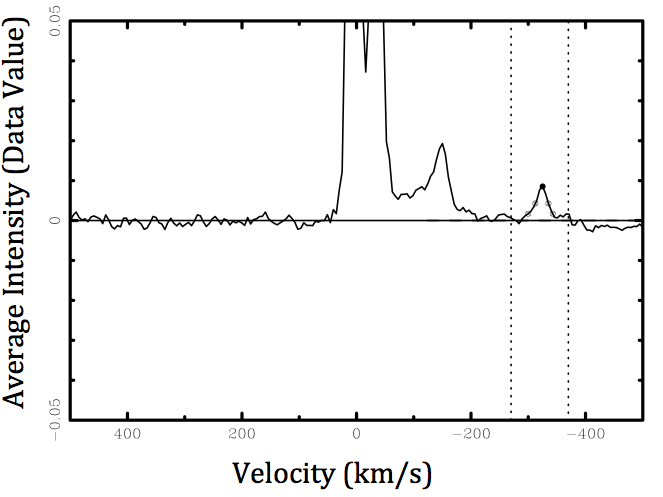}}
\caption{AGESM33-29 }
\label{ok29_all}
\end{figure}

\begin{figure}
\subfloat[]{\includegraphics[width = 3in]{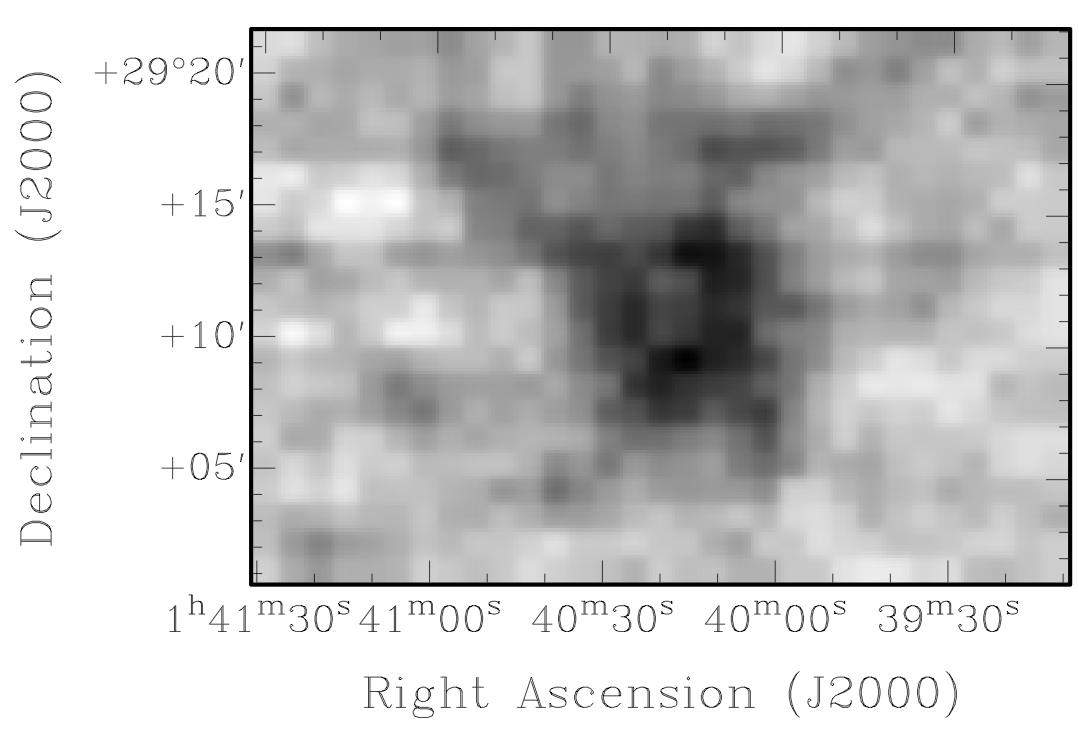}}\
\subfloat[]{\includegraphics[width = 3.4in]{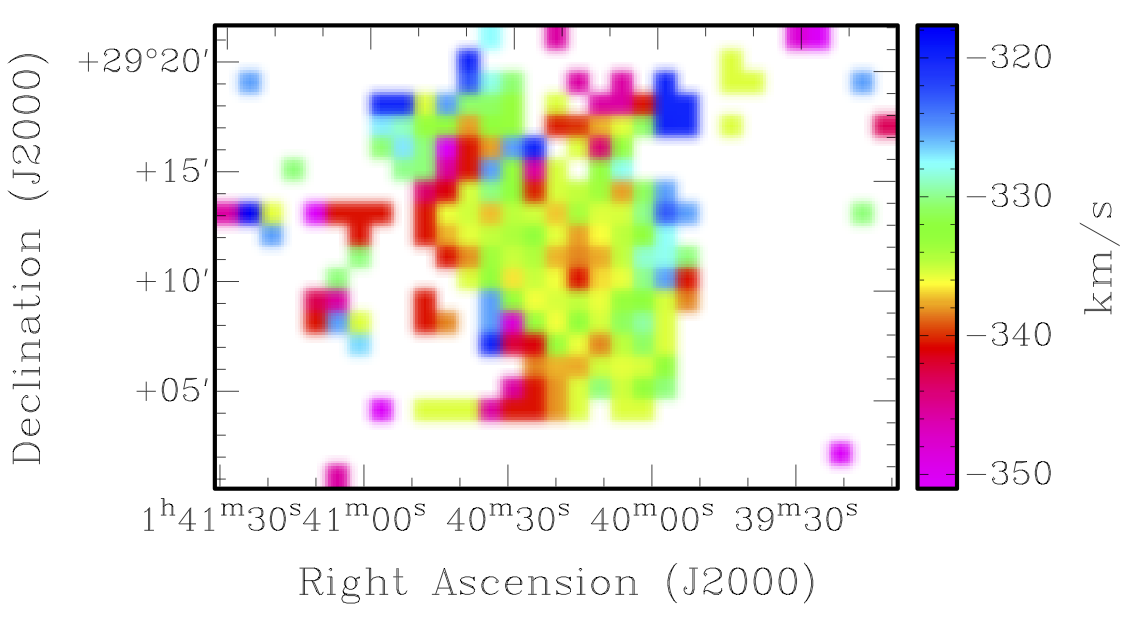}}
\newline
\centering
\subfloat[]{\includegraphics[width = 2.5in]{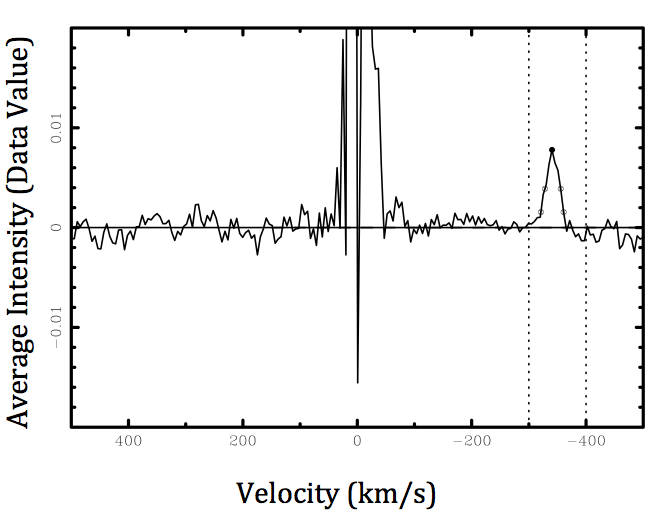}}
\caption{AGESM33-30 }
\label{ok30_all}
\end{figure}

\begin{figure}
\subfloat[]{\includegraphics[width = 2.5in]{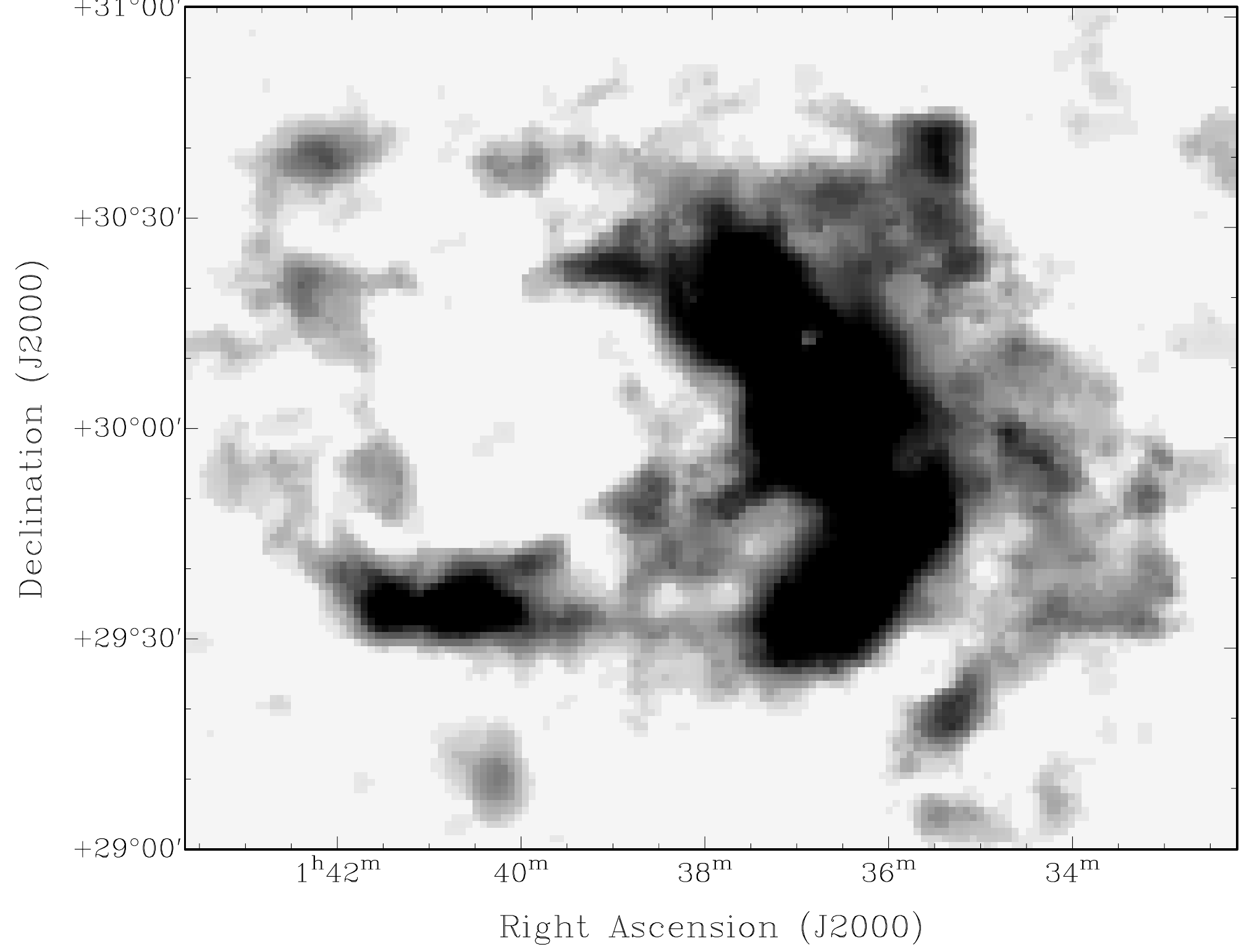}}\
\subfloat[]{\includegraphics[width = 3.4in]{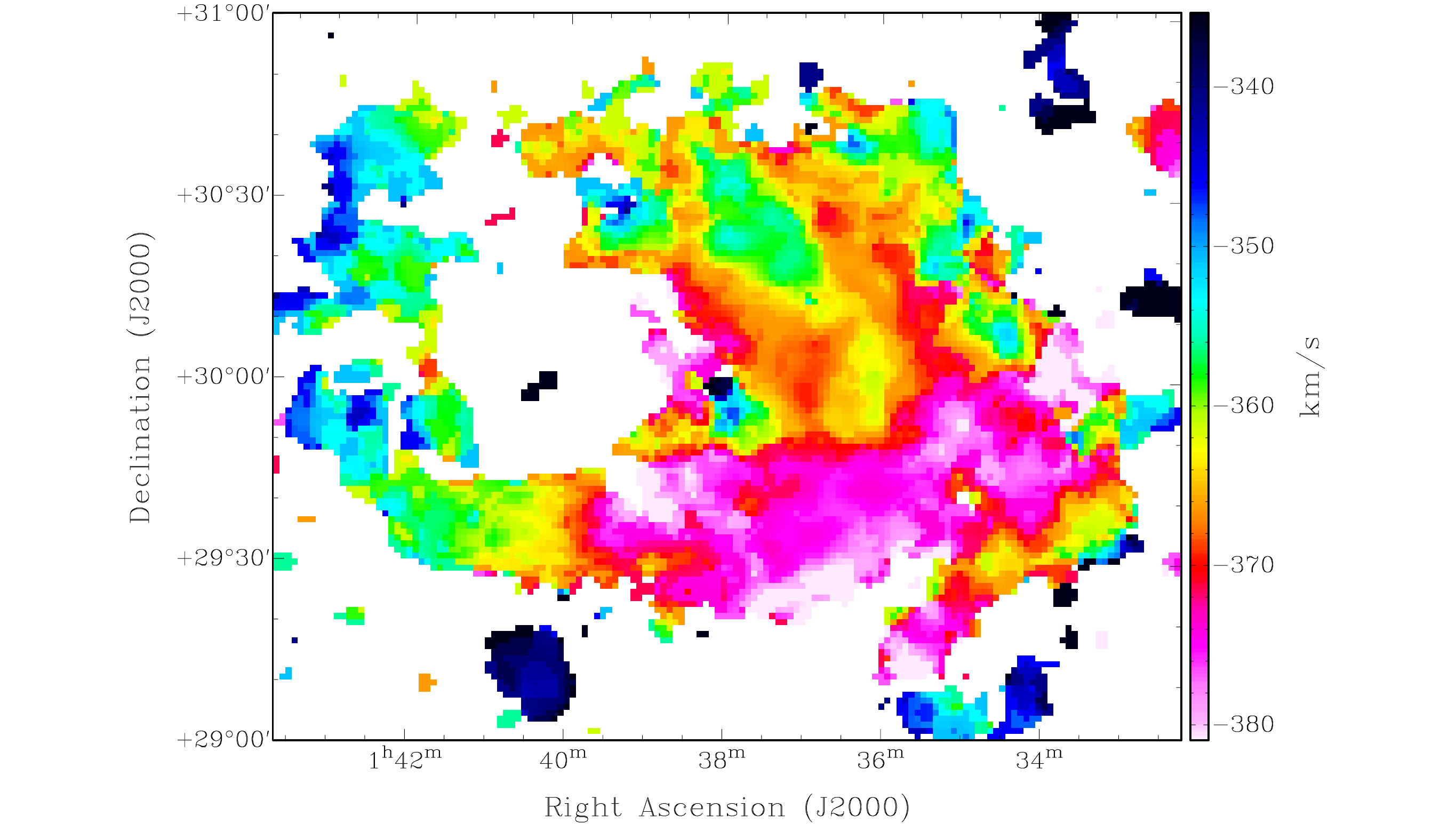}}
\newline
\centering
\subfloat[]{\includegraphics[width = 2.5in]{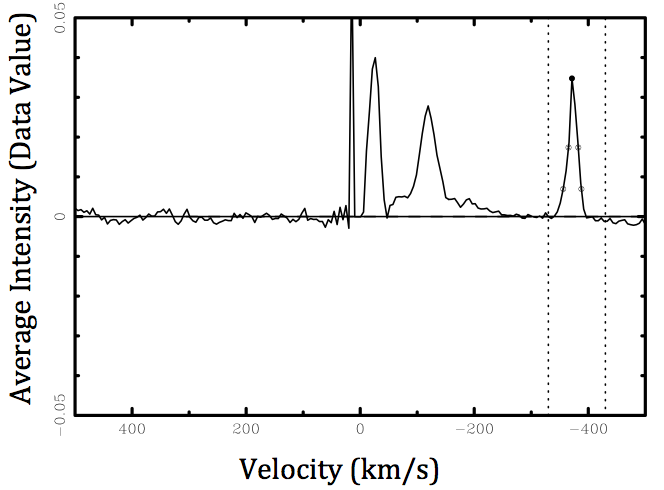}}
\caption{AGESM33-31 }
\label{ok31_all}
\end{figure}

\begin{figure}
\subfloat[]{\includegraphics[width = 3in]{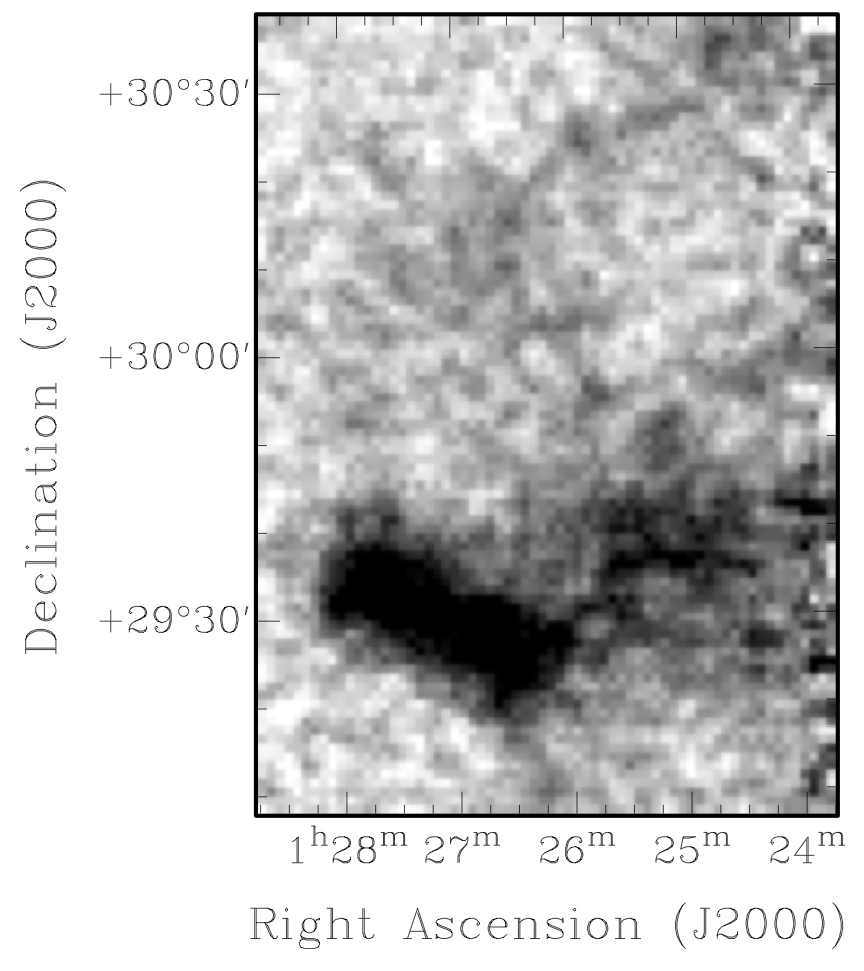}}\
\subfloat[]{\includegraphics[width = 3.4in]{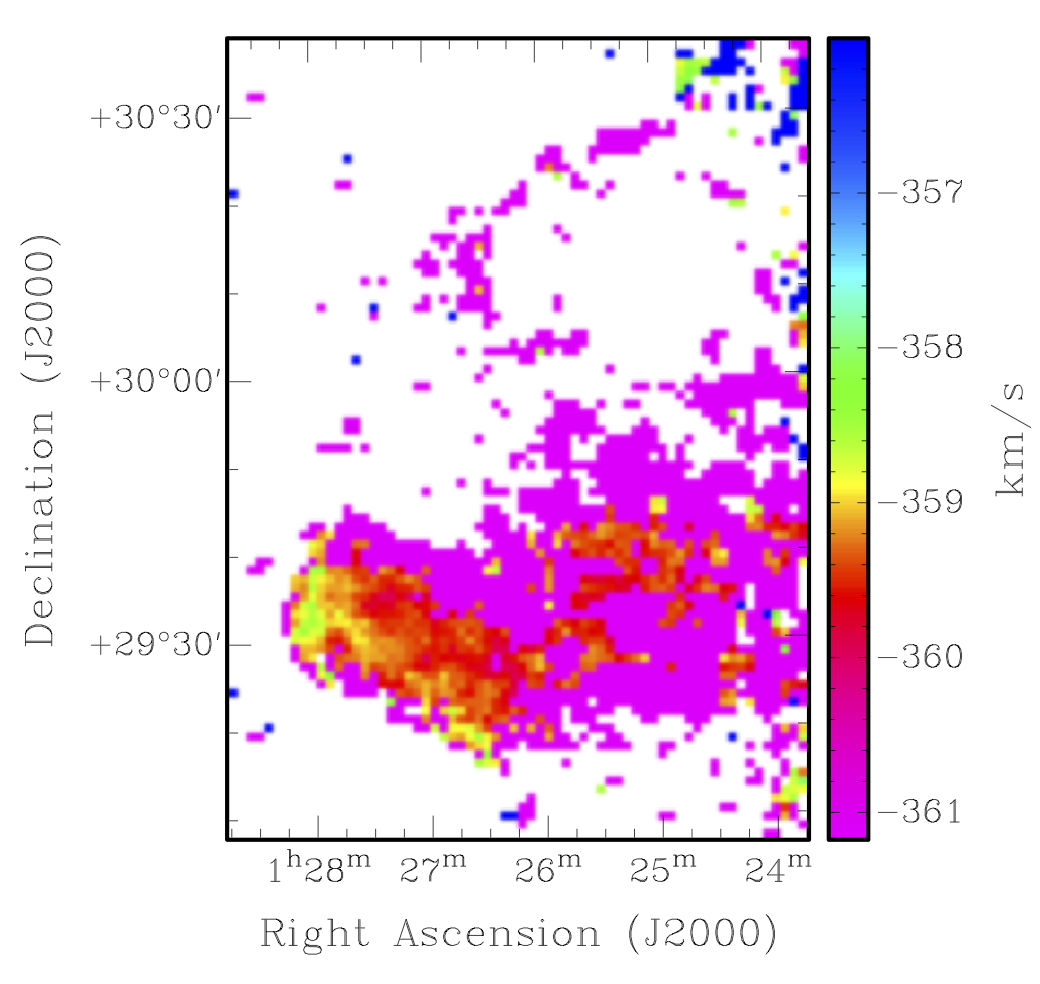}}
\newline
\centering
\subfloat[]{\includegraphics[width = 2.5in]{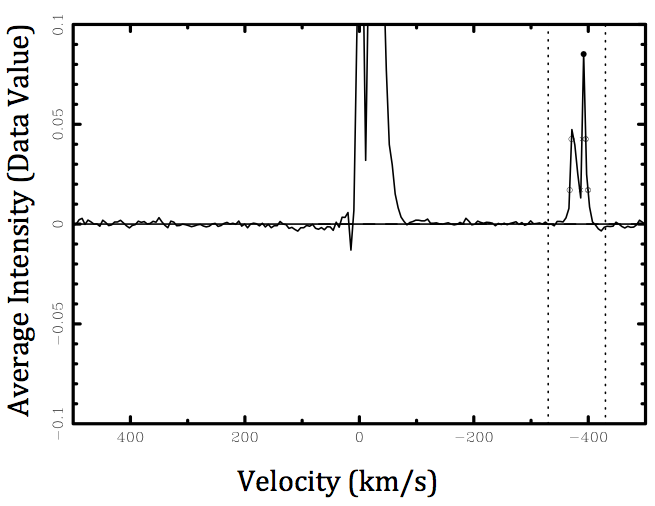}}
\caption{AGESM33-32/ Wright's Cloud }
\label{ok32_all}
\end{figure}

\end{document}